\documentclass[pre,preprintnumbers,showpacs,floatfix,aps]{revtex4}
\usepackage{amsmath}
\usepackage{amsfonts}
\usepackage{amssymb}
\usepackage{physymb}
\usepackage{graphicx}
\usepackage[hang,nooneline]{subfigure}

\newcounter{fig}

\begin{document}

\title{Dark-bright solitons in coupled nonlinear Schr\"odinger equations with unequal dispersion
coefficients}
\author{E. G. Charalampidis\footnote{
Email: charalamp@math.umass.edu}}
\affiliation{School of Civil Engineering, Faculty of Engineering, Aristotle University of
Thessaloniki, Thessaloniki 54124, Greece }
\affiliation{Department of Mathematics and Statistics, University of Massachusetts Amherst,
Amherst, Massachusetts 01003-4515, USA}
\author{P. G. Kevrekidis\footnote{%
Email: kevrekid@math.umass.edu}}
\affiliation{Department of Mathematics and Statistics, University of Massachusetts Amherst,
Amherst, Massachusetts 01003-4515, USA}
\author{D. J. Frantzeskakis\footnote{%
Email: dfrantz@phys.uoa.gr}}
\affiliation{Department of Physics, University of Athens, Panepistimiopolis, Zografos,
Athens 15784, Greece}
\author{B. A. Malomed\footnote{%
Email: malomed@post.tau.ac.il}}
\affiliation{Department of Physical Electronics, School of Electrical Engineering,
Faculty of Engineering, Tel Aviv University, Tel Aviv 69978, Israel }
\date{\today}

\begin{abstract}
We study a two-component nonlinear Schr{\"{o}}dinger system with equal, repulsive
cubic interactions and different dispersion coefficients in the two
components. We consider states that have a dark solitary wave in one
component. Treating it as a frozen one, we explore the possibility of the
formation of bright-solitonic structures in the other component. We identify
bifurcation points at which such states emerge in the bright component in
the linear limit and explore their continuation into the nonlinear regime.
An additional analytically tractable limit is found to be that of vanishing
dispersion of the bright component. We numerically identify regimes of
potential stability, not only of the single-peak ground state (the
dark-bright soliton), but also of excited states with one or more zero
crossings in the bright component. When the states are identified as
unstable, direct numerical simulations are used to investigate the outcome
of the instability development. Although our principal focus is on 
the homogeneous setting, we also briefly touch upon the counterintuitive
impact of the potential presence of a parabolic trap on the states of interest.
\end{abstract}


\maketitle

\section{Introduction}


Atomic Bose-Einstein condensates (BECs) \cite{book2a,book2} constitute a
platform that is ideal for the study of numerous nonlinear-wave phenomena
(see, e.g., reviews in \cite{emergent,revnonlin,rab,djf}). One of the
particularly interesting directions in that regard is the study of
multicomponent BEC systems and solitary waves in them. This is a subject of
broad interest, not only in the realm of atomic BECs, but also in nonlinear
optics~\cite{yuri} as well as in studies of integrable systems in
mathematical physics~\cite{ablowitz}. Arguably, one of the most intriguing
coherent structures in the multicomponent settings in the presence of
defocusing nonlinearities (in terms of optical systems) or repulsive
interatomic interactions (in BECs) are dark-bright (DB) solitons. In
particular, exact solutions for such states are available in the well-known
integrable self-defocusing two-component Manakov system~\cite{manakov}.
Generally, the DB solitons are ubiquitous in two-component systems, i.e.,
the nonlinear Schr{\"{o}}dinger equations (NLSEs) or Gross-Pitaevskii 
equations (GPEs), in which both the self-phase-modulation (SPM) and 
cross-phase-modulation (XPM) terms are represented by cubic terms.

In the DB structures, the customary dark soliton of the defocusing NLSE
induces an effective potential, via the XPM interaction, in the other
component, which in turn gives rise to self-trapping of bright-soliton
states in the latter component. This possibility has been studied
extensively in atomic BECs (see, e.g., Refs.~\cite%
{buschanglin,DDB,kanna,rajendran,val,berloff,VB,Alvarez,vaspra,vasnjp}). Such
waveforms have been reported in experiments both in two-component BEC
mixtures, which make use of two different atomic states of $^{87}$Rb \cite%
{sengdb,peterprl,peter1,peter2,peterpra,peter3}, and in nonlinear optics
\cite{seg1,seg2,seg3}. The BEC experimental studies have examined the
dynamics of a single DB soliton in a trap~\cite{sengdb,peter1}, the
generation of multiple DB solitons in a counterflow experiment~\cite%
{peterprl} (see also the theoretical work of Ref.~\cite{fot}), 
DB soliton interactions~\cite{peter2}, and the creation of SU(2)-rotated DB 
solitons, in the form of beating dark-dark solitons \cite{peterpra,peter3}.

Our aim in the present work is to extend this fundamental structural idea
for the existence of DB solitary waves beyond the previously studied
integrable or close-to-integrable limit of the Manakov model. In fact, the
nearly integrable setting has been especially useful and relevant to the
experiment due to the fact that the ratios of interspecies and intraspecies
interactions between different hyperfine atomic states of $^{87}$Rb in the
BEC mixtures ($|1,0\rangle $ and $|2,0\rangle $, as well as $|1,-1\rangle $
and $|2,-2\rangle $) are very close to unity 
\cite{sengdb,peterprl,peter1,peter2,peterpra,peter3}. The dispersion coefficients
in this setting are equal too, as they are determined by the same atomic
weight. It is relevant to note in passing that, quite recently~\cite{ourzb},
the studies of spin-orbit-coupled BECs~\cite{dalibard_rmp} have led to a
set of coupled GPEs (via a multiple-scale reduction scheme), where the
effective dispersion coefficients could differ substantially (and
controllably), as they depend on the curvature of the corresponding
dispersion relation of the two-component branches. A similar situation is in principle 
possible for a binary condensate loaded into a periodic spin-dependent
potential, in which case the effective mass may be altered differently by
the potential for two atomic states with different spins. 

The possibility of having different dispersion prefactors (which, of course,
are also different in the system of coupled GPEs corresponding to a
heteronuclear binary BEC) in the model producing the DB solitary waves is
the main motivation for the analysis reported below. In particular, if we
assume that the dark soliton in the one component induces an approximately
frozen effective potential in the other one (reserved for carrying a
bright-soliton structure), then varying the dispersion coefficient allows
us to modulate the depth and width of the effective potential. In so doing,
we can trap different bound modes, representing the ground state or excited
ones, at the level of the linear approximation for the bright component. The
analysis presented in Sec.~II allows us then to infer the value of the dispersion
coefficient along with the respective eigenvalue (the chemical potential) for
which such multiple states are accessible. Another intriguing case is the limit
of vanishing dispersion of the second component. Given the algebraic
[Thomas-Fermi (TF)] nature of the second equation in that limit, we can treat 
that case analytically and then test the solutions numerically (this is 
presented in Sec.~II as well). Based on these aspects of the analysis (the linear
and nonlinear TF limits), we then numerically examine the emergence of nonlinear
states from the predicted bifurcation points of the linear theory and their 
continuation (when possible, all the way to the zero-dispersion limit). Identifying
these solutions, we also explore their stability against small perturbations,
concluding, quite naturally, that higher excited states, i.e., bright solitary waves
with a larger number of nodes, are more prone to instability. For unstable states, 
we simulate the dynamical development of the instability, which often involves
mobility of the coherent structure, and possibilities of destruction of higher
excited states or their reshaping into lower ones. This computational analysis
is performed in Sec.~III. Finally, in Sec.~IV we summarize our findings and 
present conclusions and highlight directions for future studies.

The considerations reported in this paper provide us with a systematic means
for unveiling a whole series of previously unexplored families of solutions
in the two-component NLSE-GPE system. Actually, with the exception of the
fundamental DB soliton (i.e., the simplest among the considered states, with
a nodeless bright component), understanding of the existence and especially
stability of such states, as well as of their nonlinear dynamical behavior,
is presently 
very limited. It is therefore the purpose of this work to find out which
of these states are stable and in what parameter regions. For unstable
states, our intention is to reveal mechanisms through which the
instabilities manifest themselves, as well as eventual configurations into
which the unstable states are driven. These issues turn out to be by no
means trivial, involving both mobility and different scenarios of 
transformation into different types of robust configurations.

To conclude the Introduction, it is relevant to note that two-component
NLSEs with unequal dispersion coefficients give rise to other families of
unusual states, a known example being symbiotic bright solitary waves in
heteronuclear binary BECs \cite{victor}. They are supported by the interplay
of repulsive self-interactions and attractive cross interactions between the
components, which is essentially different from the setting considered in the
present work.

\section{The model and analytical considerations}


Given that the underlying model is relevant to both atomic BECs and
nonlinear optics, we present it in the general form of the coupled NLS
equations. 
To this end, we consider the coupled NLS system written in the following
dimensionless form:
\begin{subequations}
\begin{eqnarray}
i\partial _{t}{\Phi _{-}} &=&-\frac{D_{-}}{2}\partial _{xx}\Phi _{-}+\gamma
\left( g_{11}|\Phi _{-}|^{2}+g_{12}|\Phi _{+}|^{2}\right) \Phi _{-}+V(x)\,\Phi _{-}, \label{1} \\
i\partial _{t}{\Phi _{+}} &=&-\frac{D_{+}}{2}\partial _{xx}\Phi
_{+}+\gamma \left( g_{21}|\Phi _{-}|^{2}+g_{22}|\Phi
_{+}|^{2}\right) \Phi _{+}+V(x)\,\Phi _{+},  \label{2}
\end{eqnarray}
\label{start_manakov}
\end{subequations}
with dispersion coefficients $D_{\pm }$, nonlinearity strength $\gamma $, 
interaction coefficients $g_{jk}$ ($j,k=1,2$, with $g_{21}\equiv g_{12}$), 
and an external harmonic potential $V(x)$, of the form 
$V(x)=\frac{1}{2}\Omega^{2}x^{2}$, with normalized trap strength $\Omega$.
In BECs of different species, $D_{\pm }$ play the role of the
inverse masses, while in the spin-orbit BEC they may be associated with the
local curvature of different branches of the dispersion relation \cite{ourzb}. 
Fields $\Phi _{\pm }=\Phi _{\pm }(x,t)$ in Eqs.~(\ref{1}) and (\ref{2}) will
carry the dark (denoted by a minus subscript) and bright (denoted by a plus 
subscript) soliton components, respectively. We fix $g_{jk}=1$ for all $j,k=1,2$ 
(motivated, as indicated above, by the actual values of the interaction coefficients
for the BEC mixtures in $^{87}$Rb) and $D_{-}=\gamma =1$, which is always 
possible upon rescaling, defining $D_{+}\equiv D\,(\geq 0)$. Stationary solutions
to Eqs.~(\ref{1}) and (\ref{2})
with chemical potentials $\mu _{\pm}$ are found using the ansatz $%
\Phi _{\pm }(x,t)=\phi _{\pm }(x)\exp (-i\mu _{\pm }t)$, where $\phi _{\pm
}(x)$ are real-valued functions. Then Eqs.~(\ref{1}) and (\ref{2}) reduce 
to the coupled system of stationary equations
\begin{subequations}
\begin{eqnarray}
\mu _{-}\phi _{-} &=&-\frac{1}{2}\left( \phi _{-}\right) ^{\prime \prime
}+\left( \phi _{-}^{2}+\phi _{+}^{2}\right) \phi _{-}+V(x)\,\phi _{-},
\label{-} \\
\mu _{+}\phi _{+} &=&-\frac{D}{2}\left( \phi _{+}\right) ^{\prime \prime
}+\left( \phi _{-}^{2}+\phi _{+}^{2}\right) \phi _{+}+V(x)\,\phi _{+},
\label{+}
\end{eqnarray}
\label{coupled_manakov}
\end{subequations}
with the prime standing for $d/dx$. In the majority of cases studied below, 
we will be considering both Eqs.~(\ref{start_manakov}) and (\ref{coupled_manakov})
in the absence of the trapping potential; thus we set $V(x)=0$ from now on (unless 
explicitly noted otherwise). 

In the absence of the bright component, i.e., $\phi _{+}=0$, an obvious
dark-soliton solution of Eq.~(\ref{-}) is
\begin{equation}
\phi _{-}(x)=\sqrt{\mu _{-}}\tanh \left( \sqrt{\mu _{-}}x\right).
\label{dark}
\end{equation}%
With this solution playing the role of the background for the weak bright
component $\phi _{+}$, the linearized form of Eq.~(\ref{+}) for given $%
\mu _{-}$ amounts to an eigenvalue problem
\begin{equation}
\mathcal{L}\,\phi _{+}=\lambda \,\phi _{+},  \label{quantum}
\end{equation}%
where $\displaystyle{\mathcal{L}=\frac{D}{2}}\frac{d^{2}}{dx^{2}}{+\mu _{-}%
\mathrm{sech}^{2}\left( \sqrt{\mu _{-}}x\right) }$ is a linear operator and
$\left( \lambda ,\phi _{+}\right) $ is the eigenvalue-eigenvector pair with $%
\lambda \equiv \mu _{-}-\mu _{+}$. According to commonly known results from
quantum mechanics \cite{LL} for the respective P{\"{o}}schl-Teller potential~%
\cite{teller}, Eq.~(\ref{quantum}) gives rise to bound states of order $n$ ($%
n=0$ corresponds to the ground spatially even state, $n=1$ to the first odd
state, etc.) that exist under the following condition:
\begin{equation}
D<D_{\mathrm{crit}}^{(n)}=\frac{2}{n\left( 1+n\right)}.  \label{D}
\end{equation}%
In particular, the ground state is always present, while the first odd state
exists at $D<1$, the first excited even state ($n=2$) exists at $D<1/3$, the
next excited odd state ($n=3$) exists at $D<1/6$, and so on. This feature was fully
confirmed by our numerical computations [see, in particular, the range of $D$ 
considered below in Figs.~\ref{fig1}-\ref{fig4}, which is in accordance with 
the bound given by Eq.~(\ref{D}) for $\mu _{-}=1$].

It can thus be expected that nonlinear solutions corresponding to the ground
and excited states in the linear limit bifurcate at these critical values of
$D$ with the corresponding eigenvalues $\mu _{+}$ of the linear problem (\ref%
{quantum}); on the other hand, $\mu _{-}$ is a given amplitude of the
background for the dark soliton, which is set to be $\mu _{-}=1$ in our
numerical computations below.

The present calculation and its connection to the P{\"{o}}schl-Teller
potential also provides a lower bound for the values of $\mu _{+}$ in the
nonlinear system. In particular, considering the known properties of the
exact solution of the linear problem~\cite{LL}, nonlinear states may exist
above the level of the chemical potentials
\begin{equation}
\mu _{+}=\mu _{-}\left[ 1-\frac{D}{8}\left( \sqrt{1+\frac{8}{D}}%
-(2n+1)\right) ^{2}\right].
\label{landau_limit}
\end{equation}%
This bound has been also fully confirmed in our numerical
computations discussed below.
Furthermore, these computations demonstrate that, with the increase of $\mu
_{+}$, the bright component grows wider, progressively 
expanding to the size of the
computational domain. The latter determines the upper bound of
$\mu_+$ values considered herein.

As explained in the Introduction, $D=0$ is an additional case that can be
treated analytically. In this case, Eqs. (\ref{-}) and (\ref{+}) become
\begin{subequations}
\begin{gather}
\frac{1}{2}\phi _{-}^{\prime \prime }+\left( \mu _{-}-\mu _{+}\right) \phi
_{-}=0,~~\phi _{+}^{2}(x)=\mu _{+}-\phi _{-}^{2}(x)~~\mathrm{at}~~\phi
_{-}^{2}<\mu _{+},  \label{<} \\
\mu _{-}\phi _{-}=-\frac{1}{2}\phi _{-}^{\prime \prime }+\phi _{-}^{3},~~\phi
_{+}=0~~\mathrm{at}~~\phi _{-}^{2}>\mu _{+},  \label{>}
\end{gather}%
which resemble the TF approximation~\cite{book2a,book2,emergent} for $\phi
_{-}$ in the context of atomic BECs, with the difference that the role of
the potential here is played by the component $\phi _{+}$. Solutions to Eqs.~%
(\ref{<}) and (\ref{>}) exist for $\mu _{-}>\mu _{+}$, like in the case of
Eq. (\ref{quantum}). The {odd TF-like solutions} are built as
\end{subequations}
\begin{eqnarray}
\phi _{-}(x) &=&\phi _{0}\sin \left[ \sqrt{2\left( \mu _{-}-\mu _{+}\right) }%
x\right],~~\phi _{+}^{2}(x)=\mu _{+}-\phi _{-}^{2}(x)~~\mathrm{at}~~|x|<\xi,
\notag \\
\phi _{-}(x) &=&\mathrm{sgn}(x)\sqrt{\mu _{-}}\tanh \left[ \sqrt{\mu _{-}}%
\left( |x|-x_{0}\right) \right],~~\phi _{+}=0~~\mathrm{at}~~|x|>\xi,
\label{odd}
\end{eqnarray}%
with constants $\phi _{0}$, $\xi $, and $x_{0}$ determined by the three
matching conditions
\begin{eqnarray}
\phi _{0}\sin \left[\sqrt{2\left( \mu _{-}-\mu _{+}\right) }\xi \right] &=&%
\sqrt{\mu _{+}},  \notag \\
\sqrt{\mu _{-}}\tanh \left[\sqrt{\mu _{-}}\left( \xi -x_{0}\right) \right]
&=&\sqrt{\mu _{+}},  \label{sin} \\
\phi _{0}\sqrt{2\left( \mu _{-}-\mu _{+}\right) }\cos \left[\sqrt{2\left(
\mu _{-}-\mu _{+}\right) }\xi \right] &=&\frac{\mu _{-}}{\cosh ^{2}\left[
\sqrt{\mu _{-}}\left( \xi -x_{0}\right) \right] }.  \notag
\end{eqnarray}%
An exact analytical solution to Eq. (\ref{sin}) can be found as
\begin{eqnarray}
\phi _{0} &=&\sqrt{\frac{1}{2}\left( \mu _{-}+\mu _{+}\right) },  \notag \\
\xi -x_{0} &=&\left( 1/\sqrt{\mu _{-}}\right) \atanh\left( \sqrt{\frac{\mu
_{+}}{\mu _{-}}}\right),  \label{arcsin} \\
\xi &=&\frac{1}{\sqrt{2\left( \mu _{-}-\mu _{+}\right) }}\left[ \asin\left(
\sqrt{\frac{2\mu _{+}}{\mu _{\_}+\mu _{+}}}\right) +2\pi n\right],  \notag
\end{eqnarray}%
where $n=0$ represents the single DB soliton, while higher values of $n$
correspond to a larger number of DB solitons, e.g., $n=1$ corresponds to five solitons, etc.
Via this approach, exact analytical formulas can be derived for various
branches of solutions at $D=0$ (although, given the cumbersome nature of the
formulas, we will not discuss the corresponding higher-order analytical
expressions here).

\section{The Computational Analysis}


\subsection{Numerical methods}


Throughout this section, numerical results are presented for the coupled NLS
system (\ref{start_manakov}). Our investigation addresses three basic issues:
existence, stability and dynamical evolution. The first two are considered by 
performing one-parameter continuation of steady-state solutions, varying chemical
potential $\mu _{+}$, for different values of the dispersion coefficient $D$. 
When the solutions are found to be unstable (stable), their dynamical evolution
is monitored by means of direct numerical simulations to explore (corroborate)
the outcome of the instability development (stable relaxation).

Initially, we employ a one-dimensional uniform spatial grid consisting of $N$
points labeled by $x_{j}=-L+2jL/(N+1)$ and $j=1,\dots ,N$ with lattice
spacing (resolution) $\delta x$ and half-width $L$. The left and right
boundary points are located at $j=0$ and $j=N+1$, respectively. In all the
cases studied herein we fix $\delta x=0.1$ and $L=30$ (except for the
evolution of the first excited symmetric state with $D=0.25$ and $\mu
_{+}=0.99$, where we use $L=60$). In this way, both fields $\phi _{\pm }(x)$
and $\Phi _{\pm }(x,t)$ are replaced by their discrete counterparts on the
spatial grid, i.e., $\phi _{j,\pm }=\phi _{\pm }(x_{j})$ and $\Phi _{j,\pm
}(t)=\Phi _{\pm }(x_{j},t)$, respectively. Then the second-order spatial
derivatives in Eqs.~(\ref{start_manakov}) and (\ref{coupled_manakov}) [as well
as in Eqs.~(\ref{A11}) and (\ref{A33}) in the Appendix] are
replaced by second-order central-difference formulas. Finally, the no-flux
boundary conditions are applied at the edges of the spatial grid, i.e., $%
\partial _{x}\phi _{\pm }|_{x=\pm L}=0$ and $\partial _{x}\Phi _{\pm
}(t)|_{x=\pm L}=0$, for all $t$. The latter conditions are incorporated into
the internal discretization scheme using first-order forward and backward
difference formulas at the left and right boundaries, respectively. Thus,
the no-flux boundary conditions are enforced by explicitly requiring $\phi
_{0,\pm }=\phi _{1,\pm }$ and $\phi _{N+1,\pm }=\phi _{N,\pm }$, as well as $%
\Phi _{0,\pm }(t)=\Phi _{1,\pm }(t)$ and $\Phi _{N+1,\pm }(t)=\Phi _{N,\pm
}(t)$.

As far as the existence part is concerned, steady-state solutions to 
Eqs.~(\ref{-}) and (\ref{+}) are identified by employing a Newton-Krylov method
\cite{Kelley_nsoli}, together with a suitable initial guess in order to
ensure convergence. To that end, our starting point is the eigenvalue
problem~(\ref{quantum}), which is solved numerically. In particular, we focus
on a bound state of order $n$ and pick a value of $D$ satisfying the
threshold condition (\ref{D}). Then we determine the value of $\mu _{+}$
corresponding to one of the lowest eigenvalues $\lambda $ (e.g., the bound
state of order $n=0$ corresponds to the lowest eigenvalue $\lambda$, the
bound state of order $n=1$ to the second lowest eigenvalue $\lambda$, and
so on)
and the corresponding eigenvector (or bright component) $\phi _{+}$ is obtained
afterward. As a result, the ``seed," which is fed into the nonlinear
solver, consists, essentially, of the pair $(\mu _{+},\phi _{+})$ together
with the dark component $\phi _{-}$ given by Eq.~(\ref{dark}). Next we
trace steady-state solutions, for a given value of dispersion coefficient $D$, 
by performing a single-parameter numerical continuation with respect to
chemical potential $\mu _{+}$. Our approach is based on the sequential continuation
method, i.e., using the solution for given $\mu_{+} $, found by the nonlinear 
solver, as the seed for the next continuation step. We corroborate our 
results using the pseudo-arc-length continuation method (see, for instance, Ref. \cite%
{ps_method} and references therein), while numerical results obtained by
means of the sequential method are reported throughout this section.


We investigate the stability of the steady states obtained by the Newton
solvers at each continuation step, using linearized equations for small
perturbations (see the Appendix). In particular, an eigenvalue problem 
[cf. Eq.~(\ref{eig_prob})] is obtained (at order $\varepsilon$) and solved
numerically afterward. The steady state is classified as a stable one if 
none of the eigenfrequencies $\omega =\omega_{r}+i\,\omega _{i}$ has a 
non-vanishing imaginary part $\omega _{i}$. Two types of instabilities
can be thus identified: (i) exponential instabilities characterized by a
pair of imaginary eigenfrequencies with \textit{zero} real part and (ii)
oscillatory instabilities characterized by a complex eigenfrequency quartet.

Finally, the spectral stability results obtained from the eigenvalue problem
are checked against direct dynamical evolution of the coupled NLS system~%
(\ref{start_manakov}) forward in time. To this end, the Dormand-Prince
method with an automatic time-step adaptation procedure (see the Appendix 
in Ref. \cite{Hairer}) and tolerance $10^{-13}$ is employed. We have also 
corroborated our results using the standard fourth-order Runge-Kutta method
with a fixed time step of $\delta t=10^{-4}$, although numerical results are
presented throughout this section using only the Dormand-Prince method. Thus, 
we initialize the dynamics at $t=0$ using the available steady states, with 
two distinct initialization approaches. We initialize the dynamics under the 
presence of a small (uniformly distributed) random perturbation with amplitude 
$10^{-3}$ for the class of stable steady states. An alternative approach is to
initialize the
dynamics using the linearization ansatz~(\ref{lin_ansatz}) %
for the unstable solutions with $\varepsilon =10^{-2}$ (except for the first
excited antisymmetric steady state with $D=0.2$ and $\mu _{+}=0.77$, where
$\varepsilon =10^{-1}$ is used) and eigenvector $\mathcal{V}$ corresponding
to the (complex) eigenfrequency having the largest imaginary part. The latter
approach is useful towards seeding the relevant instability and observing the
ensuing dynamics.

\subsection{Numerical results}

\begin{figure}[!pht]
\begin{center}
\vspace{-0.5cm}
\mbox{\hspace{-0.1cm}
\subfigure[][]{\hspace{-0.3cm}
\includegraphics[height=.18\textheight, angle =0]{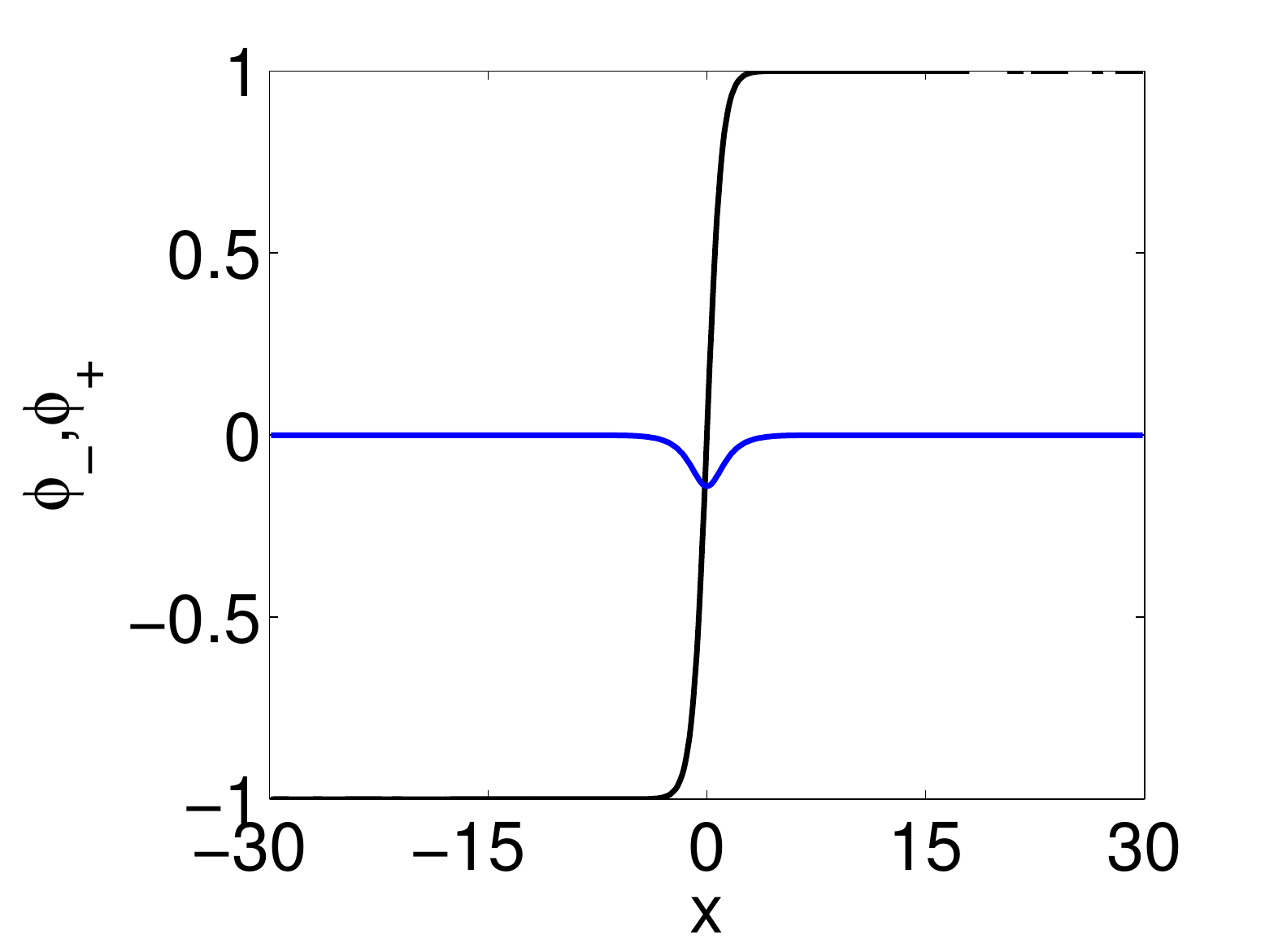}
\label{fig1a}
}
\subfigure[][]{\hspace{-0.3cm}
\includegraphics[height=.18\textheight, angle =0]{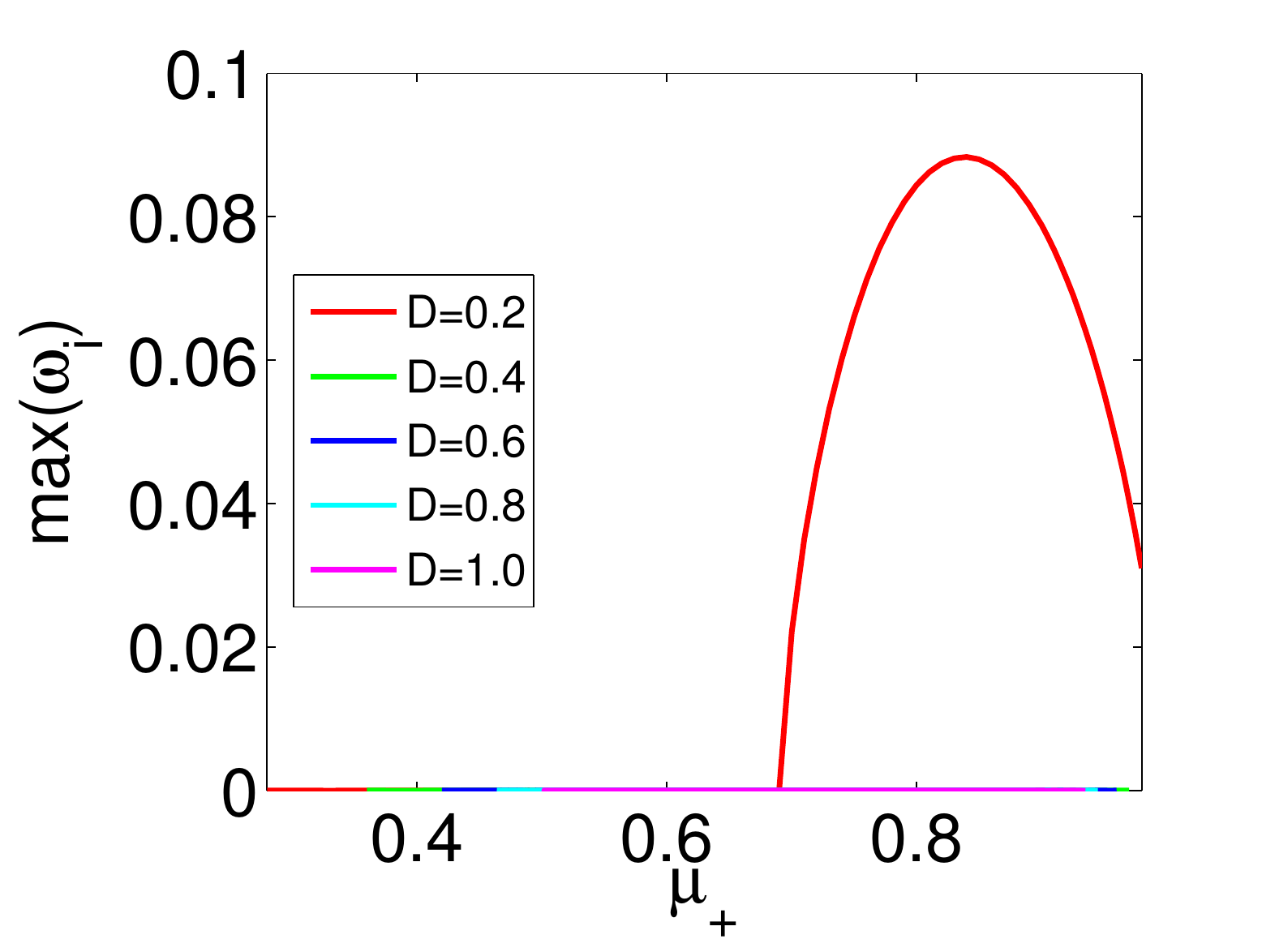}
\label{fig1b}
}
} \vspace{-0.17cm}
\mbox{\hspace{-0.1cm}
\subfigure[][]{\hspace{-0.3cm}
\includegraphics[height=.18\textheight, angle =0]{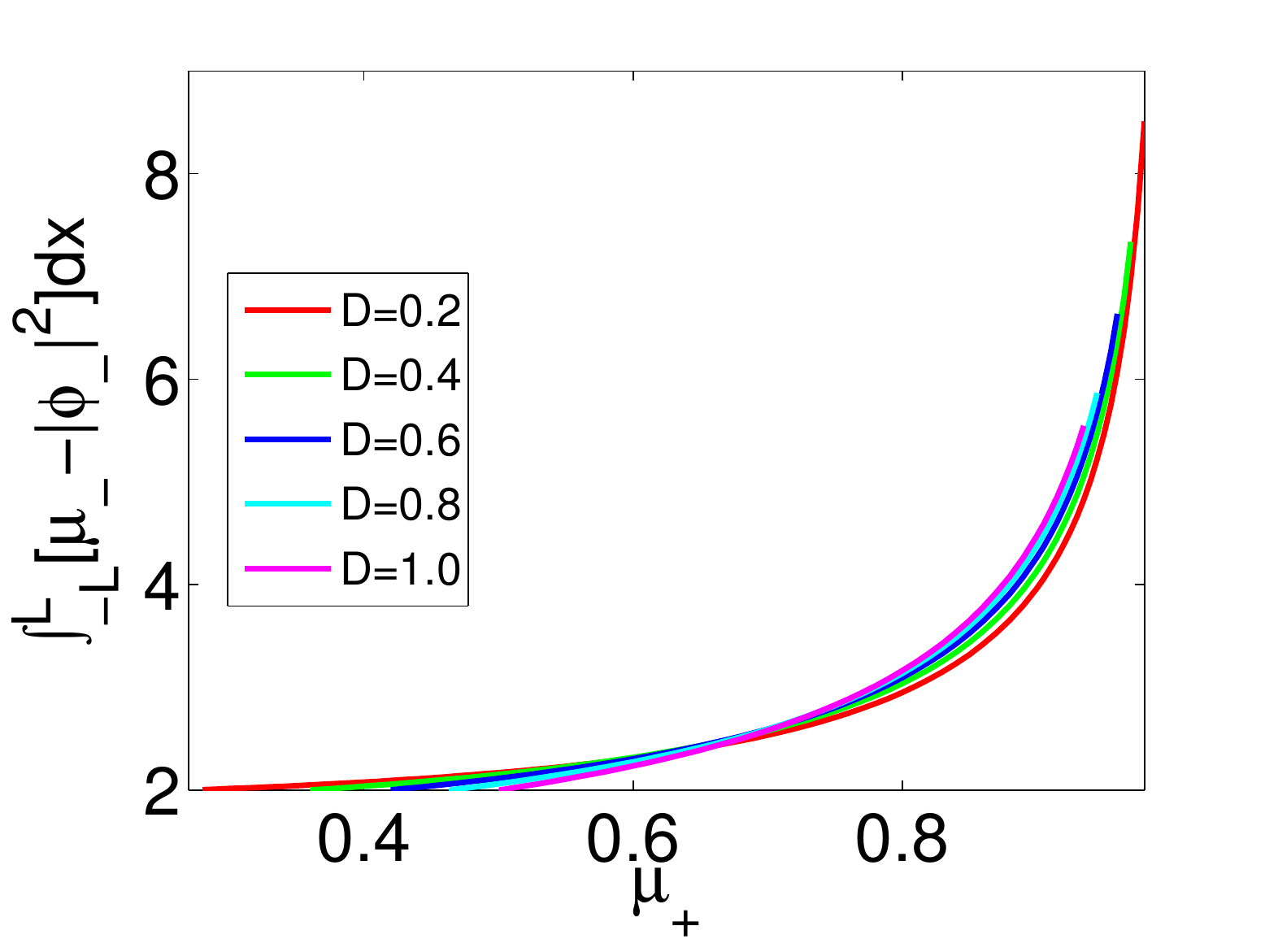}
\label{fig1c}
}
\subfigure[][]{\hspace{-0.3cm}
\includegraphics[height=.18\textheight, angle =0]{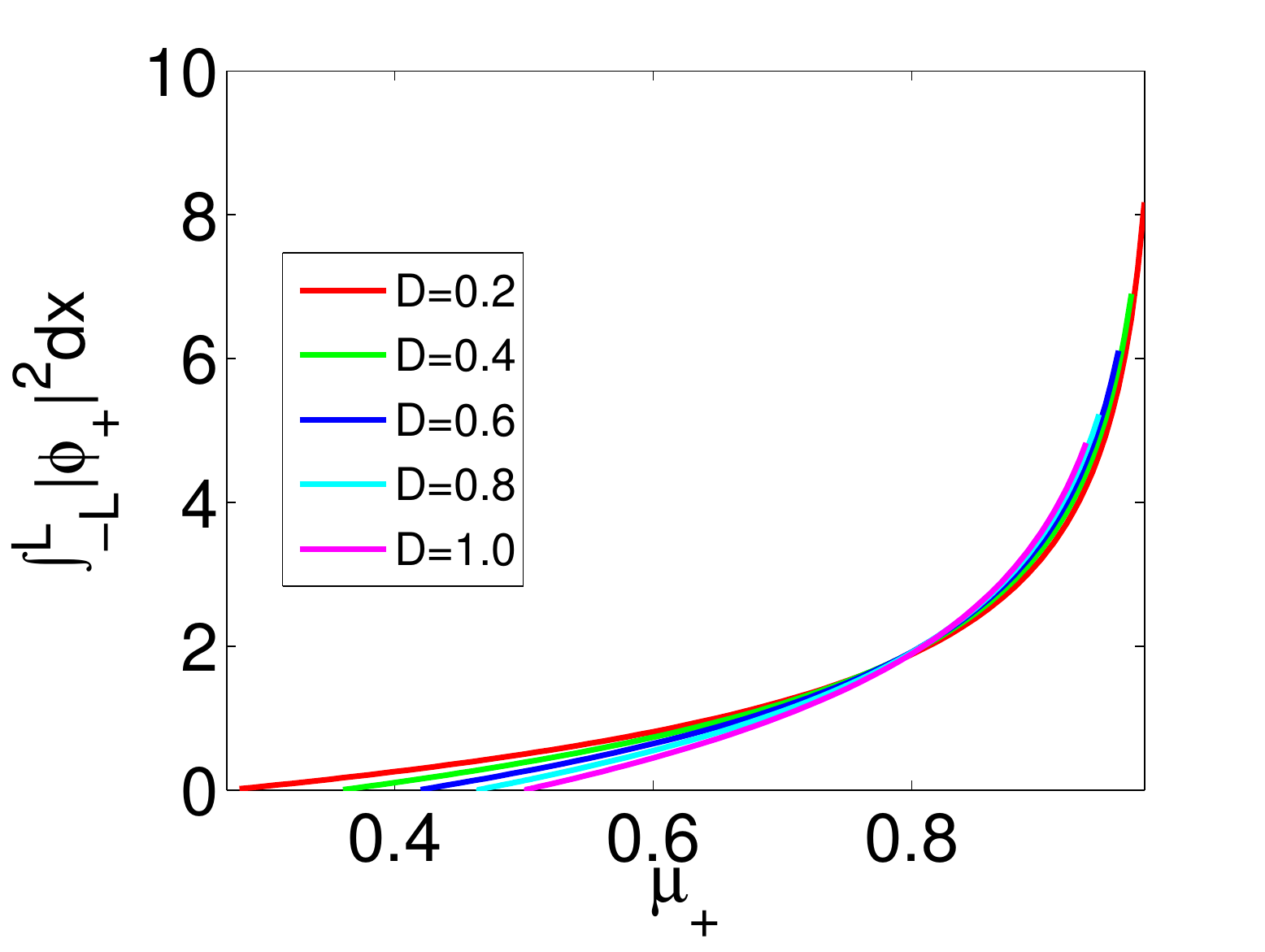}
\label{fig1d}
}
} \vspace{-0.17cm}
\mbox{\hspace{-0.1cm}
\subfigure[][]{\hspace{-0.3cm}
\includegraphics[height=.18\textheight, angle =0]{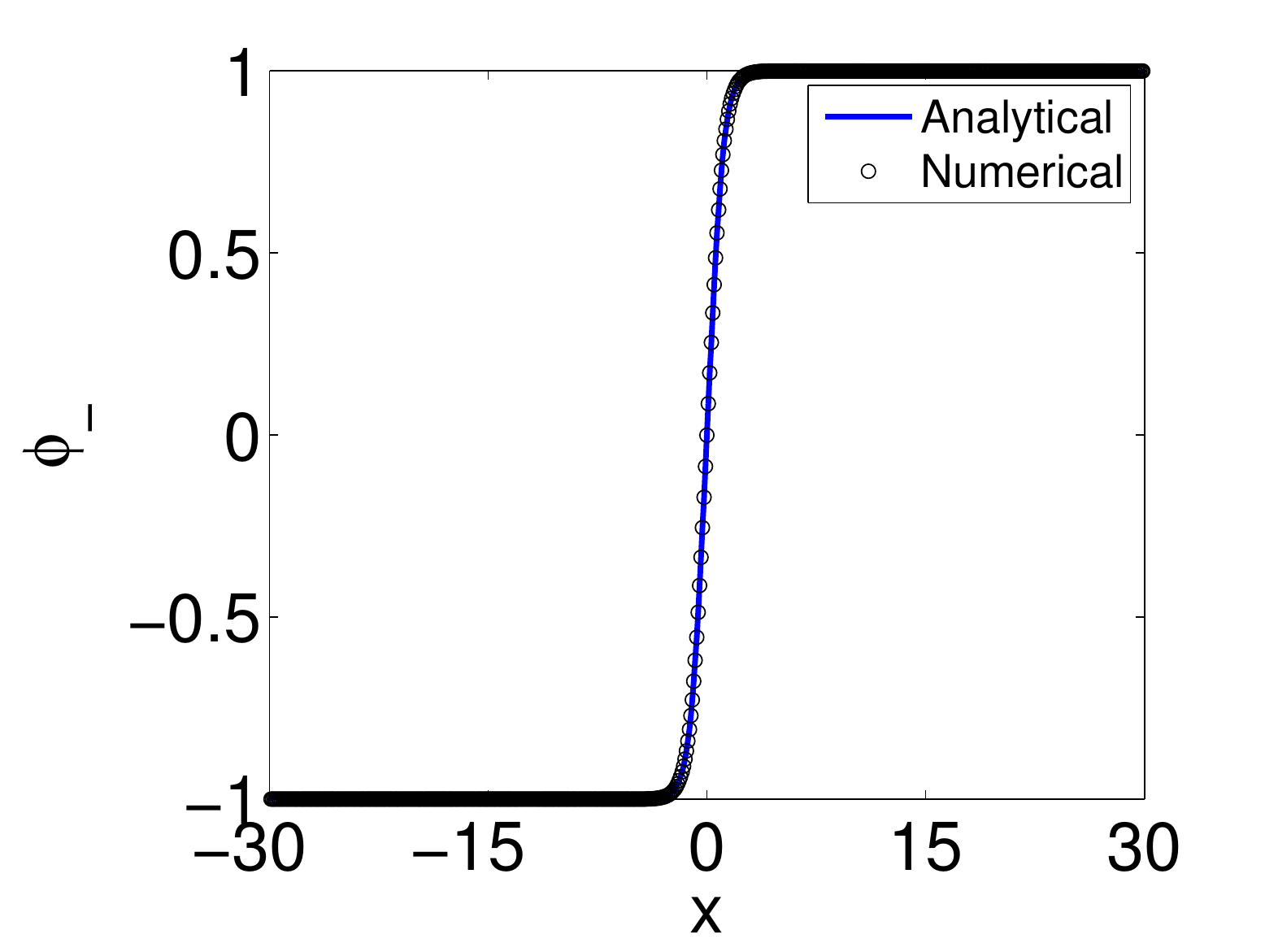}
\label{fig1e}
}
\subfigure[][]{\hspace{-0.3cm}
\includegraphics[height=.18\textheight, angle =0]{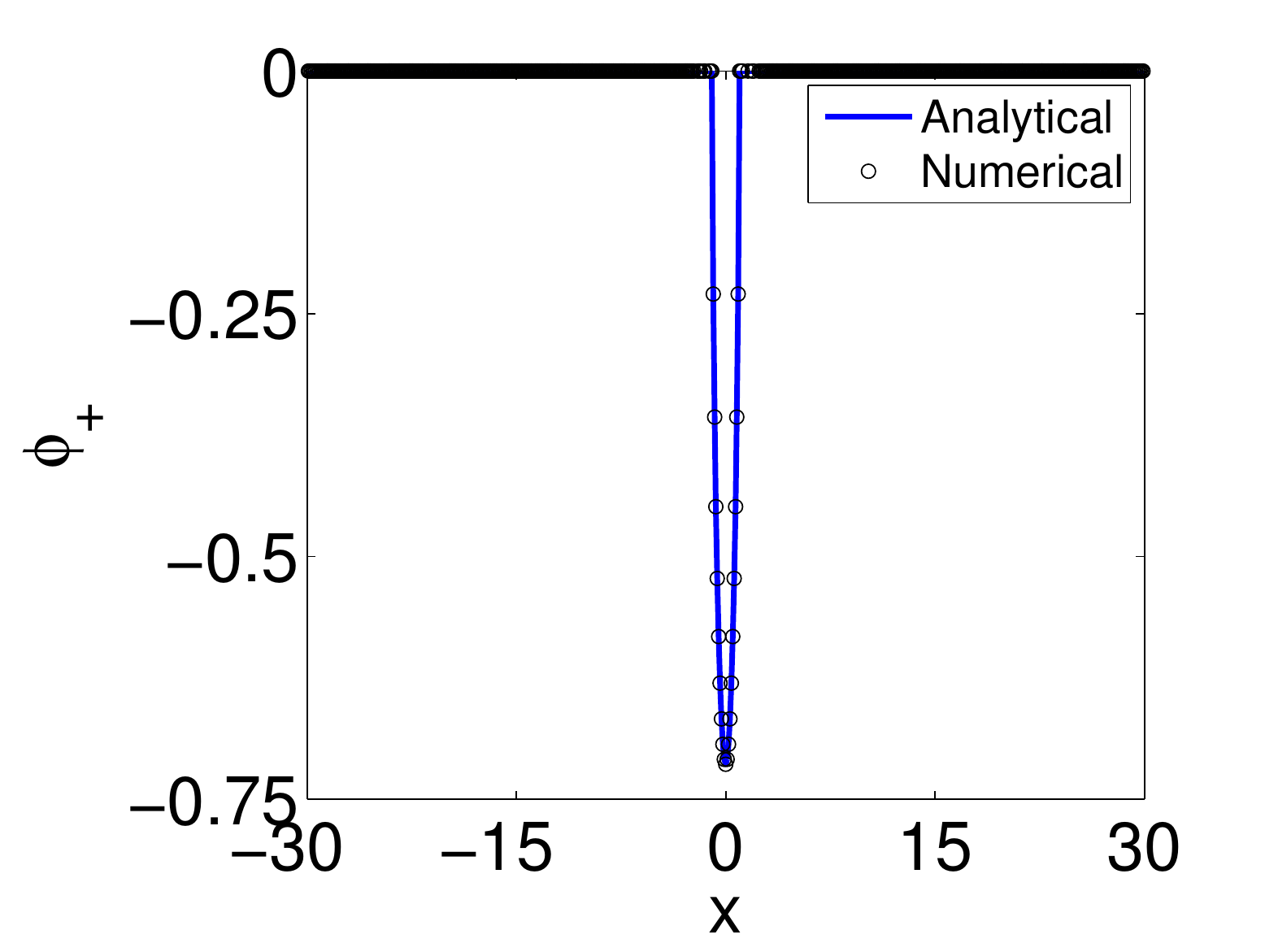}
\label{fig1f}
}
}
\end{center}
\par
\vspace{-0.55cm}
\caption{(Color online) Bound states and continuation results corresponding
to $n=0$ (i.e., the ground states). (a) Steady-state profiles of the dark 
(black line) and bright (blue line) soliton solutions for the parameters 
$D=1$ and $\protect\mu _{+}=0.51$. (b) Maximal imaginary eigenfrequency as 
a function of the continuation parameter $\protect\mu_{+}$ at various fixed
values of $D$. (c) and (d) Solution norms, i.e., integral powers 
(\protect\ref{I}) associated with the solution branches, as functions of the
continuation parameter $\protect\mu_{+}$ at various fixed values of $D$. (e) 
Dark and (f) bright soliton solutions analytically predicted by Eq. (\protect\ref{odd})
(blue line) and numerically obtained (black circles) for $D=0$ and $%
\protect\mu _{+}=0.51$.}
\label{fig1}
\end{figure}
\begin{figure}[tp]
\begin{center}
\vspace{-0.1cm}
\mbox{\hspace{-0.1cm}
\subfigure[][]{\hspace{-0.3cm}
\includegraphics[height=.18\textheight, angle =0]{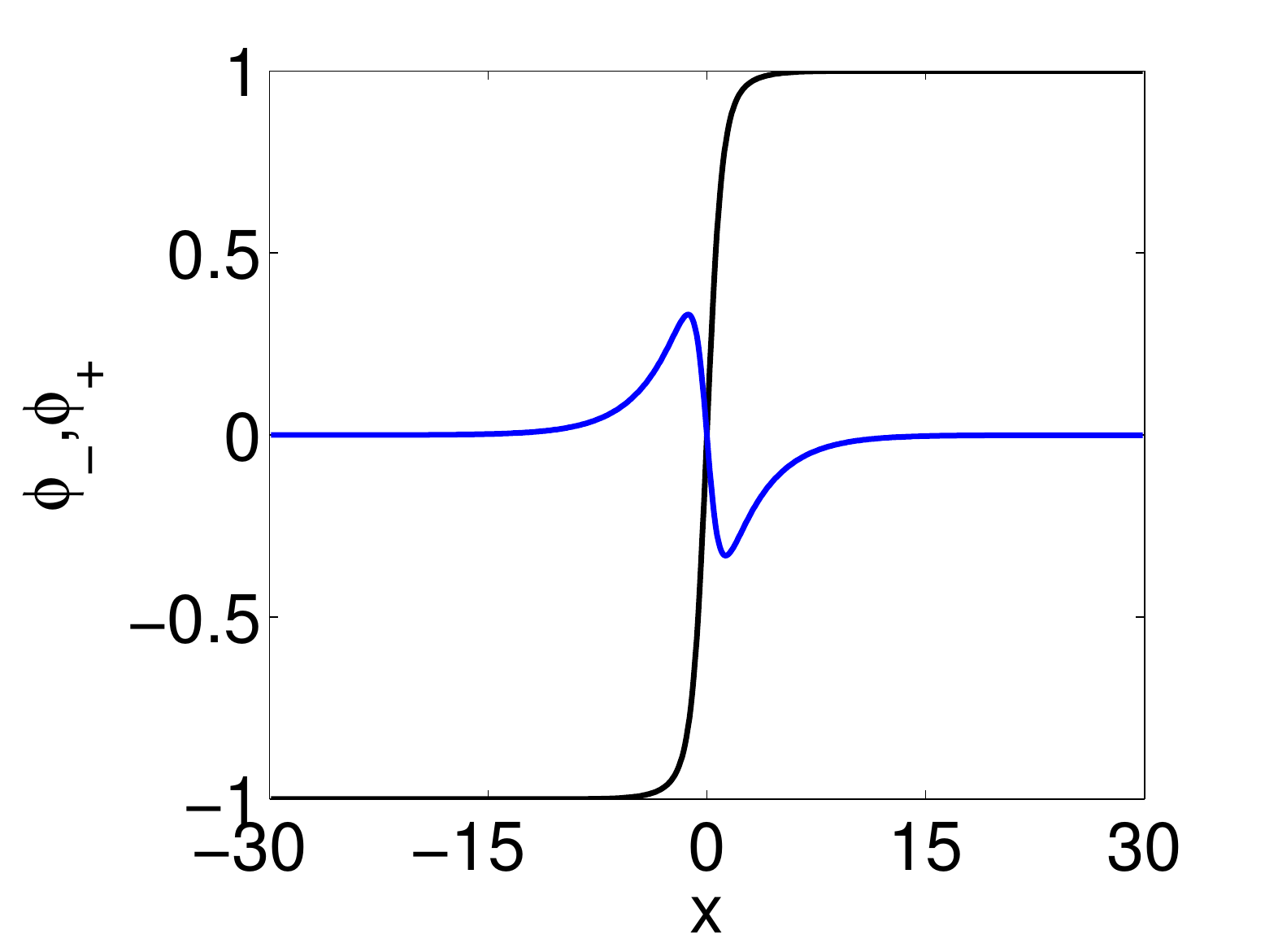}
\label{fig2a}
}
\subfigure[][]{\hspace{-0.3cm}
\includegraphics[height=.18\textheight, angle =0]{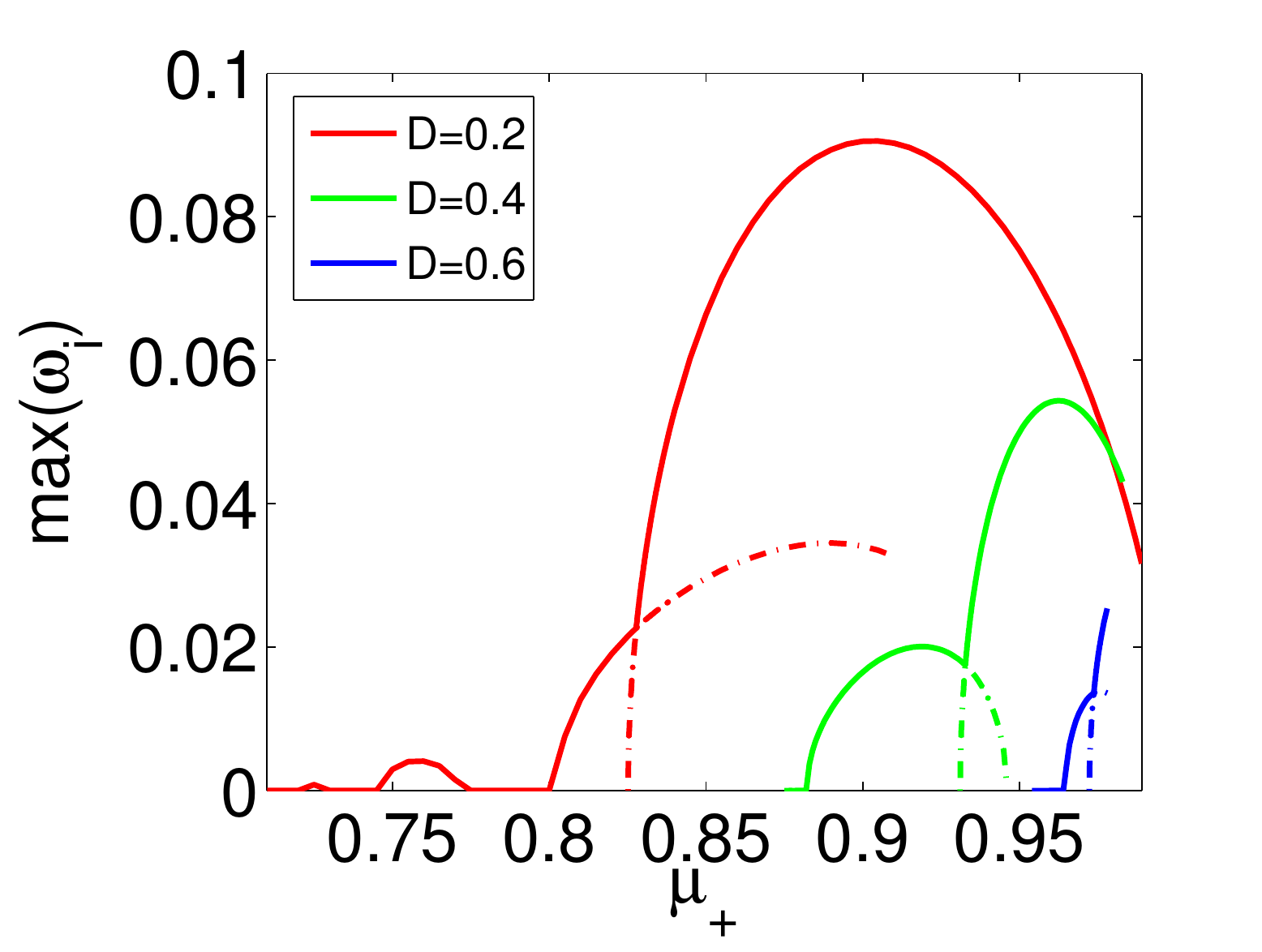}
\label{fig2b}
}
}
\mbox{\hspace{-0.1cm}
\subfigure[][]{\hspace{-0.3cm}
\includegraphics[height=.18\textheight, angle =0]{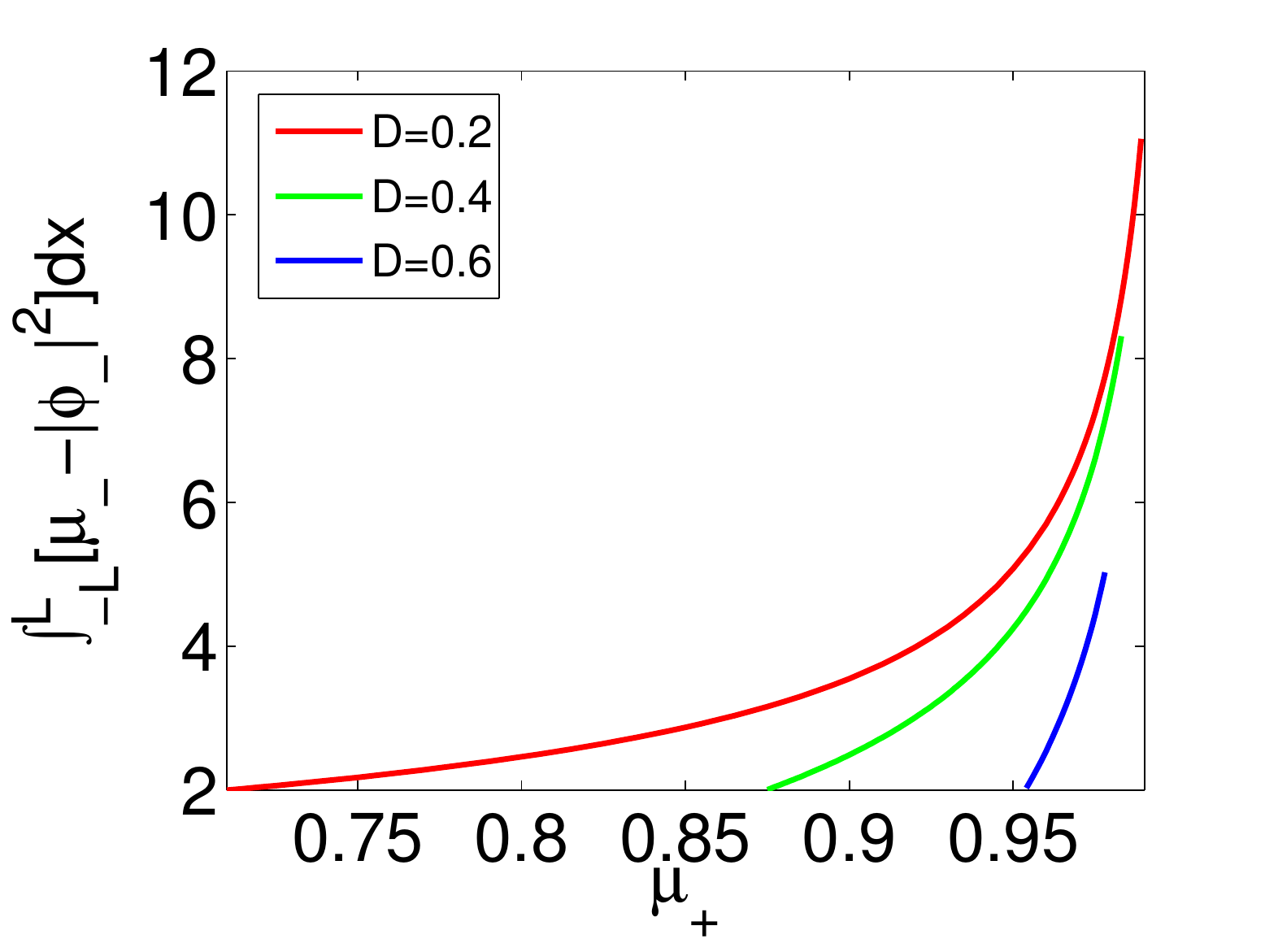}
\label{fig2c}
}
\subfigure[][]{\hspace{-0.3cm}
\includegraphics[height=.18\textheight, angle =0]{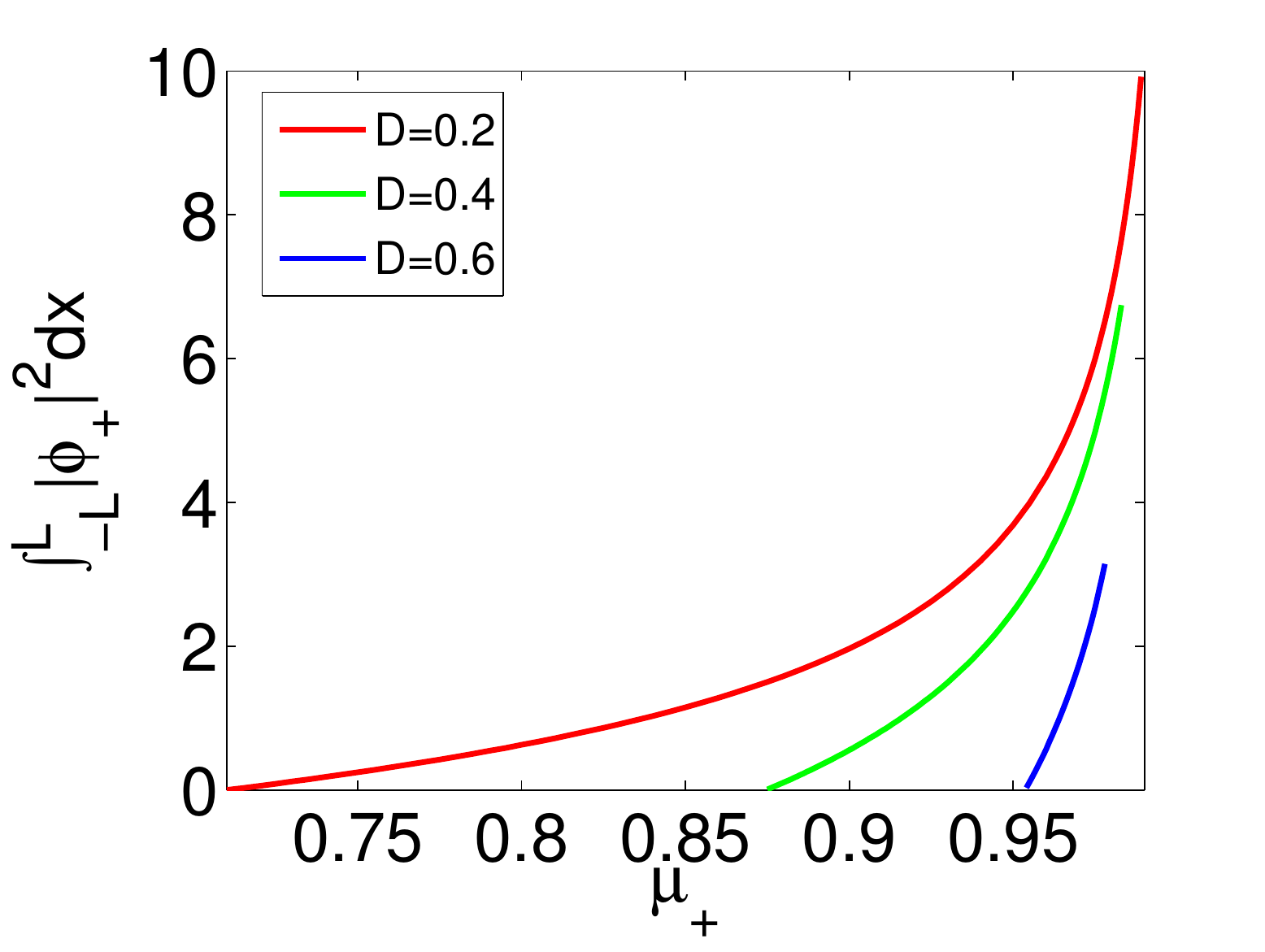}
\label{fig2d}
}
}
\end{center}
\par
\vspace{-0.7cm}
\caption{(Color online) Same as Fig.~\protect\ref{fig1}, but for bound
states of order $n=1$ (i.e., first excited spatially odd states). (a) 
Steady-state profiles of the dark (black line) and bright (blue line) 
soliton solutions for $D=0.6$ and $\protect\mu _{+}=0.96$. (b) Maximal 
imaginary eigenfrequency as a function of the continuation parameter 
$\protect\mu _{+}$ at various fixed values of $D$. Notice that the second 
highest instability growth rates are also shown by dash-dotted lines 
(see the text for details). (c) and (d) Solution norms,
given by integrals (\protect\ref{I}), as functions of the continuation
parameter $\protect\mu _{+}$ at various fixed values of $D$.}
\label{fig2}
\end{figure}
\begin{figure}[t]
\begin{center}
\vspace{-0.1cm}
\mbox{\hspace{-0.1cm}
\subfigure[][]{\hspace{-0.3cm}
\includegraphics[height=.18\textheight, angle =0]{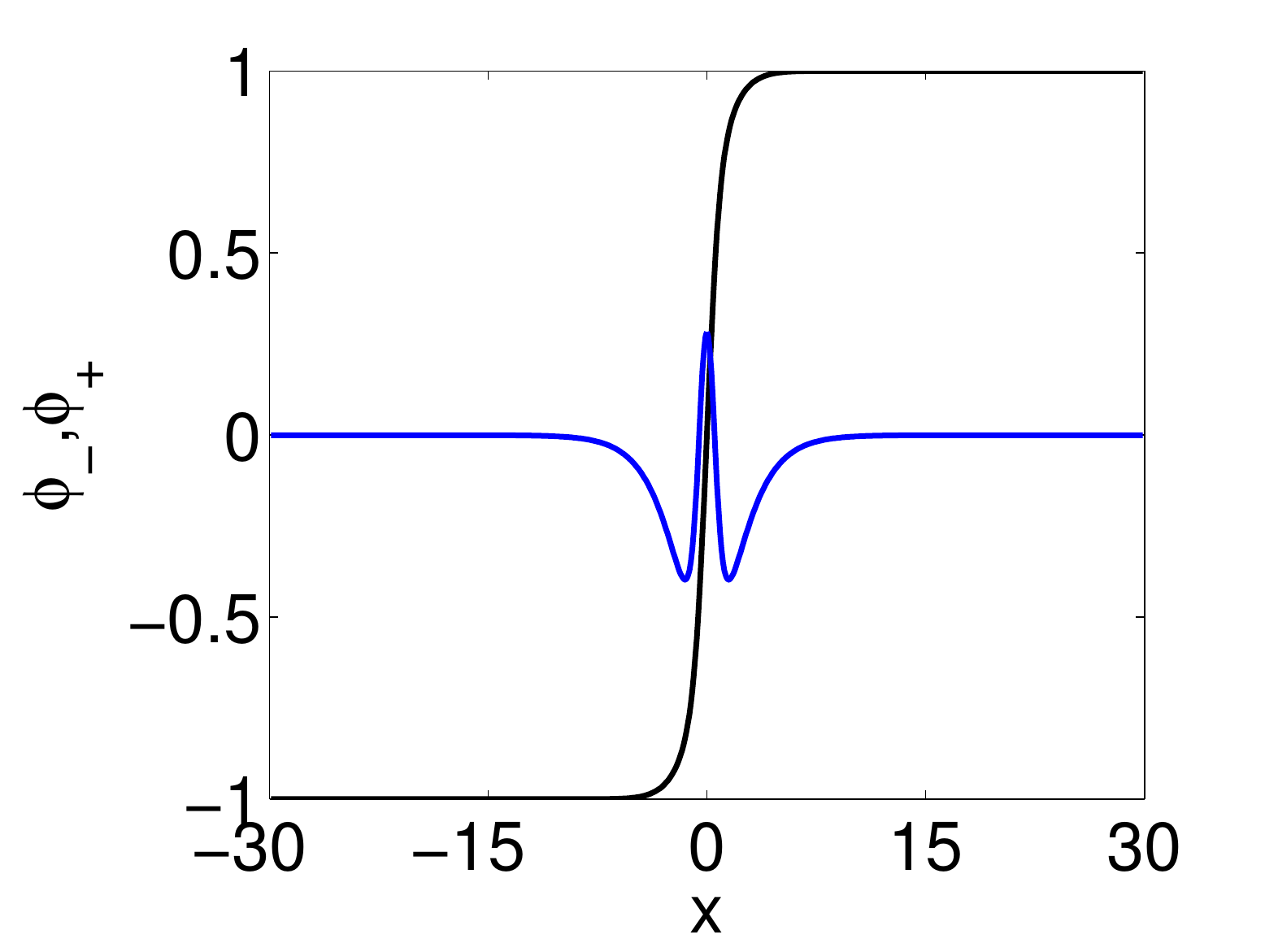}
\label{fig3a}
}
\subfigure[][]{\hspace{-0.3cm}
\includegraphics[height=.18\textheight, angle =0]{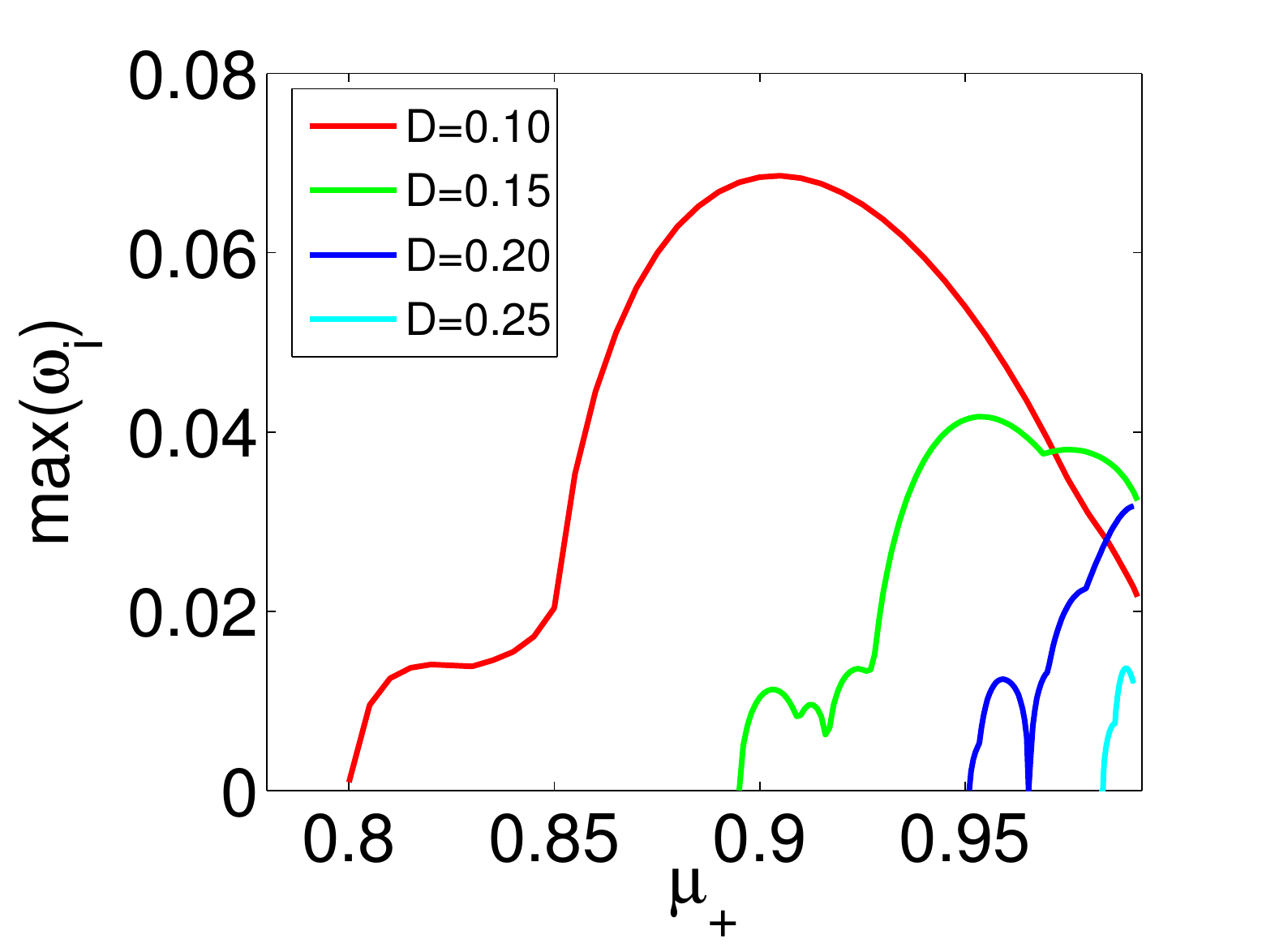}
\label{fig3b}
}
}
\mbox{\hspace{-0.1cm}
\subfigure[][]{\hspace{-0.3cm}
\includegraphics[height=.18\textheight, angle =0]{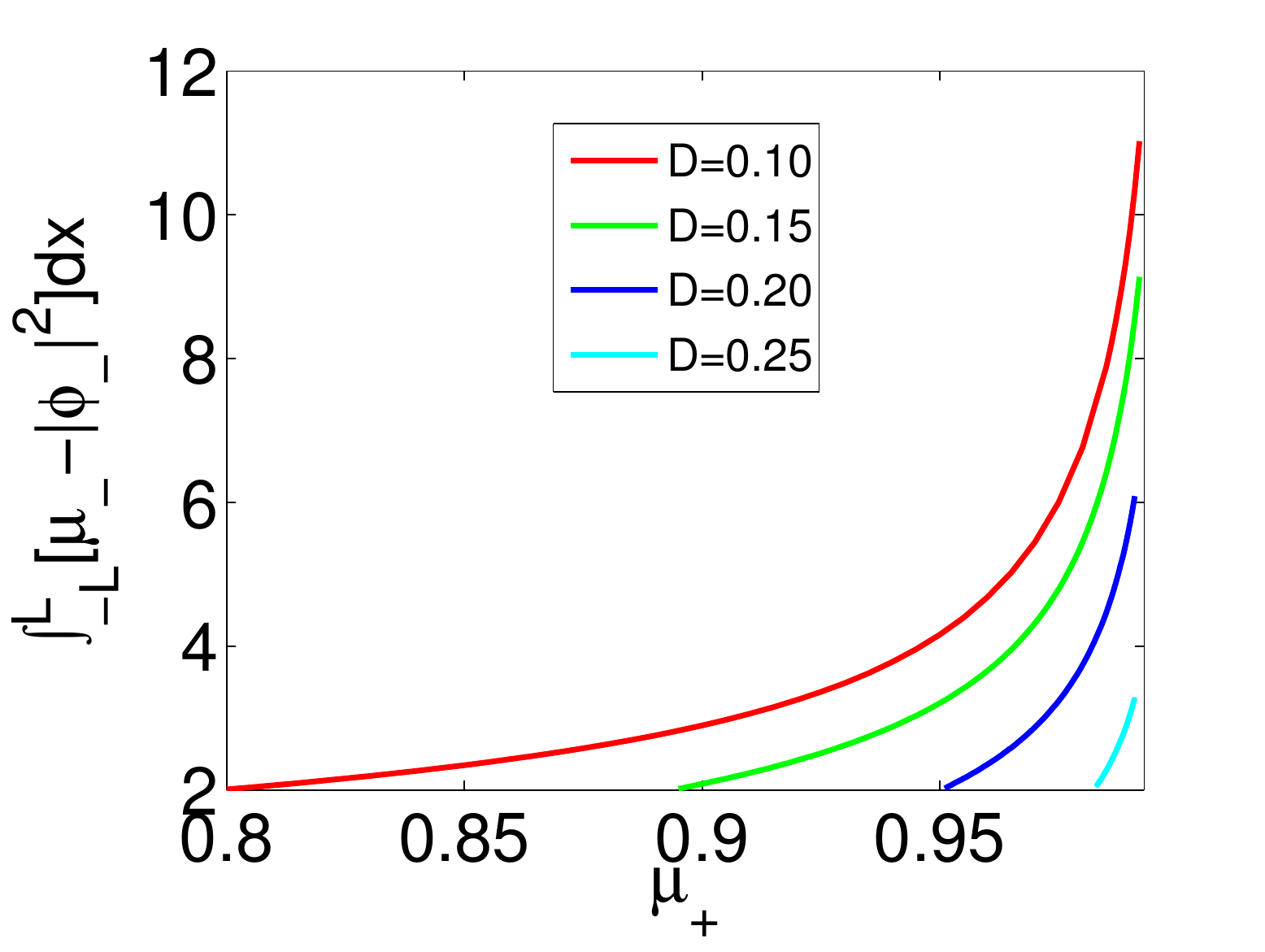}
\label{fig3c}
}
\subfigure[][]{\hspace{-0.3cm}
\includegraphics[height=.18\textheight, angle =0]{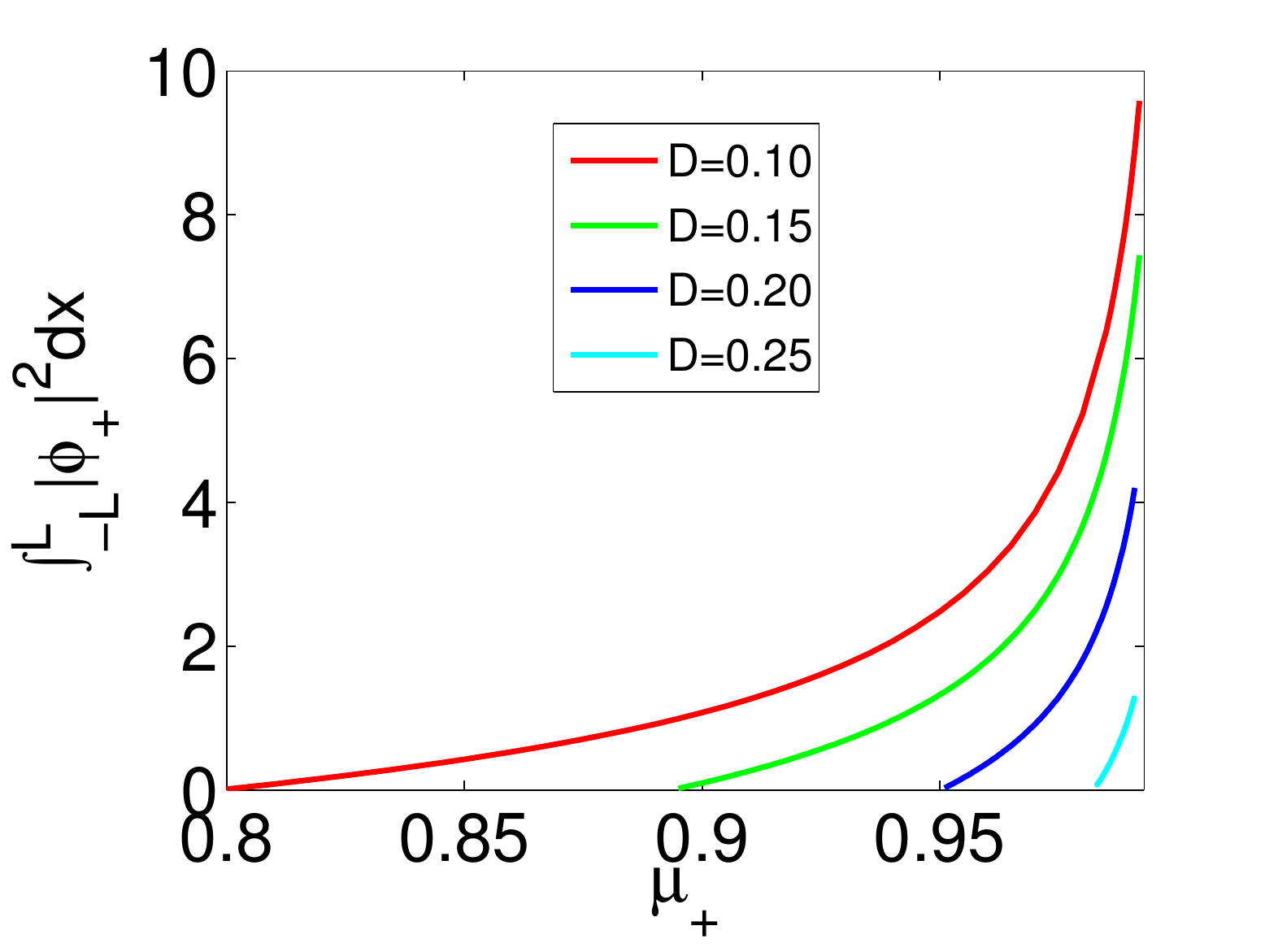}
\label{fig3d}
}
}
\end{center}
\par
\vspace{-0.7cm}
\caption{(Color online) Same as Fig.~\protect\ref{fig1}, but for bound
states of order $n=2$ (i.e., first excited spatially even states). (a) 
Steady-state profiles of the dark (black line) and bright (blue line) 
soliton solutions for $D=0.2$ and $\protect\mu _{+}=0.9655$. (b) Maximal
imaginary eigenfrequency as a function of the continuation parameter 
$\protect\mu _{+}$ at various values of $D$. (c) and (d) Powers associated 
with the solution branches [see Eq. (\protect\ref{I})] versus the continuation
parameter $\protect\mu _{+}$ at various values of $D$.}
\label{fig3}
\end{figure}
\begin{figure}[th]
\begin{center}
\vspace{-0.1cm}
\mbox{\hspace{-0.1cm} \subfigure[][]{\hspace{-0.3cm}
\includegraphics[height=.18\textheight, angle =0]{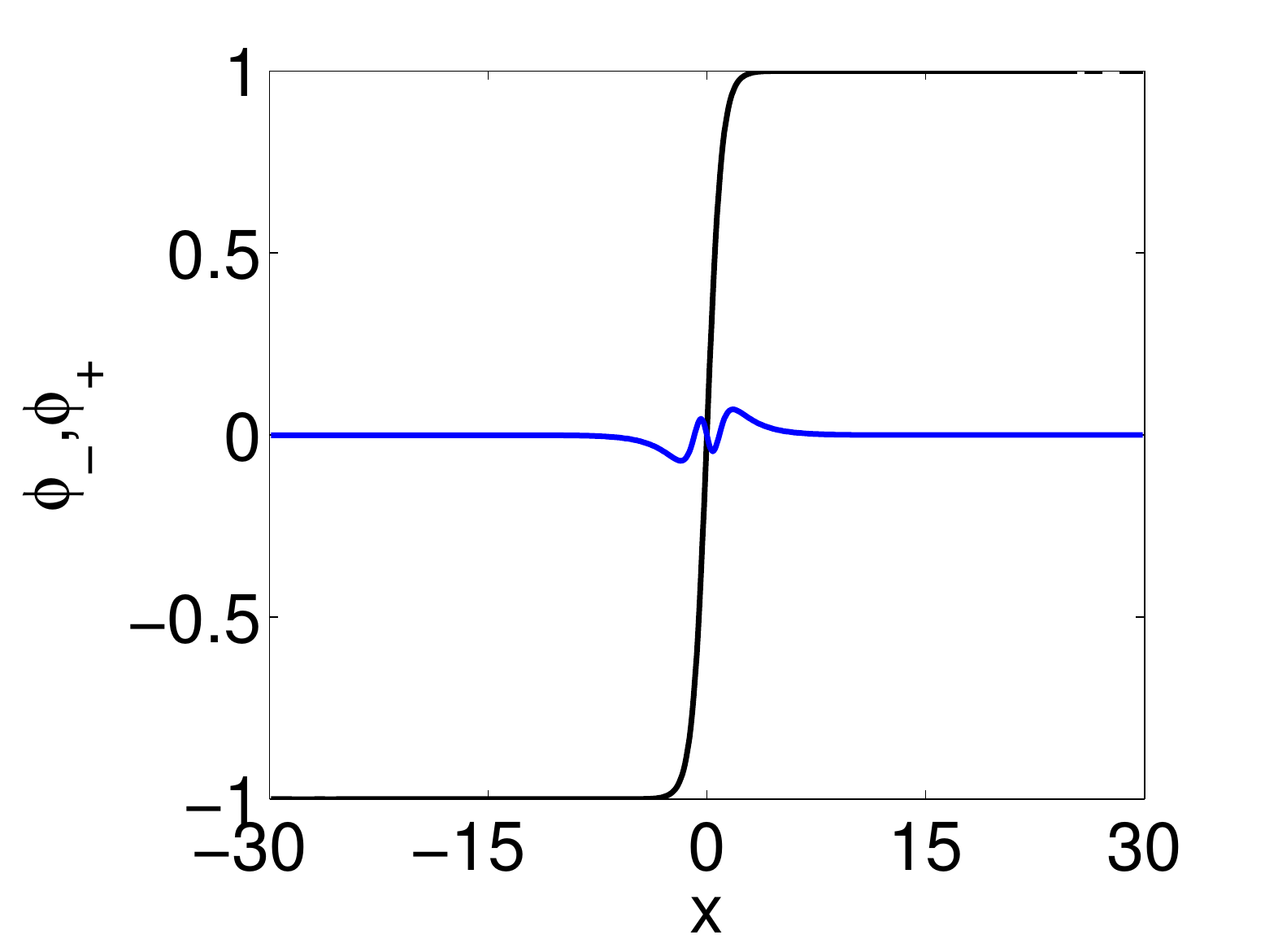}
\label{fig4a} } \subfigure[][]{\hspace{-0.3cm}
\includegraphics[height=.18\textheight, angle =0]{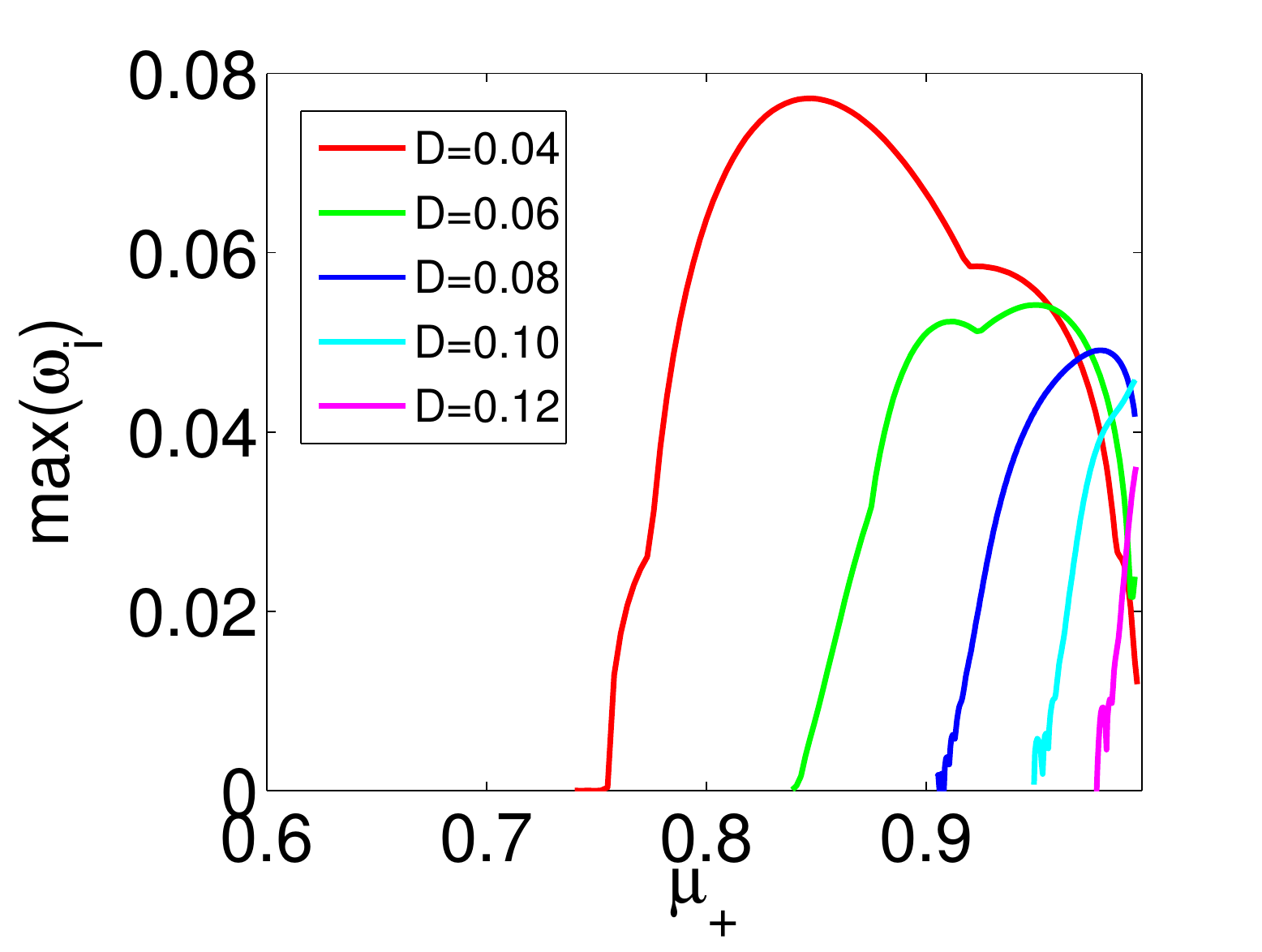}
\label{fig4b} } }
\mbox{\hspace{-0.1cm}
\subfigure[][]{\hspace{-0.3cm}
\includegraphics[height=.18\textheight, angle =0]{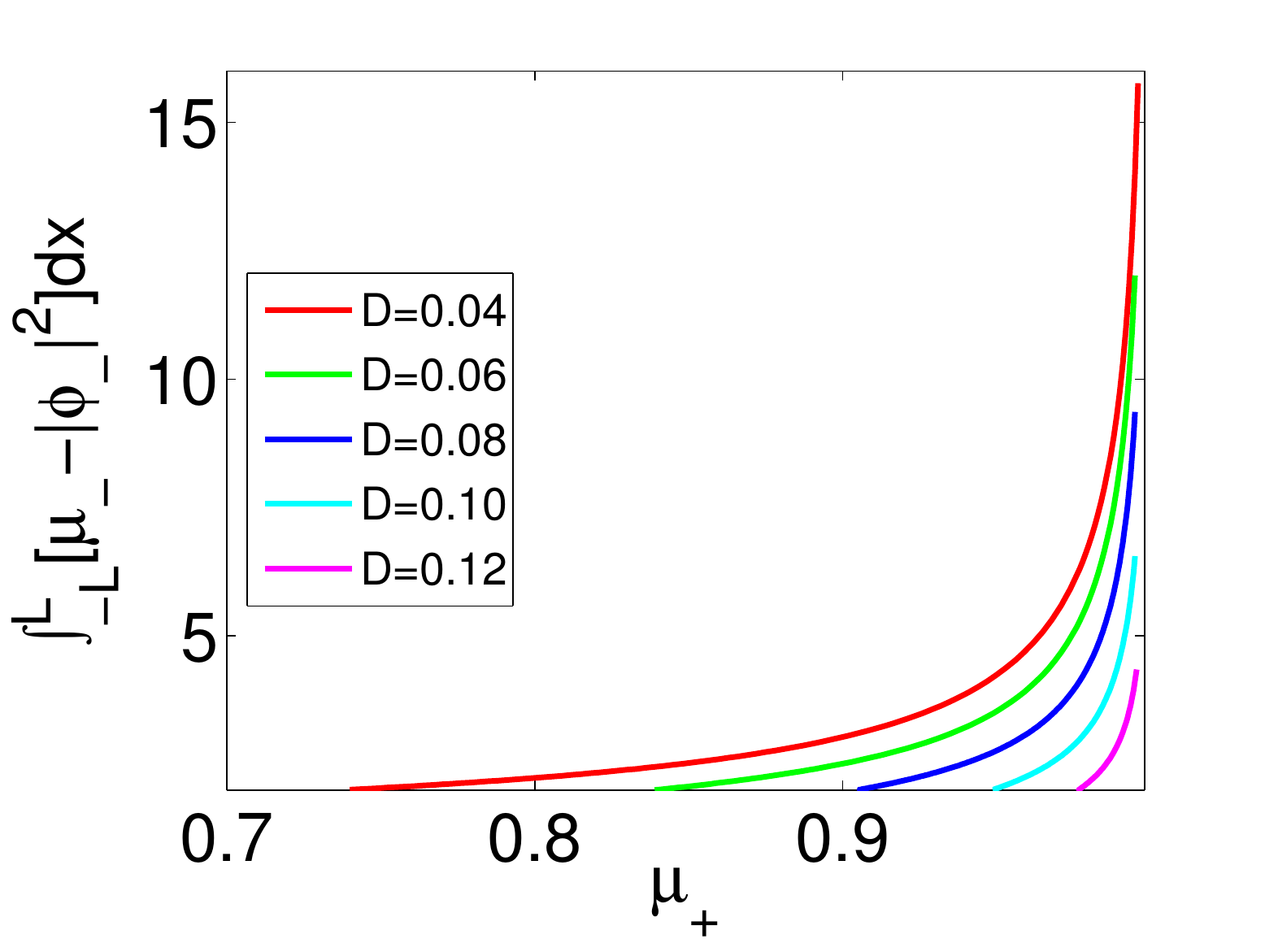}
\label{fig4c} } \subfigure[][]{\hspace{-0.3cm}
\includegraphics[height=.18\textheight, angle =0]{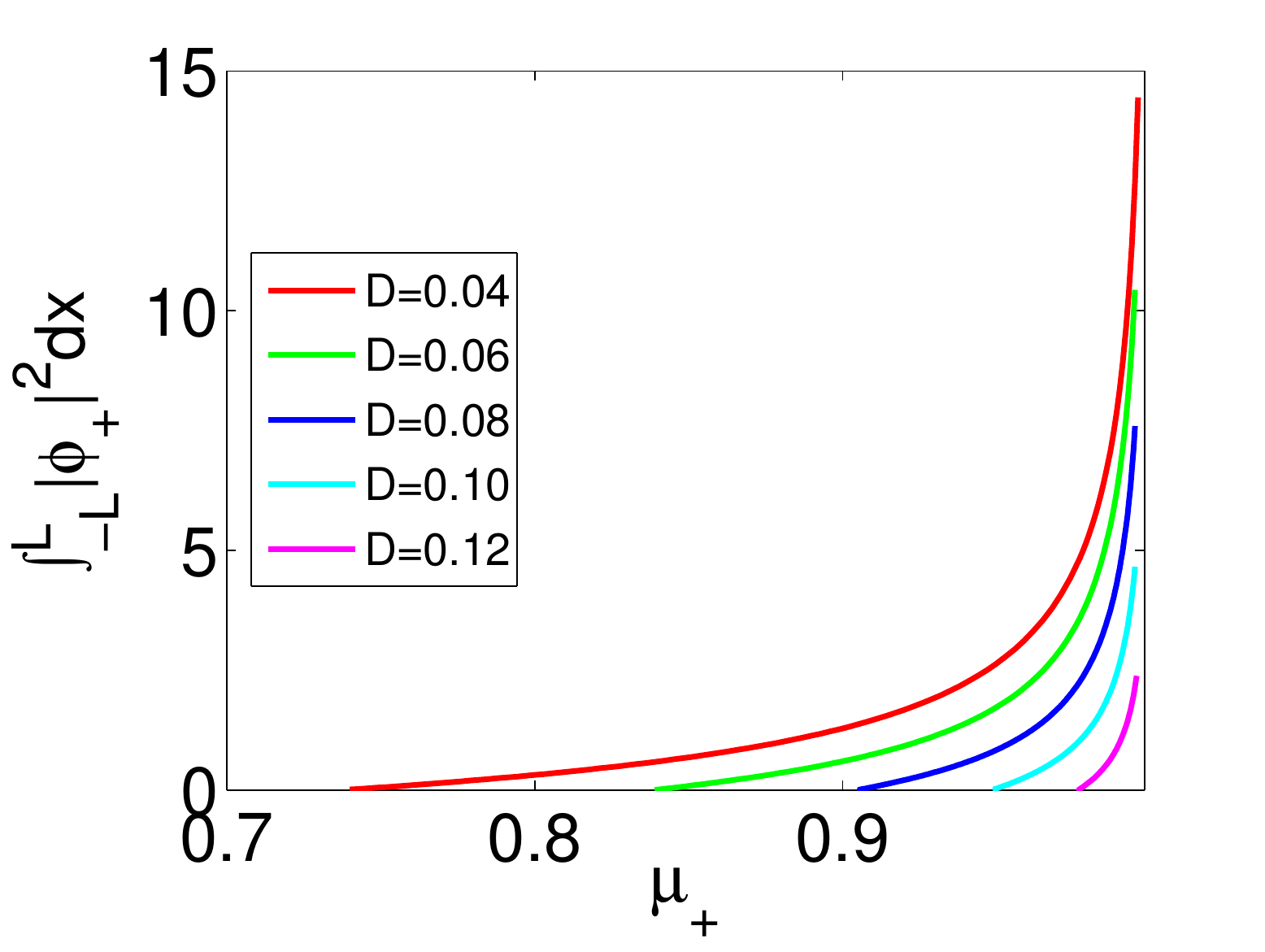}
\label{fig4d} } }
\end{center}
\par
\vspace{-0.7cm}
\caption{(Color online) Same as Fig.~\protect\ref{fig1}, but for bound
states of order $n=3$ (i.e., second excited antisymmetric states). (a) 
Steady-state profiles of the dark (black line) and bright (blue line) 
soliton solutions for $D=0.12$ and $\protect\mu _{+}=0.977$. (b) Maximal
imaginary eigenfrequency as a function of the continuation parameter 
$\protect\mu _{+}$ at various values of $D$. (c) and (d) are similar 
to Figs.~\protect\ref{fig3c} and \protect\ref{fig3d}.}
\label{fig4}
\end{figure}

In this section, numerical results are presented for the coupled NLS
system~(\ref{start_manakov}) and organized following the reasoning mentioned
in the previous section. Figures~\ref{fig1}-\ref{fig4} summarize the results 
for the existence of steady-state solutions and the corresponding parametric
continuations (using $\mu _{+}$ as the continuation parameter, at different
fixed values of $D$) for bound states of order $n=0$ (ground states with a 
single-hump bright component, corresponding to a generalization of the fundamental
DB solitary waves), $n=1 $ (the first excited odd states), $n=2$ (the second 
excited states, which are spatially even), and $n=3$ (the third excited states
overall and second excited odd ones), respectively. Figures \ref{fig1a}-\ref{fig4a}
present a typical example of the relevant profiles, while Figs.~\ref{fig1b}-\ref{fig4b}
showcase the growth rate of the most unstable perturbation mode $\textrm{max}(\omega _{i})$, 
which, if positive, indicates instability for the particular pair $(\mu _{+},D)$.
Furthermore, Figs.~\ref{fig1c}-\ref{fig4c} and \ref{fig1d}-\ref{fig4d} summarize the
existence results by presenting
\begin{equation}
\int_{-L}^{L}\left[ {\mu _{-}-|\phi _{-}(x)|^{2}}\right] {dx,~~} ~~
\int_{-L}^{L}{|\phi _{+}(x)|^{2}dx},  \label{I}
\end{equation}
respectively, as functions of $\mu _{+}$ and for various fixed values of $D$. 
These integrals represent the total power in optics or the atom number in the 
BEC, considered as a function of the propagation constant or chemical potential,
respectively.

\begin{figure}[tph]
\begin{center}
\vspace{-0.1cm}
\mbox{\hspace{-0.1cm}
\subfigure[][]{\hspace{-0.3cm}
\includegraphics[height=.18\textheight, angle =0]{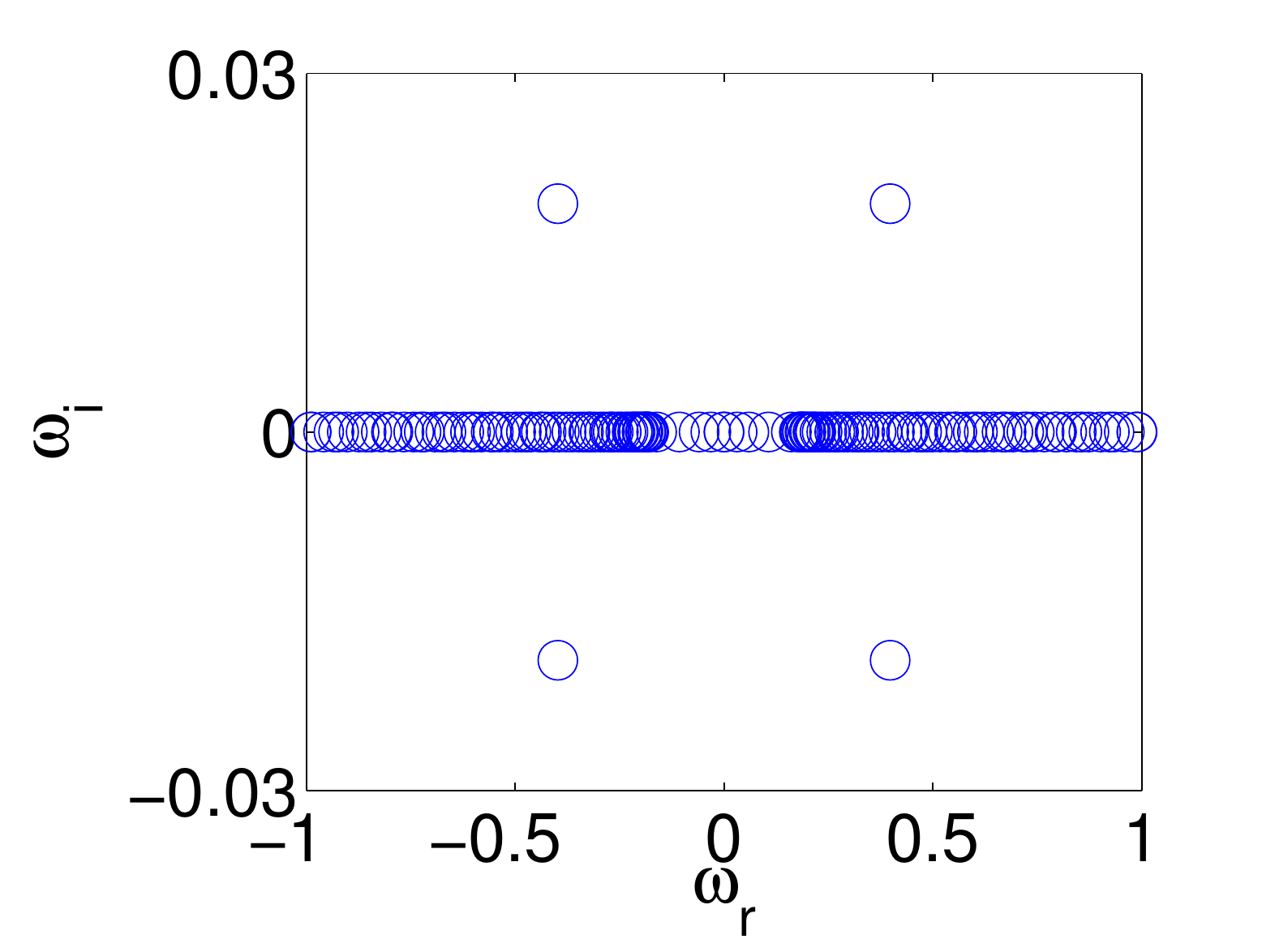}
\label{fig2_b1}
}
\subfigure[][]{\hspace{-0.3cm}
\includegraphics[height=.18\textheight, angle =0]{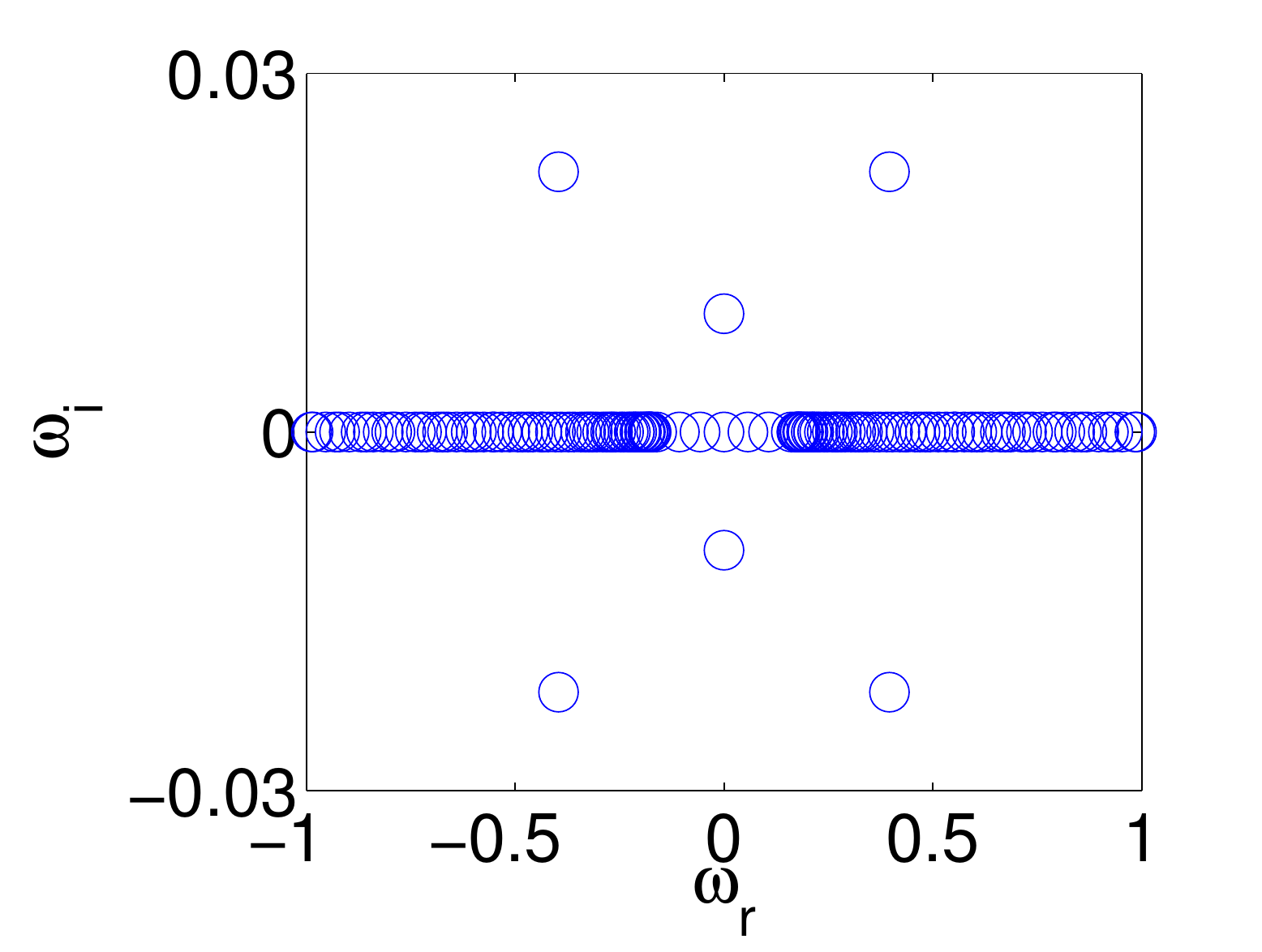}
\label{fig2_b2}
}
}
\mbox{\hspace{-0.1cm}
\subfigure[][]{\hspace{-0.3cm}
\includegraphics[height=.18\textheight, angle =0]{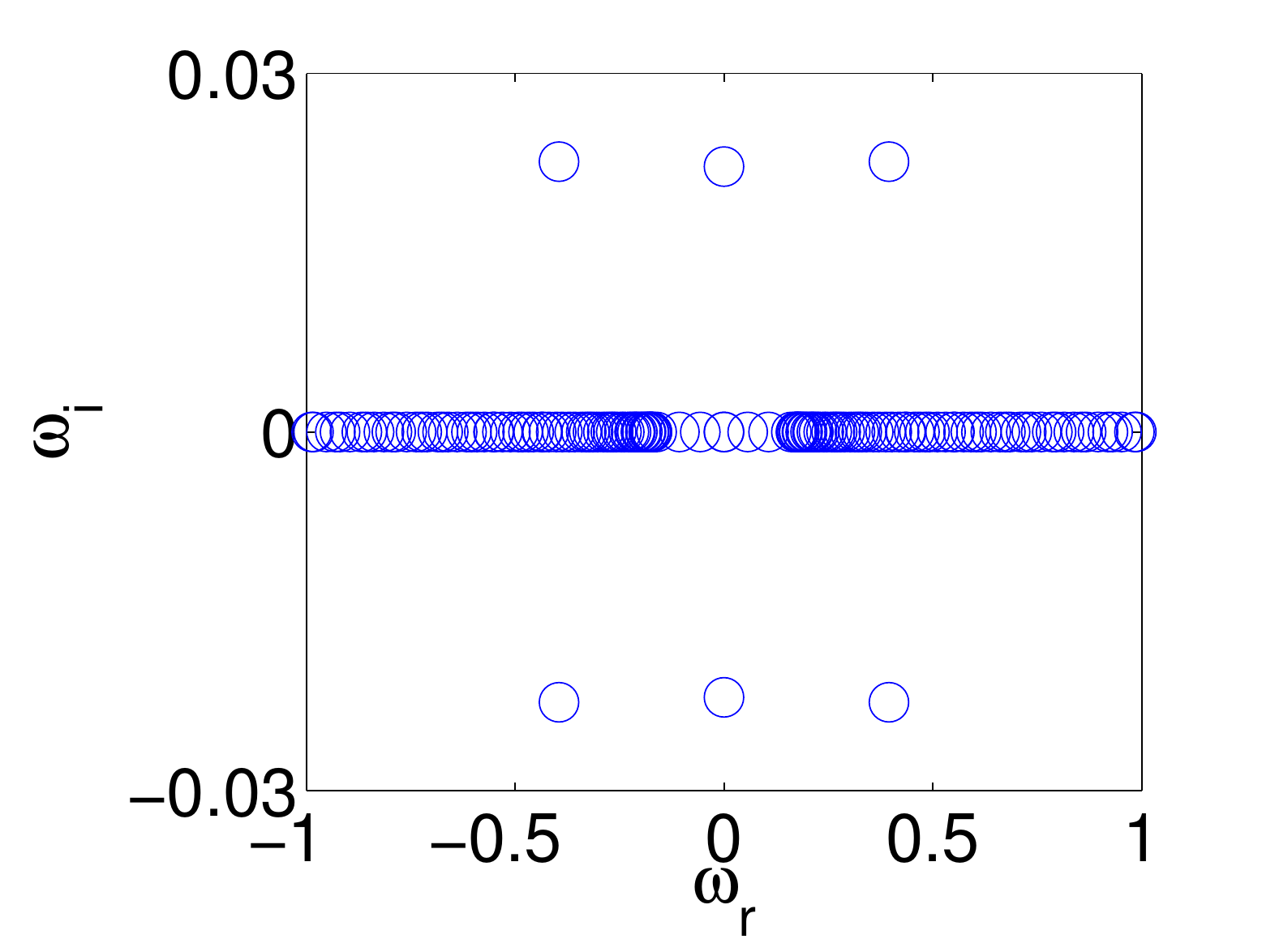}
\label{fig2_b3}
}
\subfigure[][]{\hspace{-0.3cm}
\includegraphics[height=.18\textheight, angle =0]{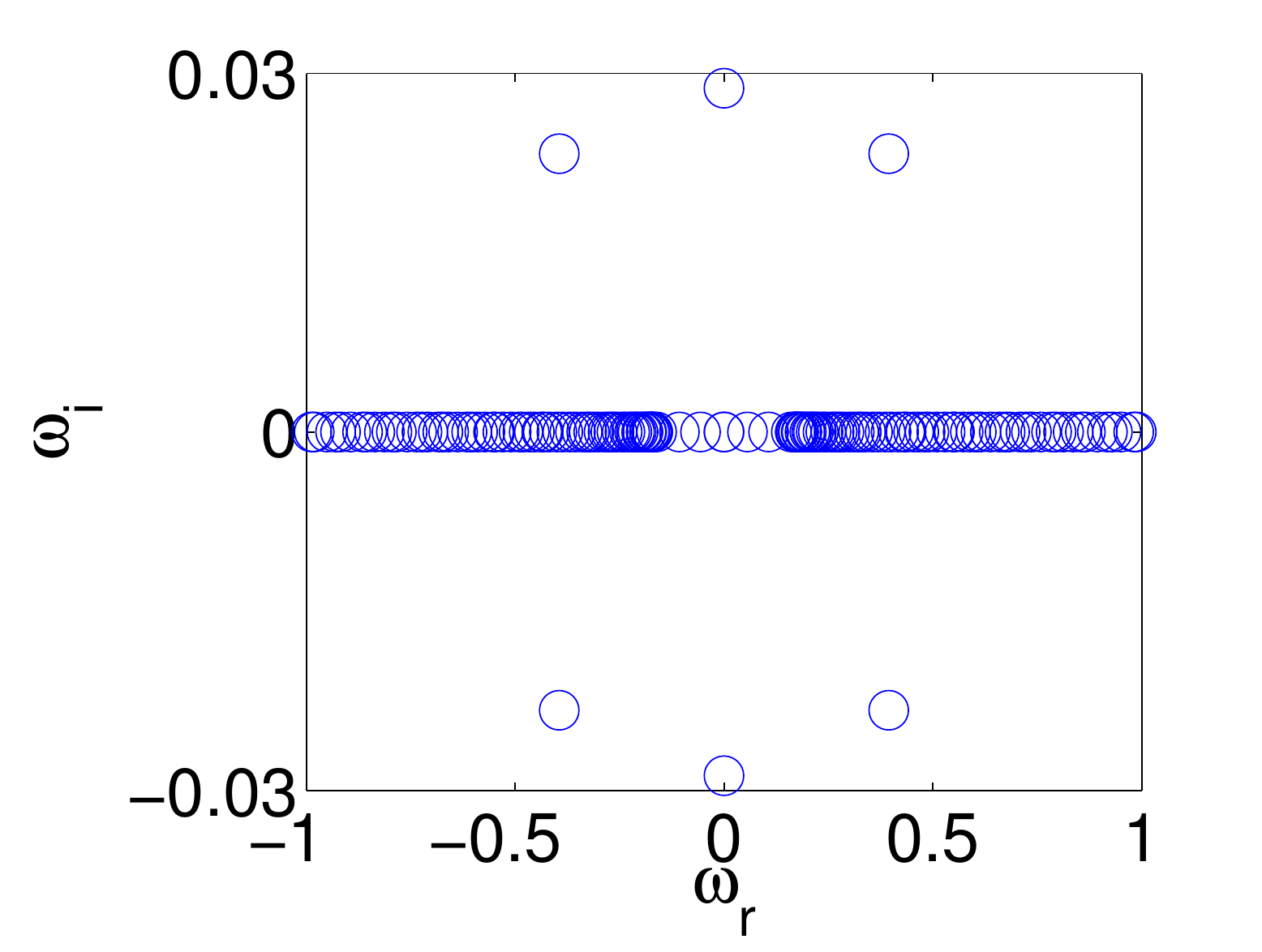}
\label{fig2_b4}
}
}
\end{center}
\par
\vspace{-0.7cm}
\caption{(Color online) Eigenfrequency spectra corresponding to bound states
of order $n=1$ with $D=0.2$ and for (a) $\protect\mu _{+}=0.82$, (b) $\protect\mu _{+}=0.8257$, 
(c) $\protect\mu _{+}=0.8277$, and (d) $\protect\mu _{+}=0.8294$.}
\label{fig2_b}
\end{figure}

In particular, it can be inferred from Fig.~\ref{fig1b} that solutions
corresponding to the ground state are stable for $D=0.4-1.0$ and
for all values of $\mu _{+}$ within the range of interest [see, e.g., Eq.~%
(\ref{landau_limit}) and the subsequent discussion]. In contrast,
the solution branch corresponding to $D=0.2$ is stable up to a critical
point $\mu _{+}^{\mathrm{crit}}\approx 0.69$, while past this value an
exponential instability, accounted for by an imaginary eigenfrequency pair
with zero real part, emerges [see also Fig.~\ref{fig5f}]. Similar arguments
can be applied to Figs.~\ref{fig2b}-\ref{fig4b}, although the description is
somewhat different. In particular, it can be seen in Fig.~\ref{fig2b} that the 
solution branch corresponding to $D=0.2$ 
possesses a number of instability intervals for $\mu _{+}>0.71$.
However, in the present case the instability is accounted for by a complex
eigenfrequency quartet, which corresponds to an oscillatory instability
related to a Hamiltonian Hopf bifurcation [see also Fig.~\ref{fig6i} below
as a case example]; for a recent discussion of relevant bifurcations, see,
e.g.,~Ref. \cite{royg}. While instabilities of this type are present as
shown in Fig.~\ref{fig2_b1}, past the value of $\mu _{+}\approx 0.8252$ an
additional imaginary eigenfrequency pair appears too, as depicted in Fig.~%
\ref{fig2_b2} and initially marked with a dash-dotted red line in Fig.~%
\ref{fig2b}. As $\mu _{+}$ further increases, the exponentially unstable
mode grows [cf. Fig.~\ref{fig2_b3}] and becomes dominant [cf. Fig.~\ref%
{fig2_b4}], while the oscillatory one follows a smaller growth rate, as
depicted in Fig.~\ref{fig2b} by a dash-dotted red line (the crossing of the
solid line with the dash-dotted line marks the exchange of the dominant form of
the instability). This is also the case for the solution branches with 
$D=0.4$ and $0.6$ depicted in Fig.~\ref{fig2b} by dash-dotted green and blue lines, 
respectively. This transition between the exponentially and oscillatory 
unstable modes also occurs for the bound states of order $%
n=2$ [cf. Fig.~\ref{fig3b}] and $n=3$ [cf. Fig.~\ref{fig4b}]. An additional
feature that arises in the latter figures is well known in the context of
discrete systems as a finite-size effect and was introduced in Ref. \cite%
{kivshar}. This is related to the fact that the continuous spectrum of background (phonon) excitations
is discretized on our spatial grid, hence complex eigenfrequencies may
return to the real eigenfrequency axis temporarily before colliding with
another pair to exit again as quartets. It is expected (cf.~Ref. \cite%
{kivshar}) that these discrete effects gradually disappear as the spectrum 
becomes denser, i.e., in the infinite-domain limit. It is the combination of
the above-mentioned exchanges of the dominant instability type and of temporary
restabilizations that gives rise to the spikes in Figs.~\ref{fig3b} and~\ref{fig4b}. 
In such cases, only the dominant instability growth rate is shown; recall that 
Fig.~\ref{fig3} presents the results for the second excited (first excited even) 
branch and Fig.~\ref{fig4} those for the third excited (second excited odd) branch.

It is relevant to note that the branches with higher $n$ are generally more
prone to instability than ones with lower $n$. The ground-state single-hump
solution is generally fairly robust, as is suggested by the observability of
the fundamental DB soliton in both atomic and optical settings~\cite%
{sengdb,peterprl,peter1,seg1}. Our results reveal that only at very low
values of $D$ does an instability arise for this state. On the other hand,
branches with $n=1-3$ are less robust. Among them, our results
suggest that the $n=1$ branch attains the highest instability growth rates.
However, examining the parametric intervals of the instability (i.e., widths
of the intervals of $\mu _{+}$ over which the branches remain stable), we
observe that the higher the $n$, the narrower the corresponding stability
interval becomes. Actually, the instability growth rates are relatively
weak, typically $\sim 10^{-2}$, which suggests that the solutions should be
long-lived ones, as the dynamical simulations corroborate below.

\begin{figure}[tp]
\begin{center}
\vspace{-0.2cm}
\mbox{\hspace{-0.1cm}
\subfigure[][]{\hspace{-0.3cm}
\includegraphics[height=.18\textheight, angle =0]{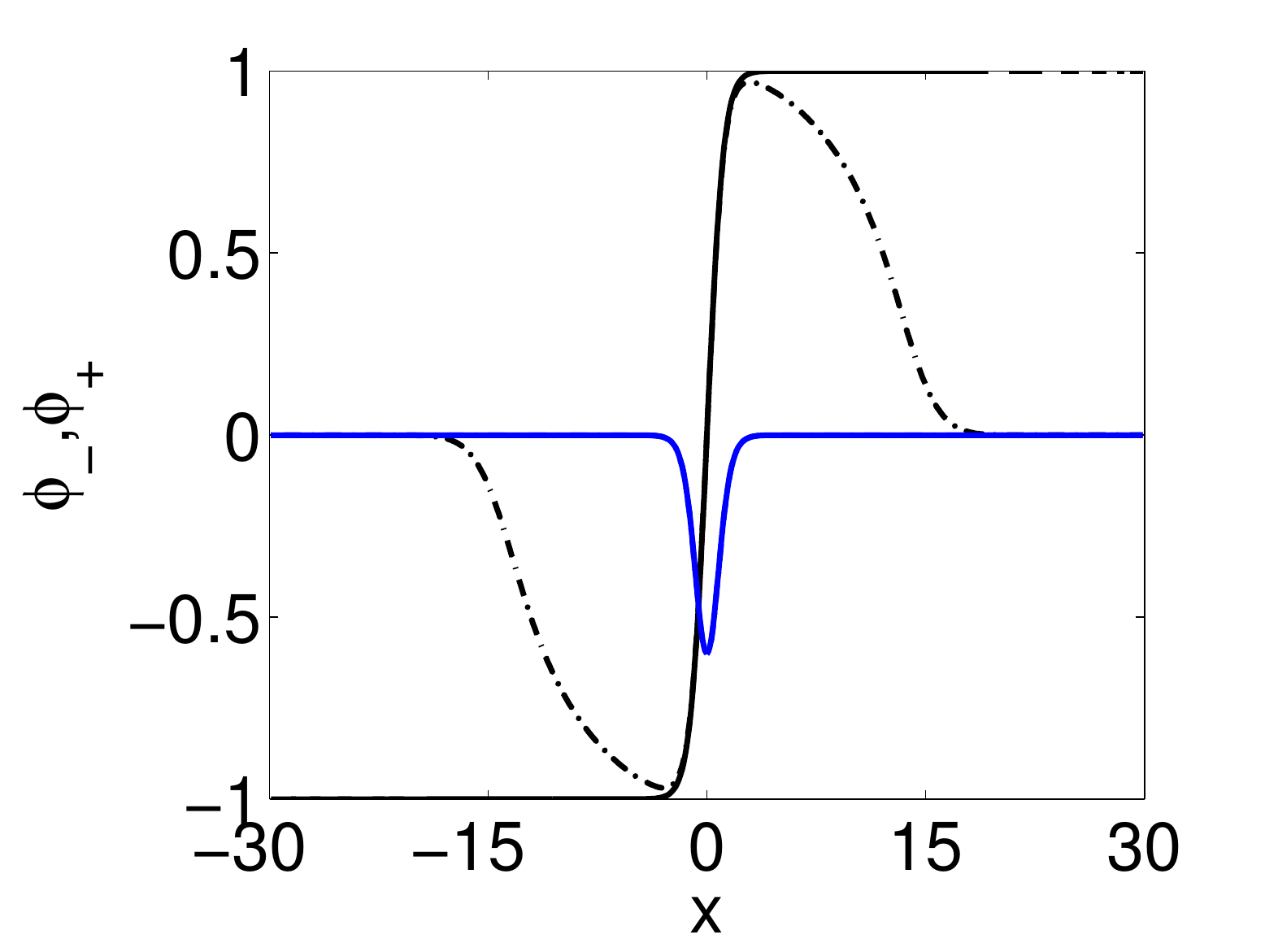}
\label{fig5a_trap}
}
\subfigure[][]{\hspace{-0.3cm}
\includegraphics[height=.18\textheight, angle =0]{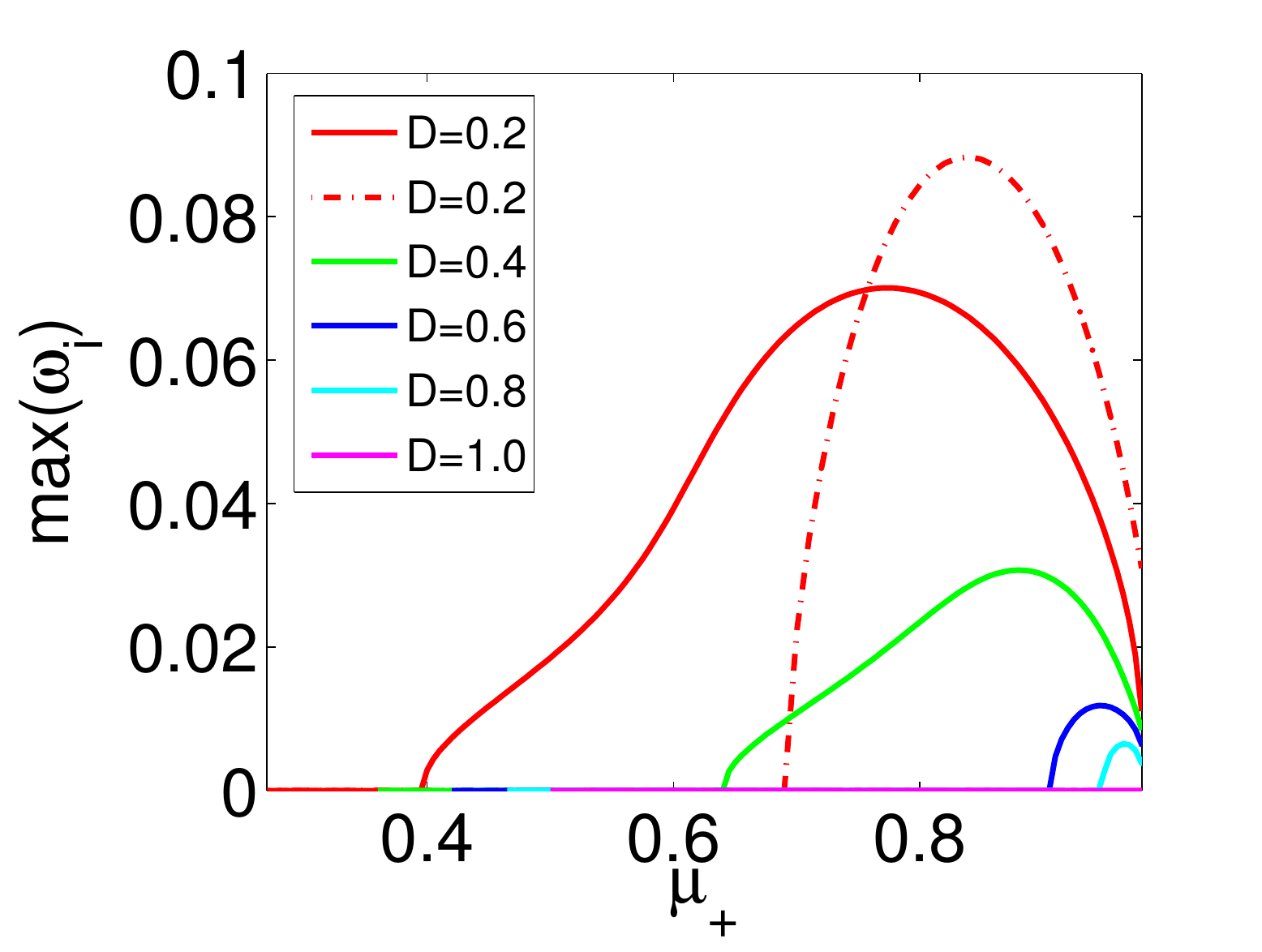}
\label{fig5b_trap}
}
}
\mbox{\hspace{-0.1cm}
\subfigure[][]{\hspace{-0.3cm}
\includegraphics[height=.18\textheight, angle =0]{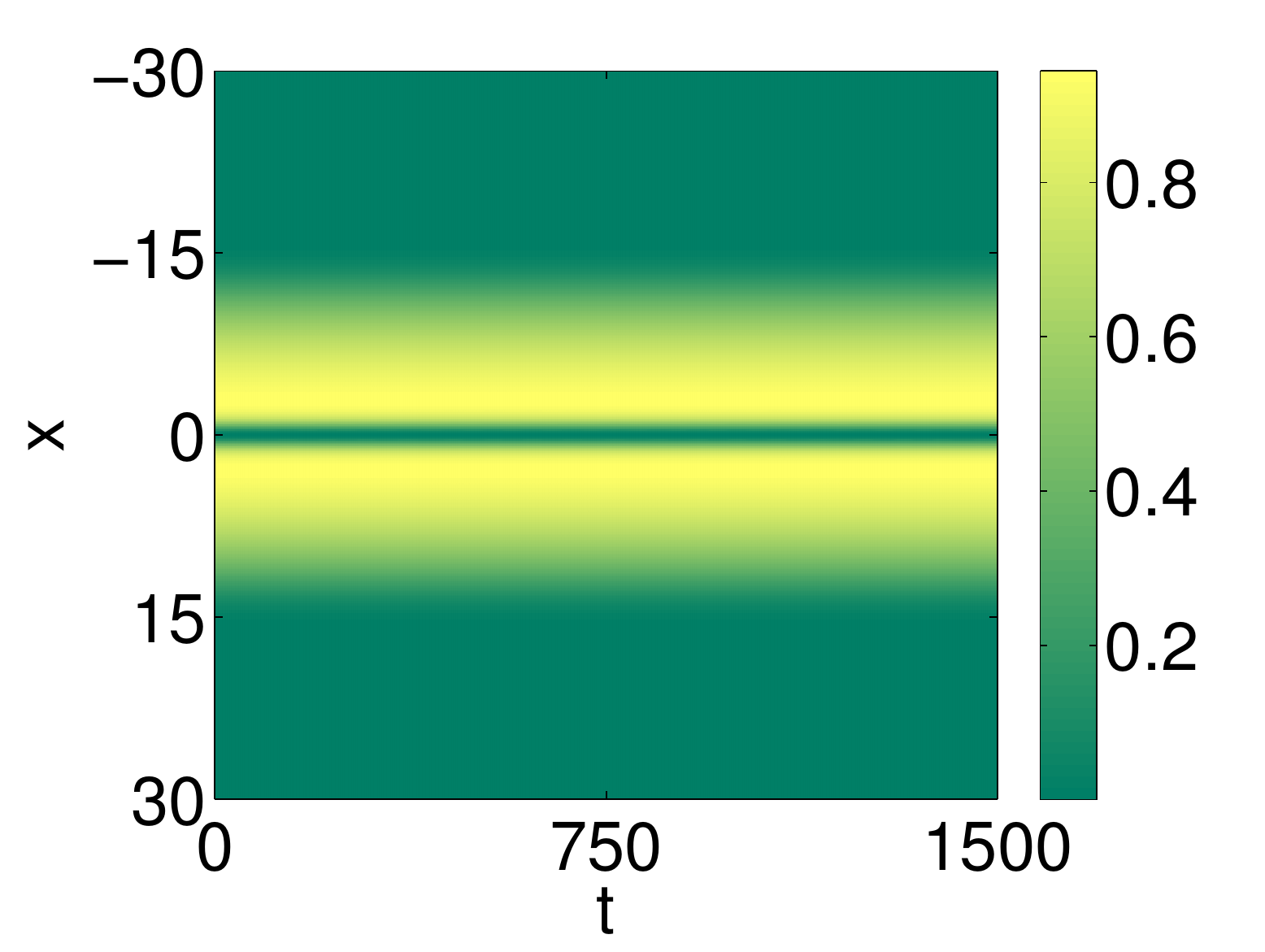}
\label{fig5c_trap}
}
\subfigure[][]{\hspace{-0.3cm}
\includegraphics[height=.18\textheight, angle =0]{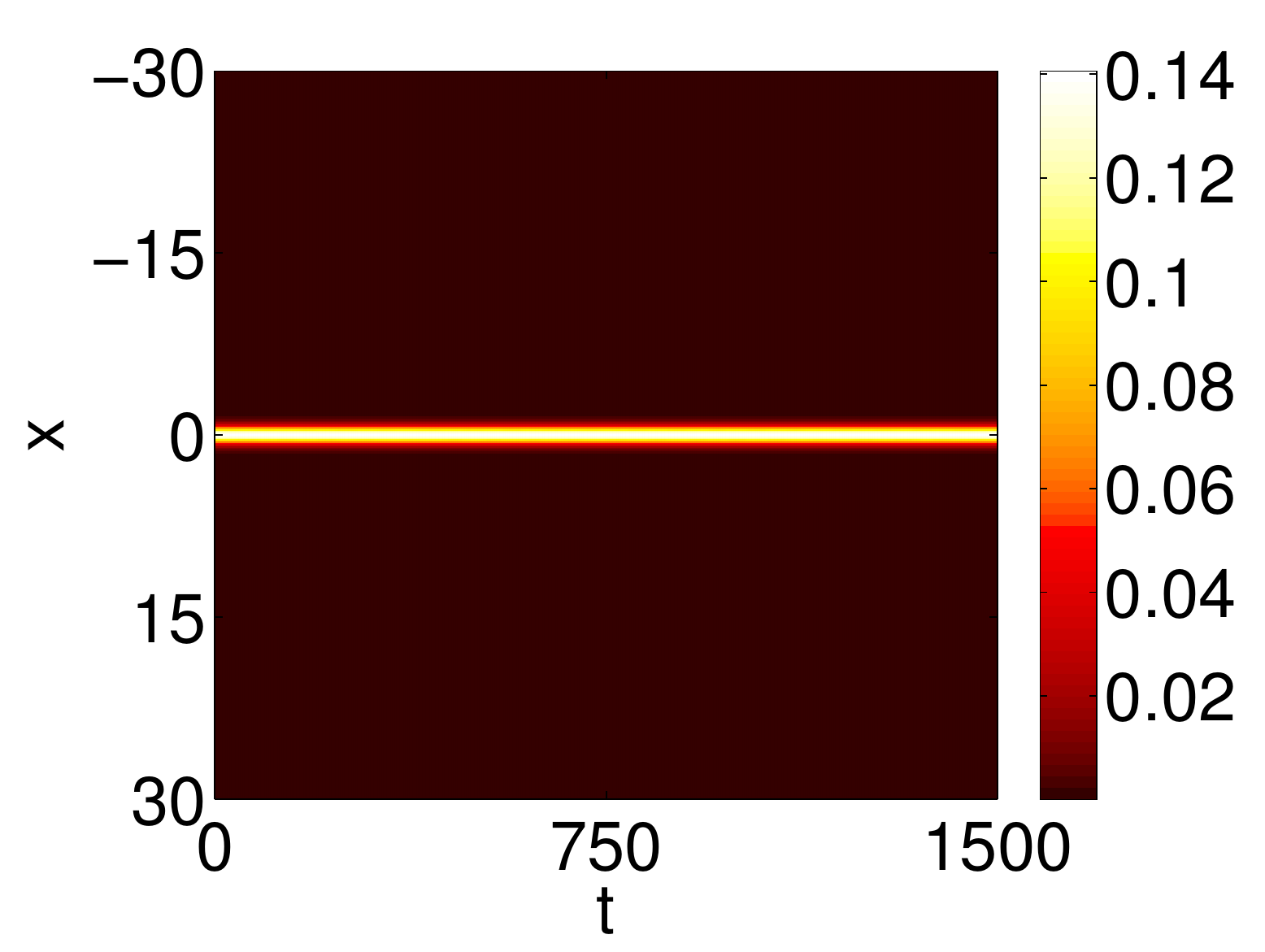}
\label{fig5d_trap}
}
\subfigure[][]{\hspace{-0.3cm}
\includegraphics[height=.18\textheight, angle =0]{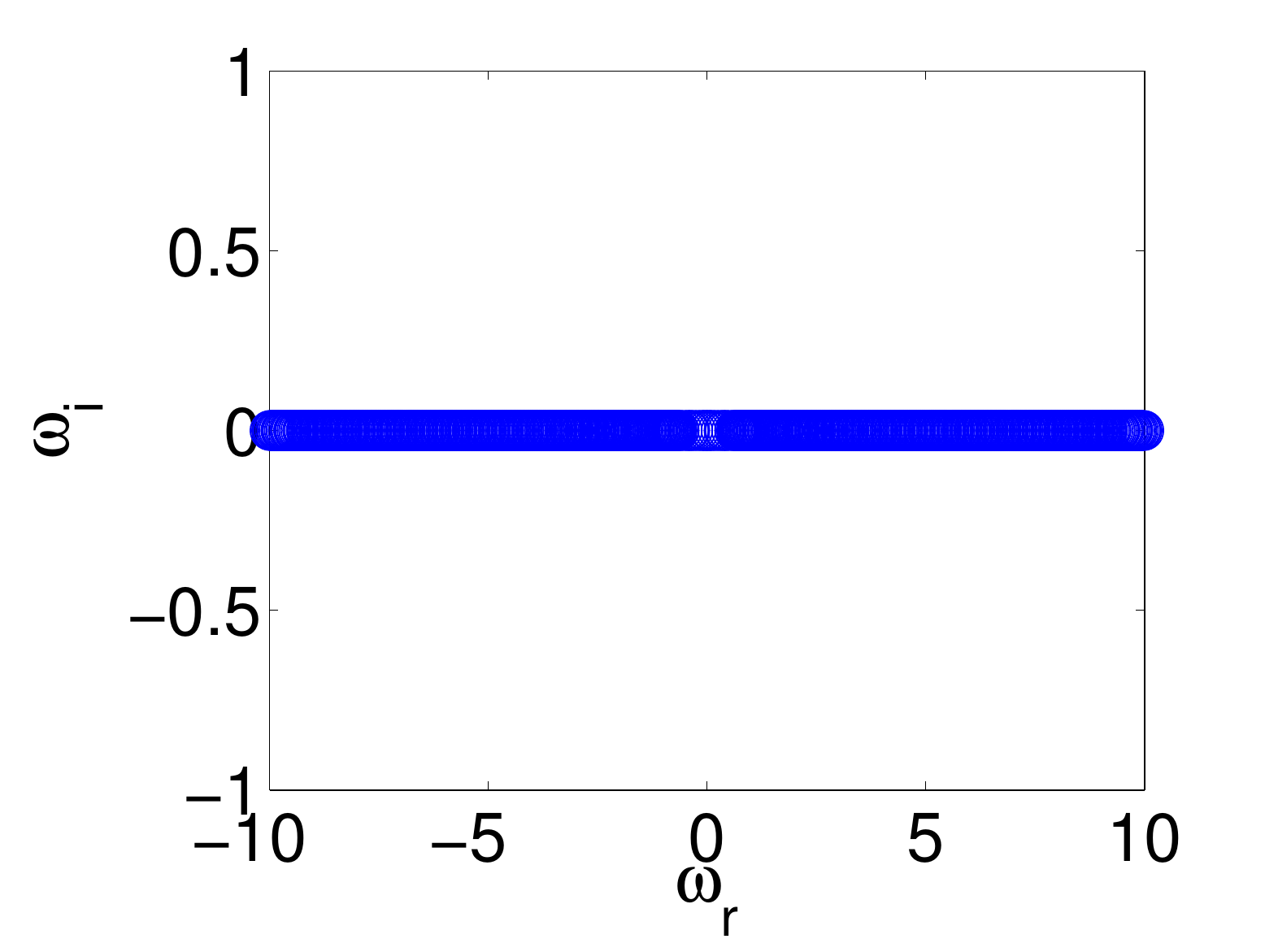}
\label{fig5e_trap}
}
}
\mbox{\hspace{-0.1cm}
\subfigure[][]{\hspace{-0.3cm}
\includegraphics[height=.18\textheight, angle =0]{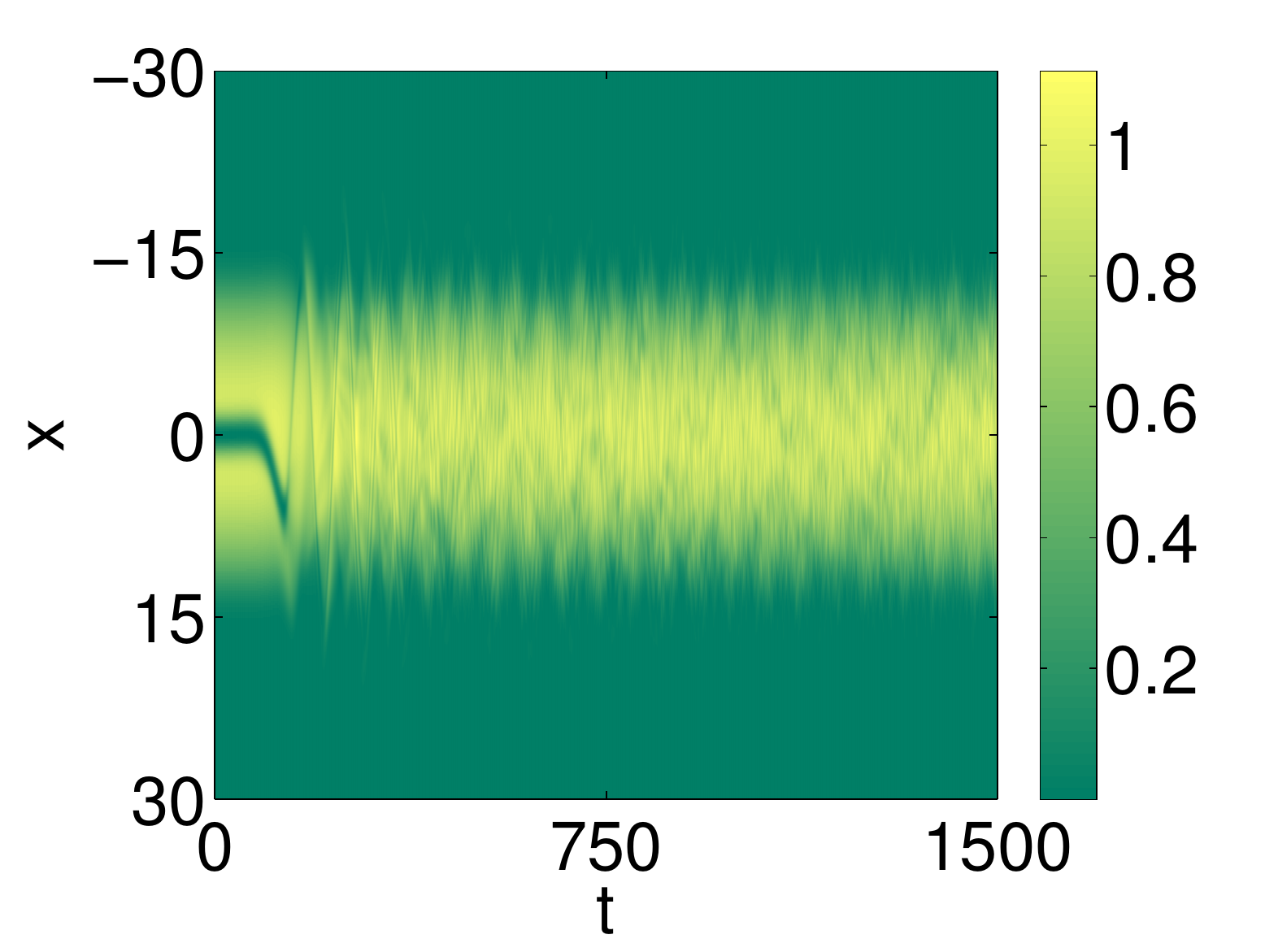}
\label{fig5f_trap}
}
\subfigure[][]{\hspace{-0.3cm}
\includegraphics[height=.18\textheight, angle =0]{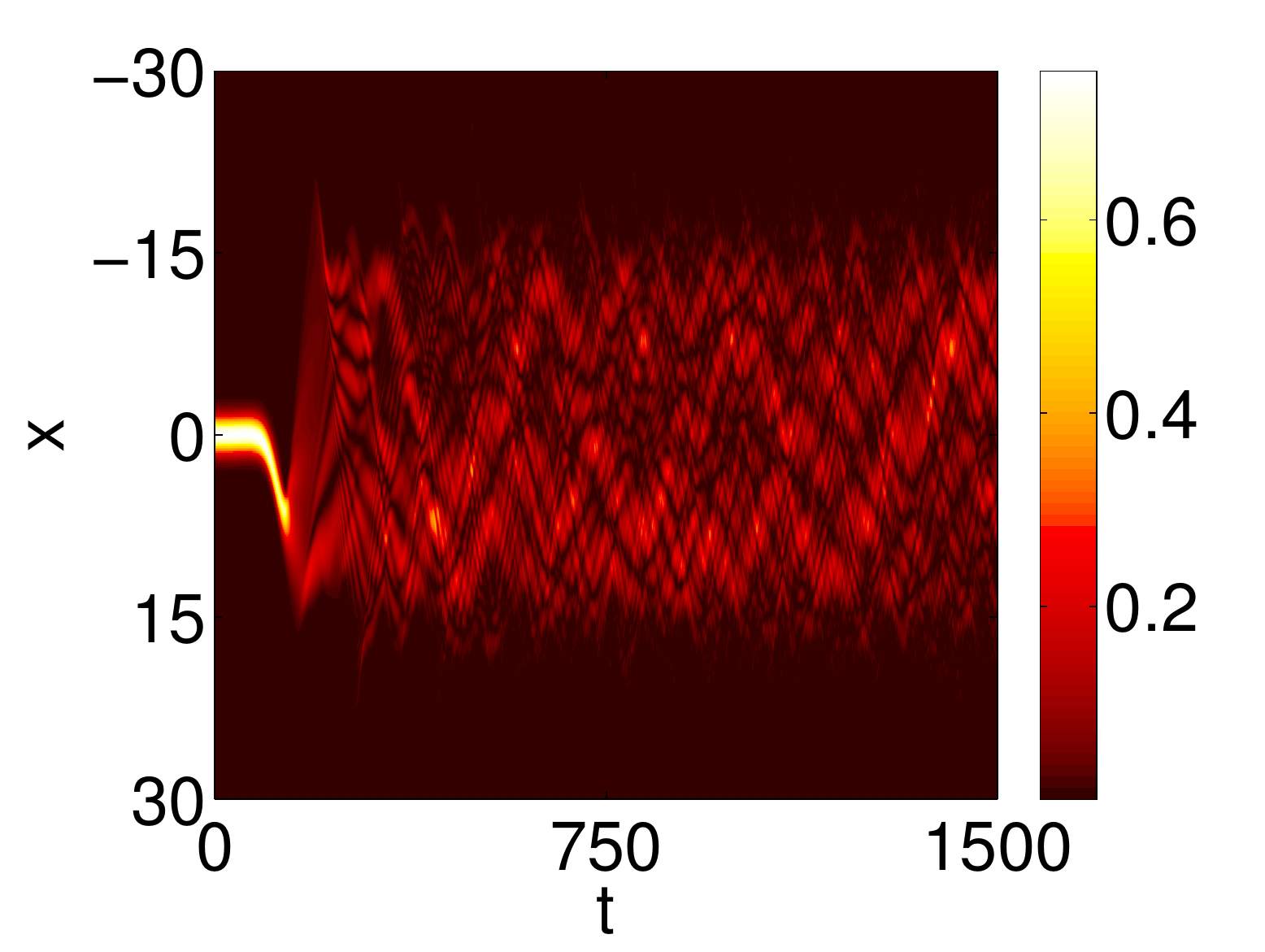}
\label{fig5g_trap}
}
\subfigure[][]{\hspace{-0.3cm}
\includegraphics[height=.18\textheight, angle =0]{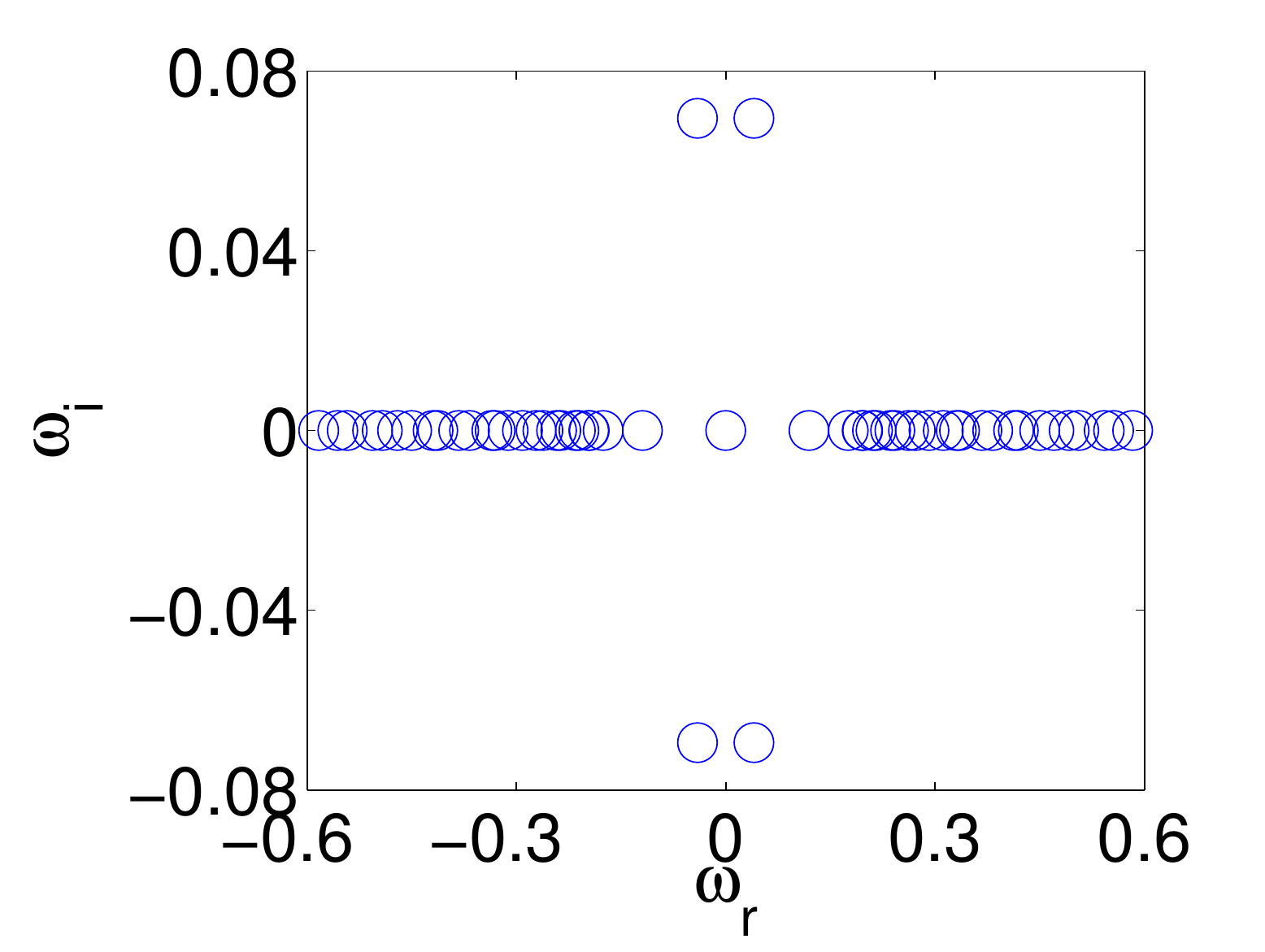}
\label{fig5h_trap}
}
}
\end{center}
\par
\vspace{-0.7cm}
\caption{(Color online) Summary of results corresponding to $n=0$
(i.e., the ground states) in the presence of the trapping potential
with $\Omega =0.1$. (a) Comparison of the steady-state profiles in
the absence (solid lines) and presence (dash-dotted lines) of the trap. 
In particular, the dark (black) and bright (blue) soliton solutions are
depicted for $D=0.2$ and $\protect\mu _{+}=0.51$. (b) Maximal imaginary
eigenfrequency as a function of the continuation parameter $\protect\mu_{+}$
at various values of $D$; the dash-dotted red line corresponds to the $D=0.2$ 
branch of Fig.~\protect\ref{fig1b}, for comparison. Also shown is the 
spatiotemporal evolution of densities $|\Phi _{-}(x,t)|^{2}$ (c) and (f)  
$|\Phi _{+}(x,t)|^{2}$ and (d) and (g) corresponding to perturbed soliton 
solutions, along with (e) and (h) the eigenfrequency spectrum of the steady 
states for (c)-(e) $D=0.2$ and $\protect\mu_{+}=0.36$ and (f)-(h) $D=0.2$ 
and $\protect\mu _{+}=0.8$.}
\label{fig5_trap}
\end{figure}

\begin{figure}[t]
\begin{center}
\vspace{-0.2cm}
\mbox{\hspace{-0.1cm}
\subfigure[][]{\hspace{-0.3cm}
\includegraphics[height=.18\textheight, angle =0]{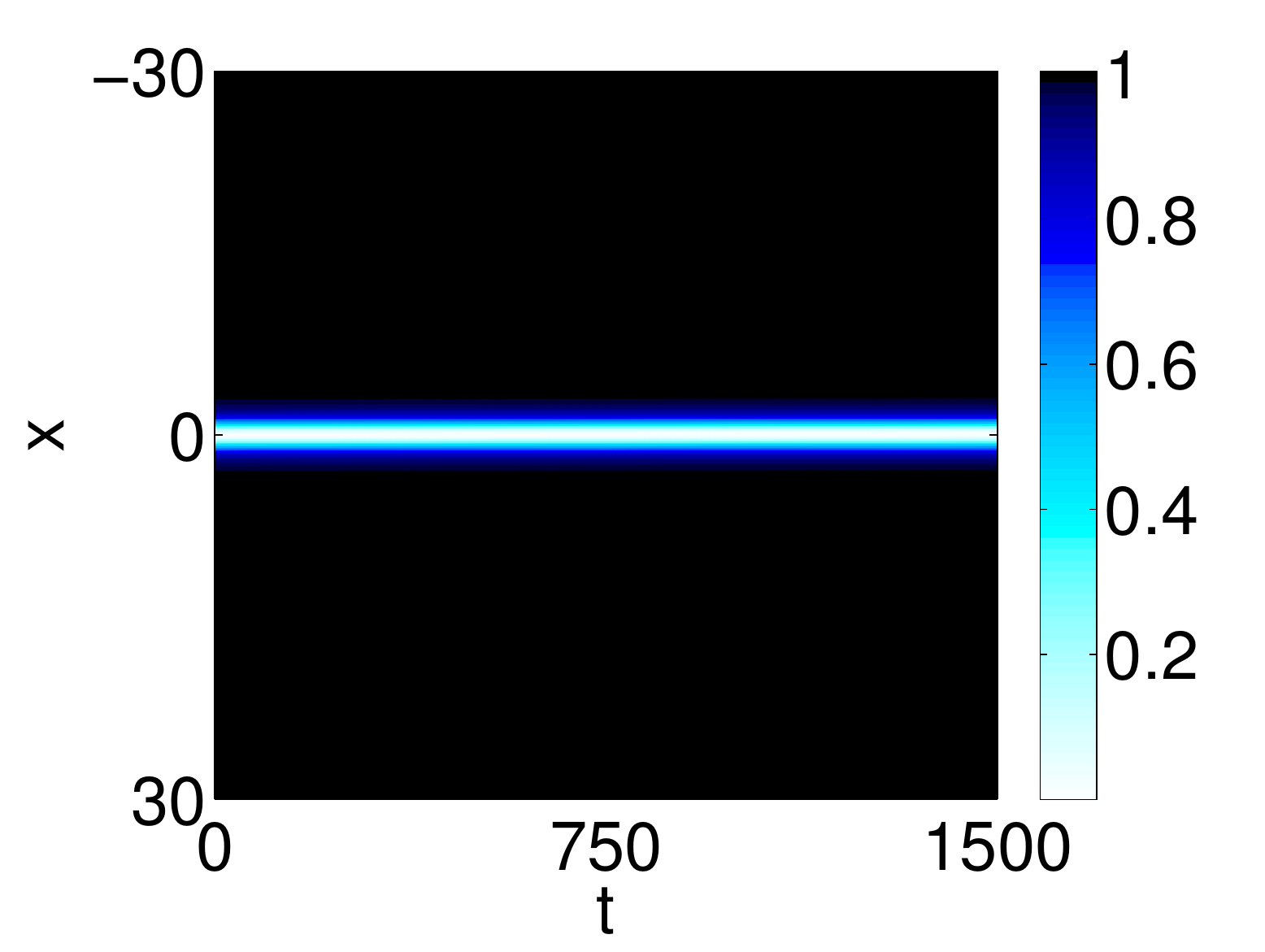}
\label{fig5a}
}
\subfigure[][]{\hspace{-0.3cm}
\includegraphics[height=.18\textheight, angle =0]{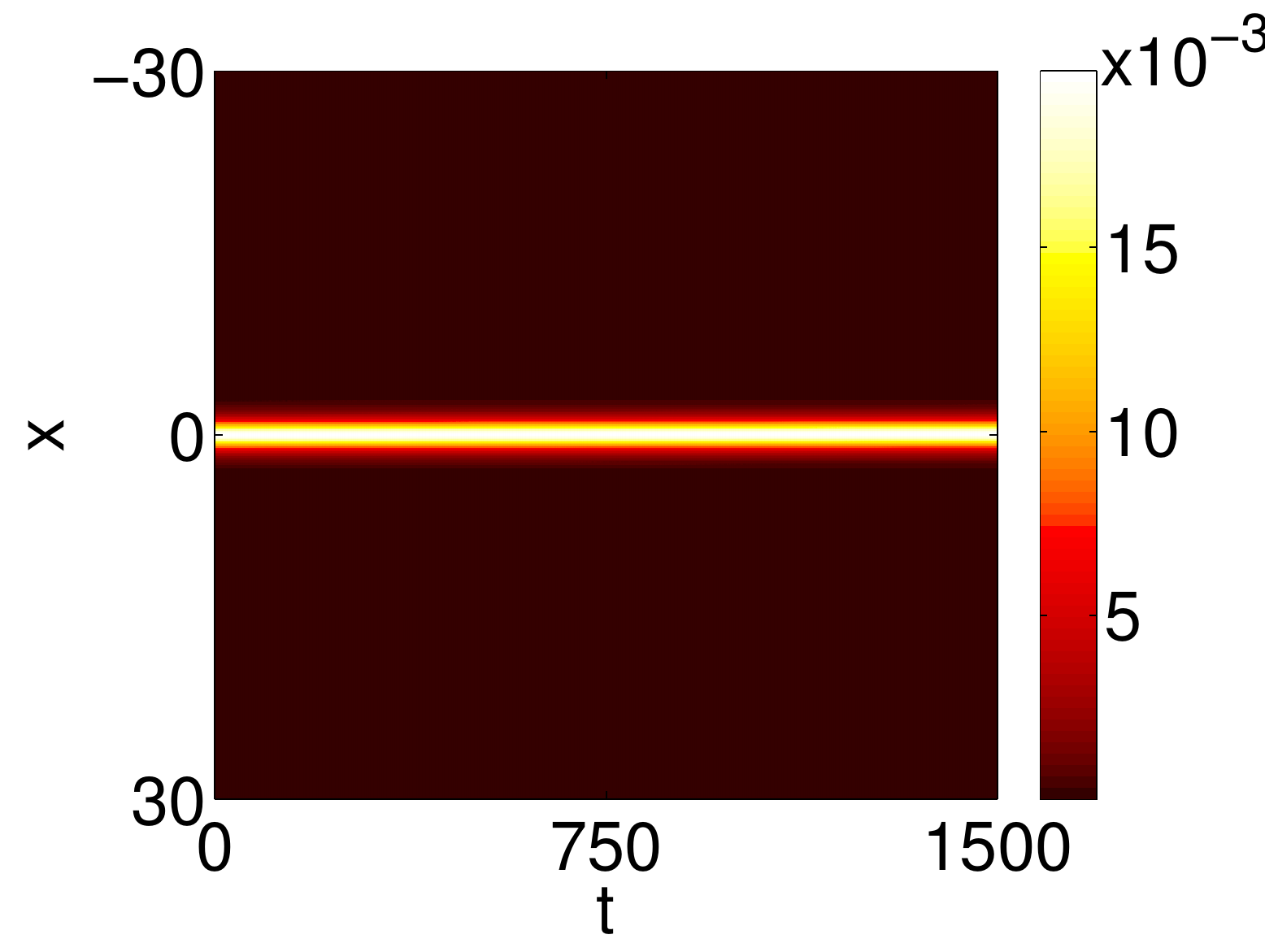}
\label{fig5b}
}
\subfigure[][]{\hspace{-0.3cm}
\includegraphics[height=.18\textheight, angle =0]{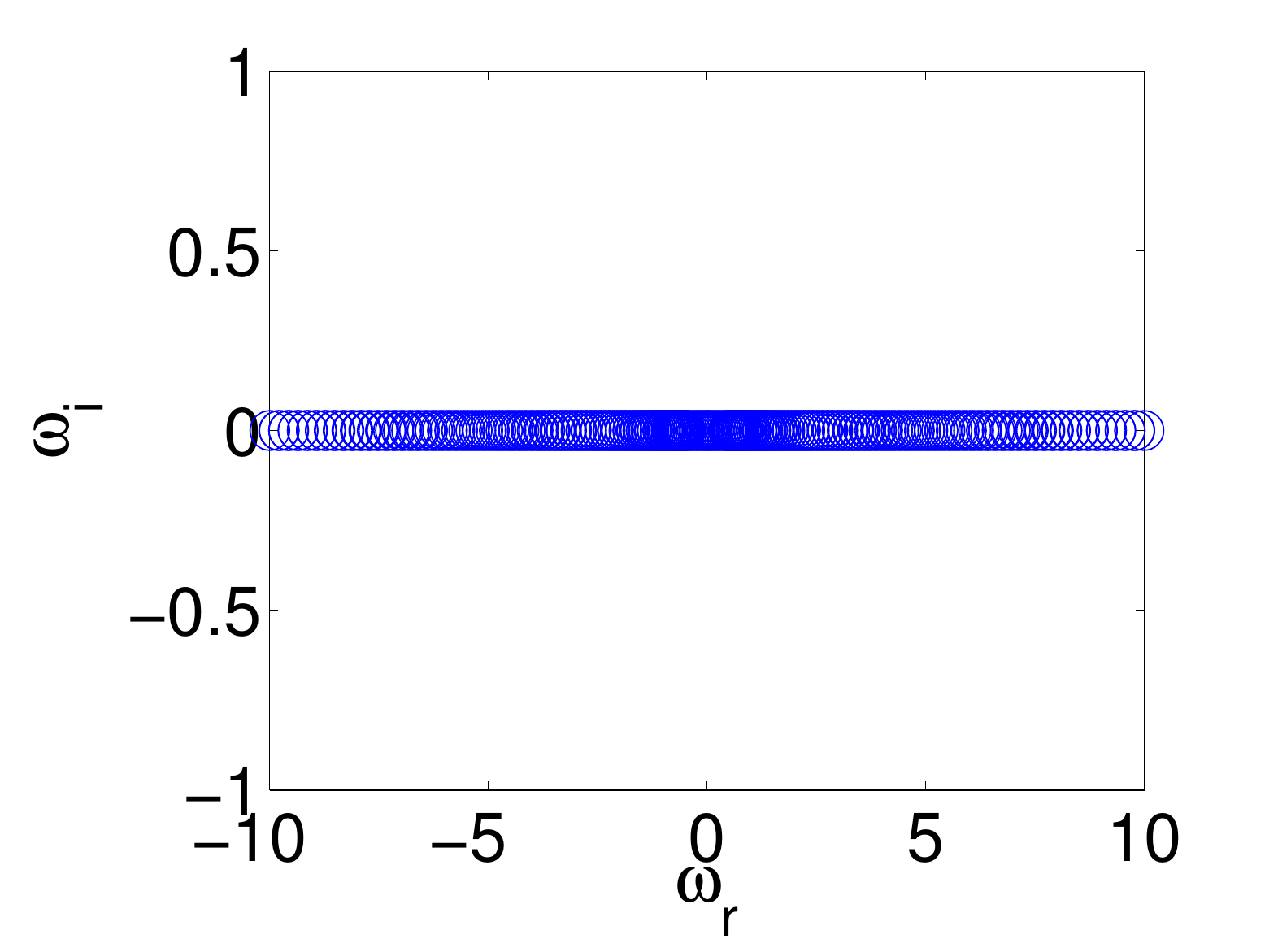}
\label{fig5c}
}
}
\mbox{\hspace{-0.1cm}
\subfigure[][]{\hspace{-0.3cm}
\includegraphics[height=.18\textheight, angle =0]{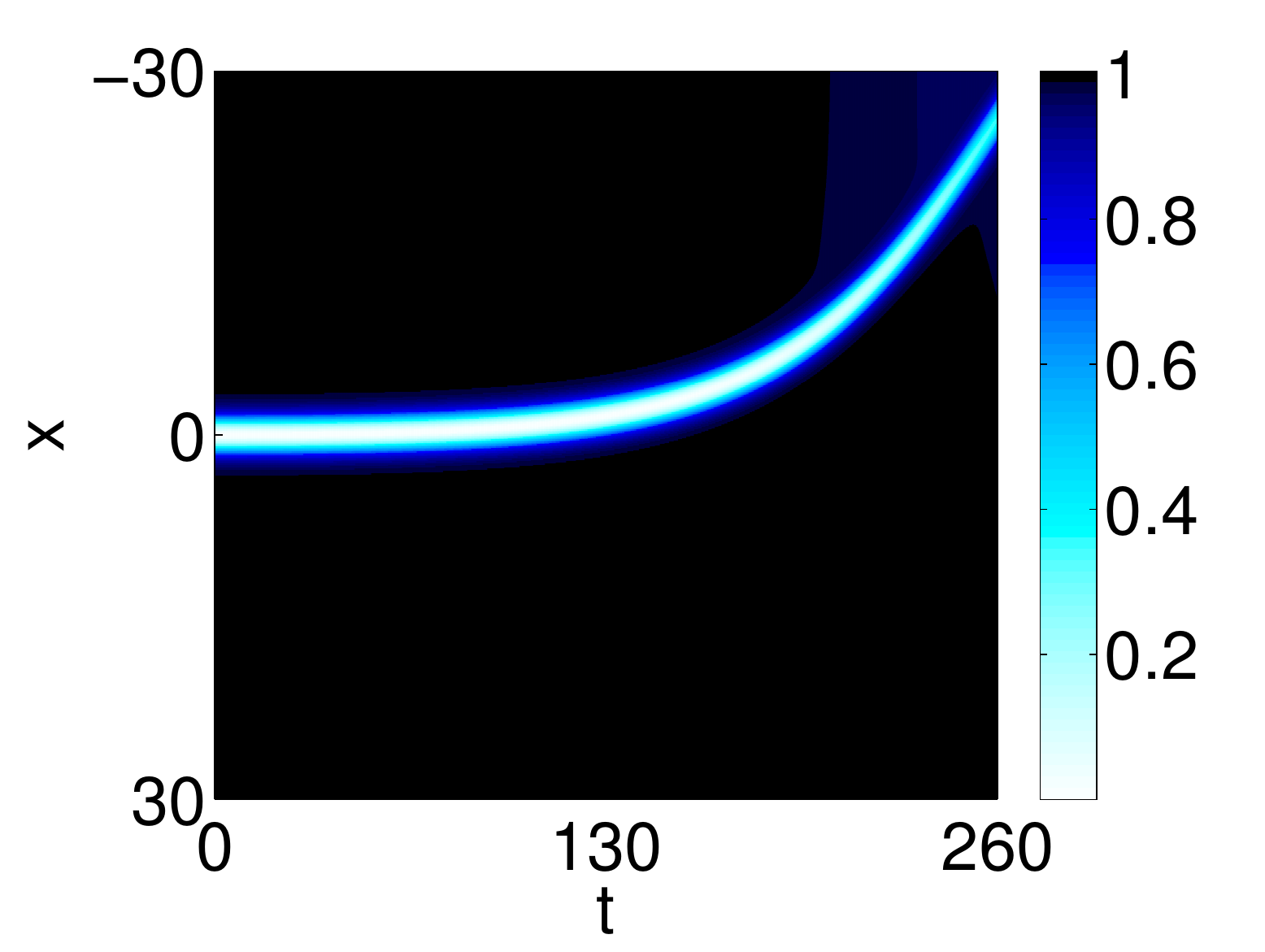}
\label{fig5d}
}
\subfigure[][]{\hspace{-0.3cm}
\includegraphics[height=.18\textheight, angle =0]{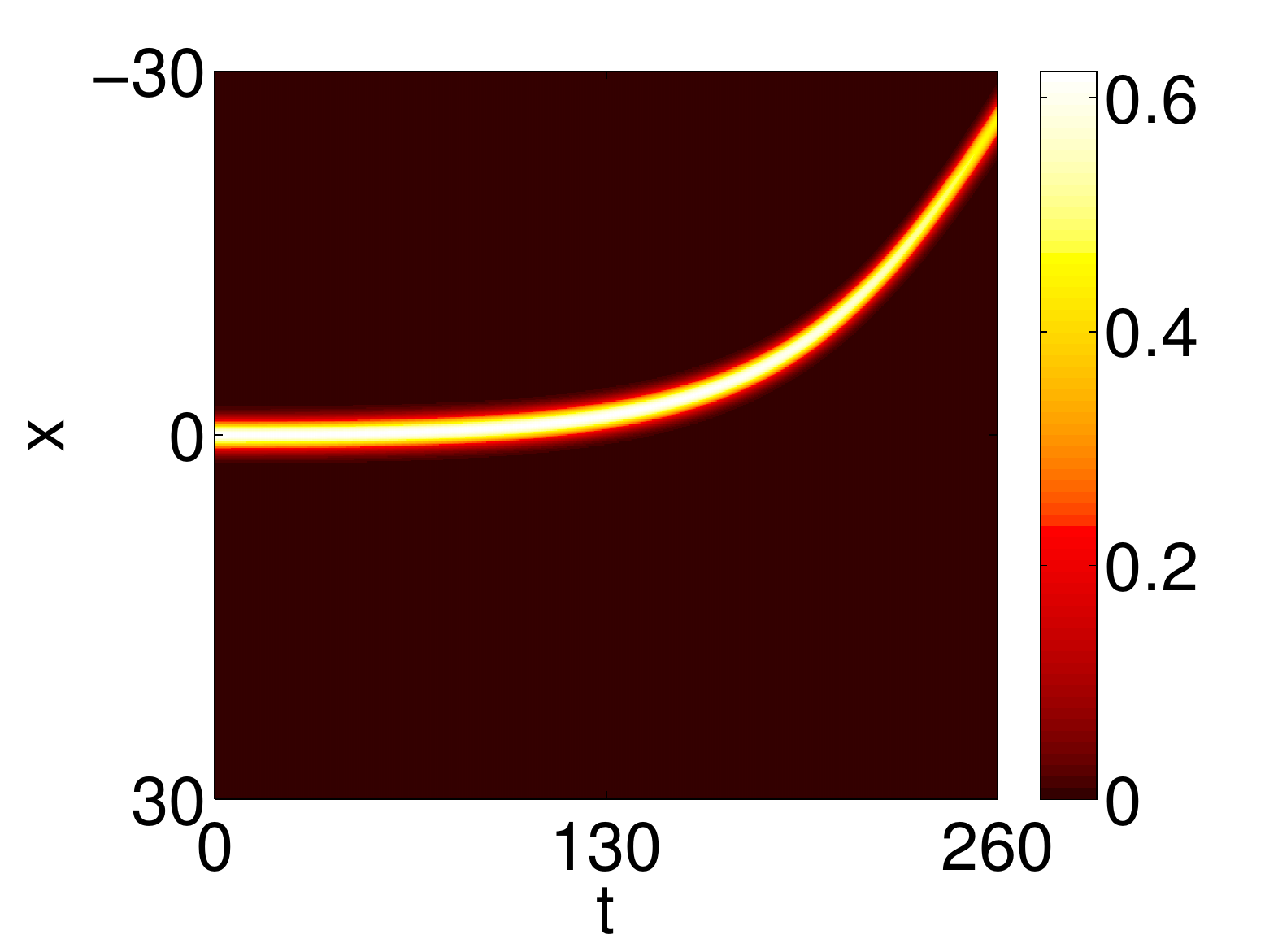}
\label{fig5e}
}
\subfigure[][]{\hspace{-0.3cm}
\includegraphics[height=.18\textheight, angle =0]{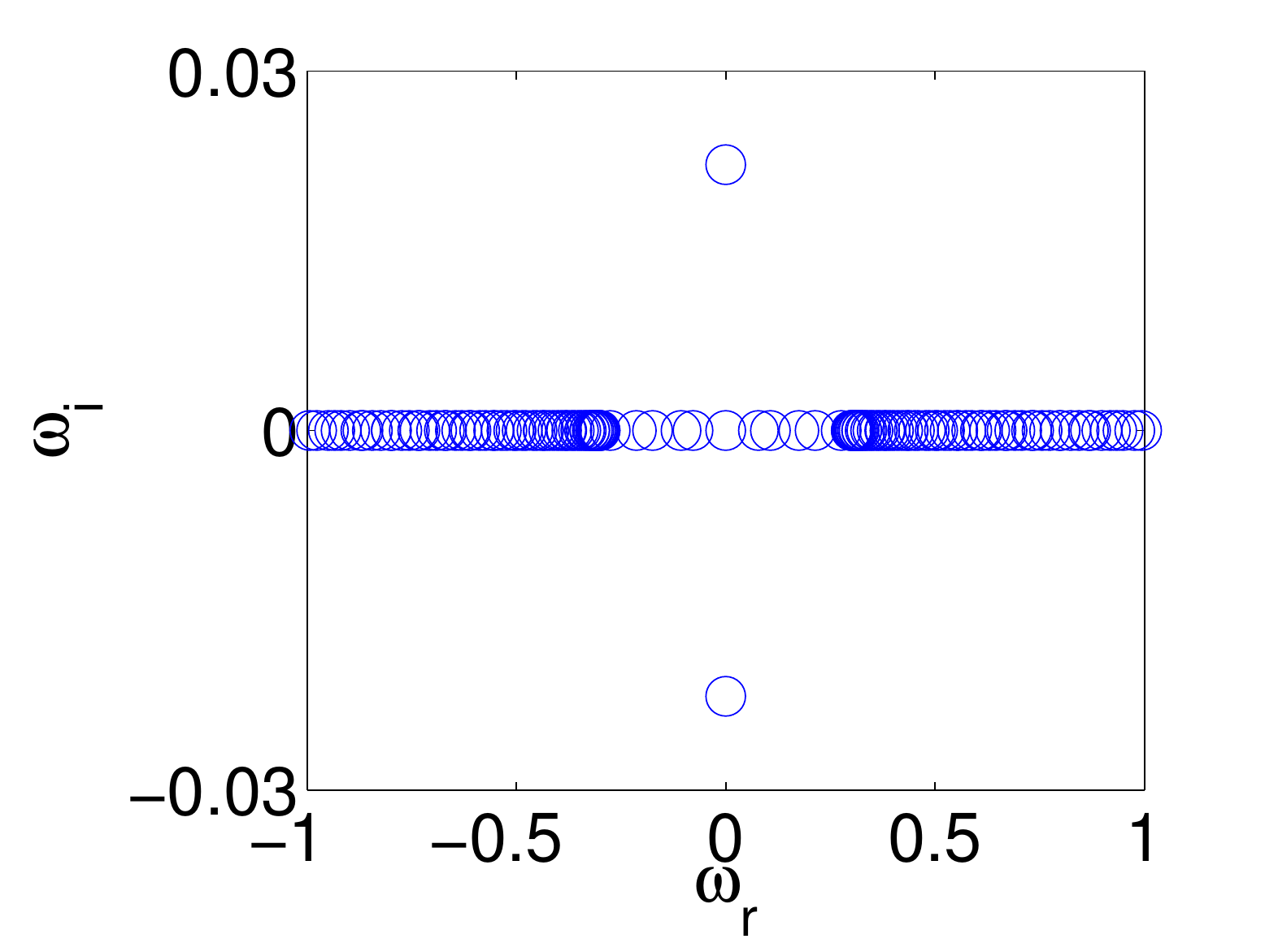}
\label{fig5f}
}
}
\end{center}
\par
\vspace{-0.7cm}
\caption{(Color online) Spatiotemporal evolution of densities $|\Phi_{-}(x,t)|^{2}$
(a) and (d), and $|\Phi _{+}(x,t)|^{2}$ (b) and (e) corresponding to 
(perturbed) soliton solutions of order $n=0$, as well as the eigenfrequency 
spectra (c) and (f) of the steady states for (a)-(c) $D=1$ and 
$\protect\mu _{+}=0.51$ and (d)-(f) $D=0.2$ and $\protect\mu _{+}=0.7$.}
\label{fig5}
\end{figure}

\begin{figure}[th]
\begin{center}
\vspace{-0.1cm}
\mbox{\hspace{-0.1cm}
\subfigure[][]{\hspace{-0.3cm}
\includegraphics[height=.18\textheight, angle =0]{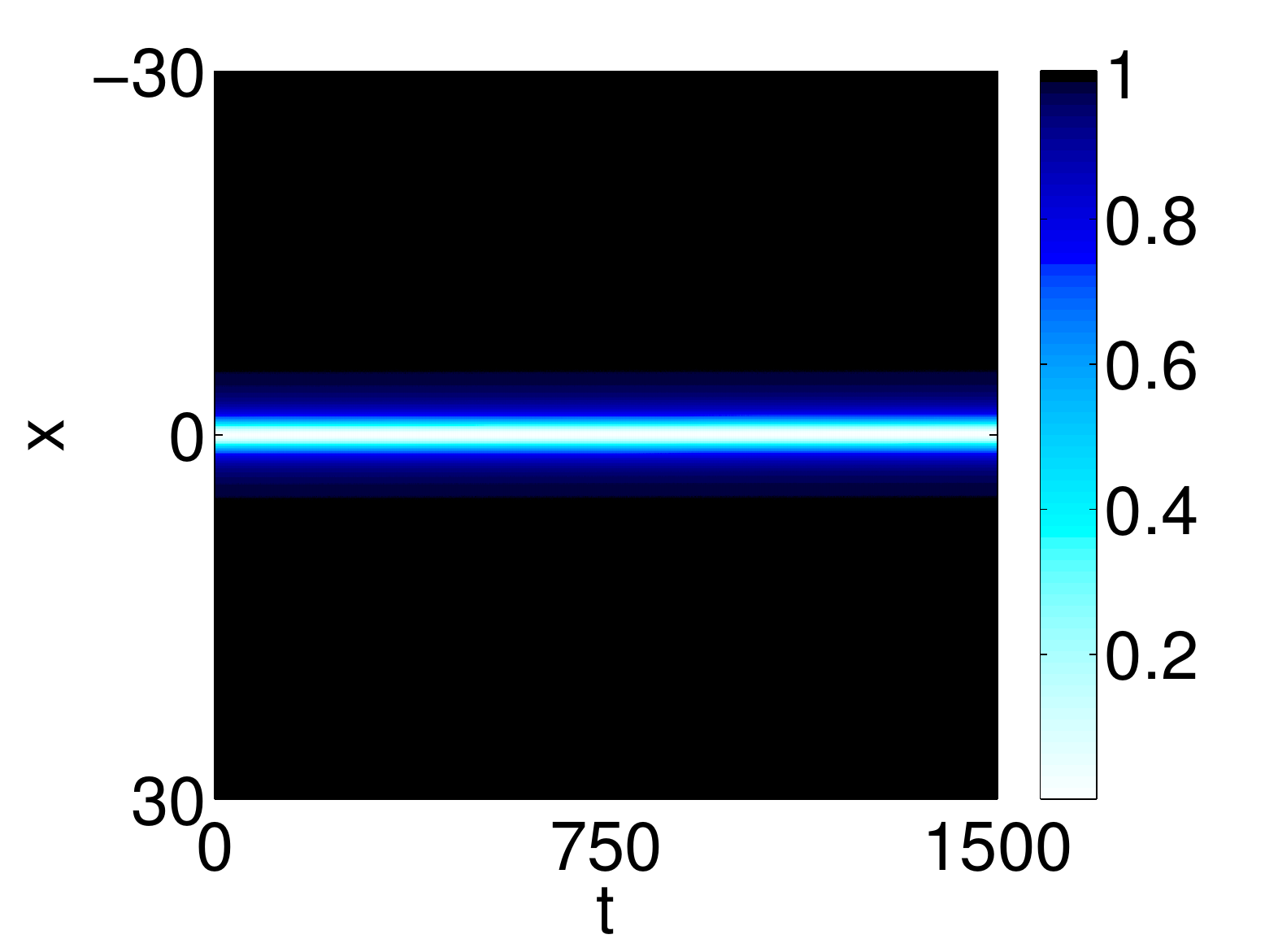}
\label{fig6a}
}
\subfigure[][]{\hspace{-0.3cm}
\includegraphics[height=.18\textheight, angle =0]{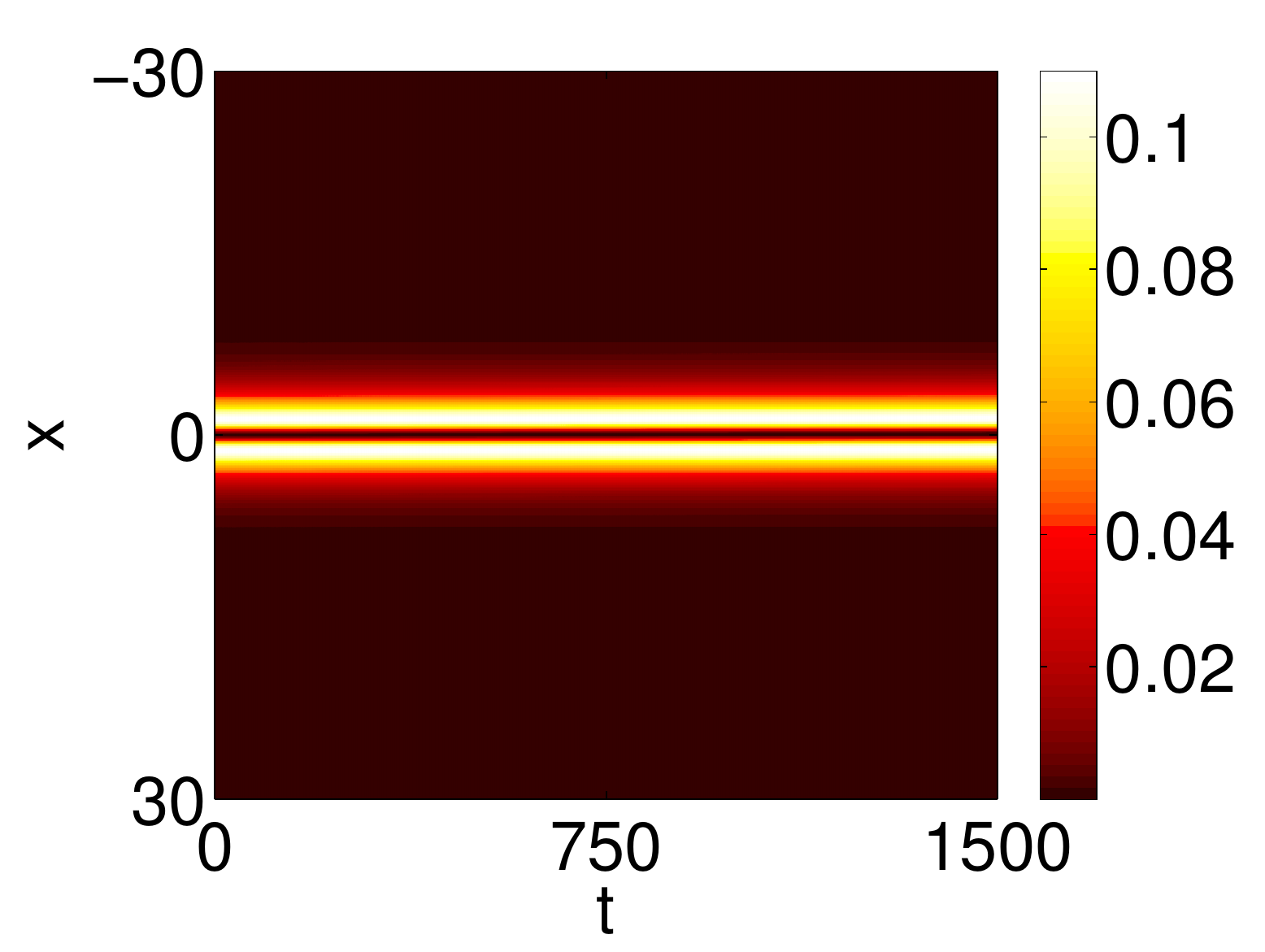}
\label{fig6b}
}
\subfigure[][]{\hspace{-0.3cm}
\includegraphics[height=.18\textheight, angle =0]{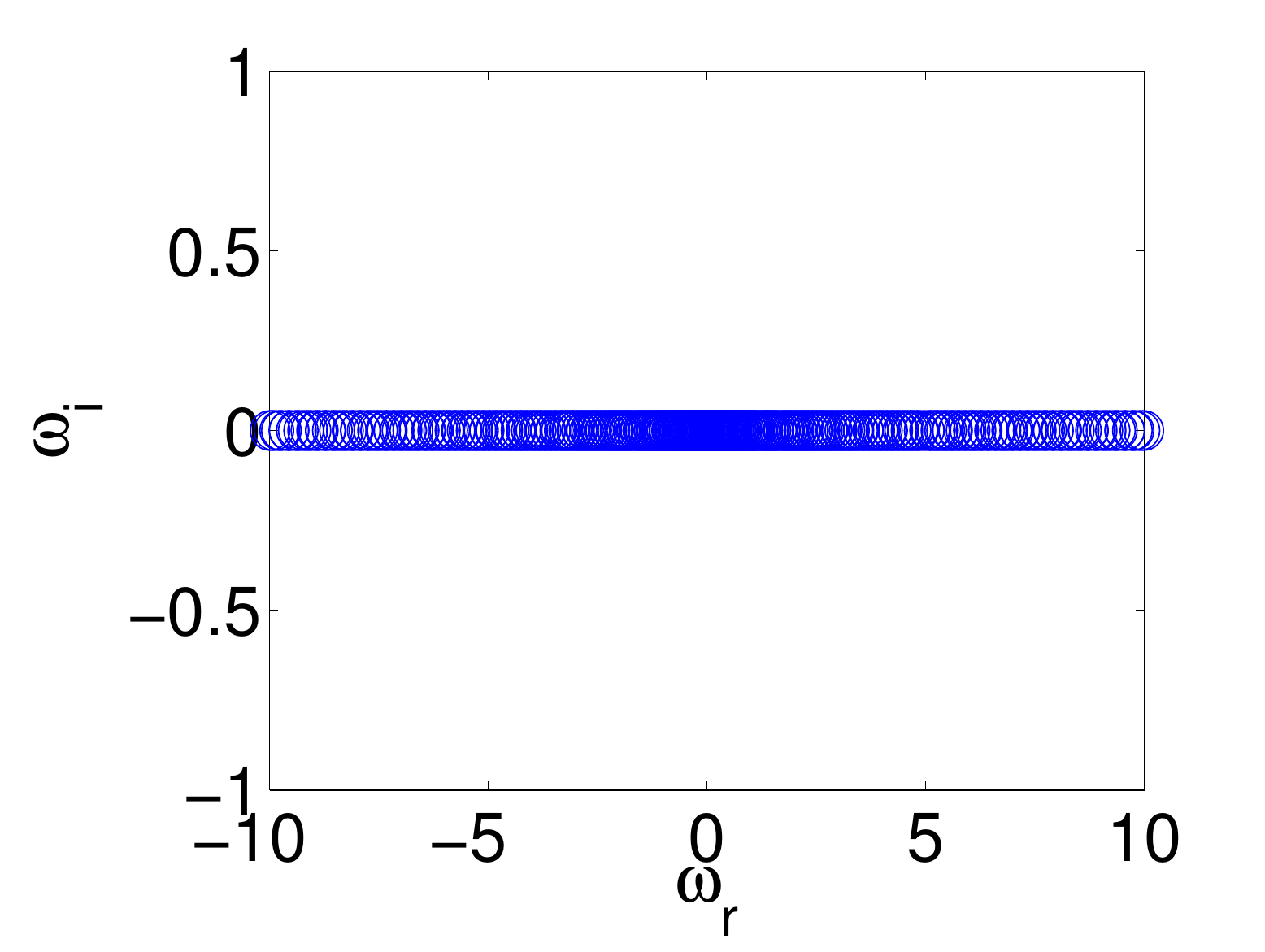}
\label{fig6c}
}
}
\mbox{\hspace{-0.1cm}
\subfigure[][]{\hspace{-0.3cm}
\includegraphics[height=.18\textheight, angle =0]{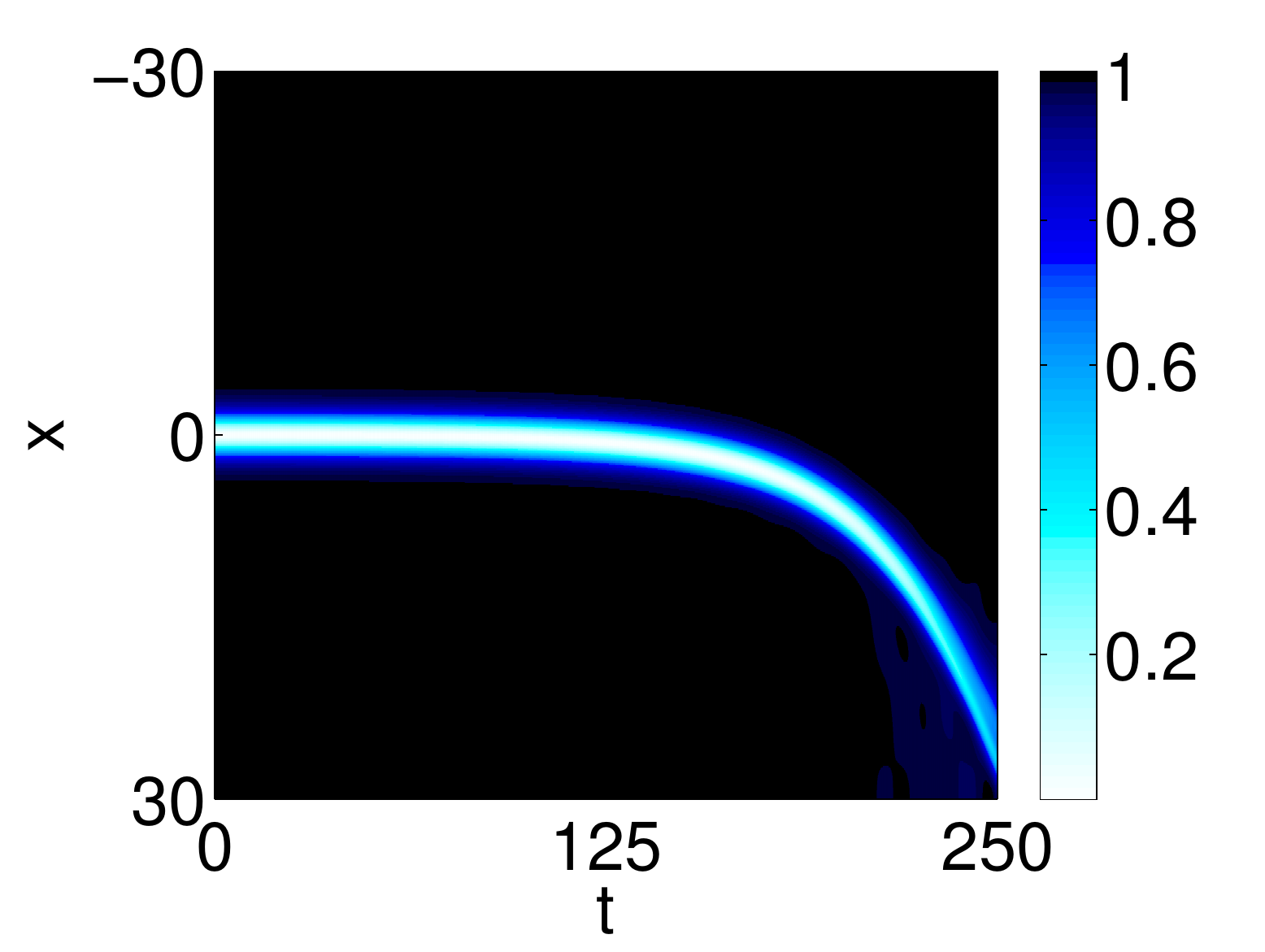}
\label{fig6d}
}
\subfigure[][]{\hspace{-0.3cm}
\includegraphics[height=.18\textheight, angle =0]{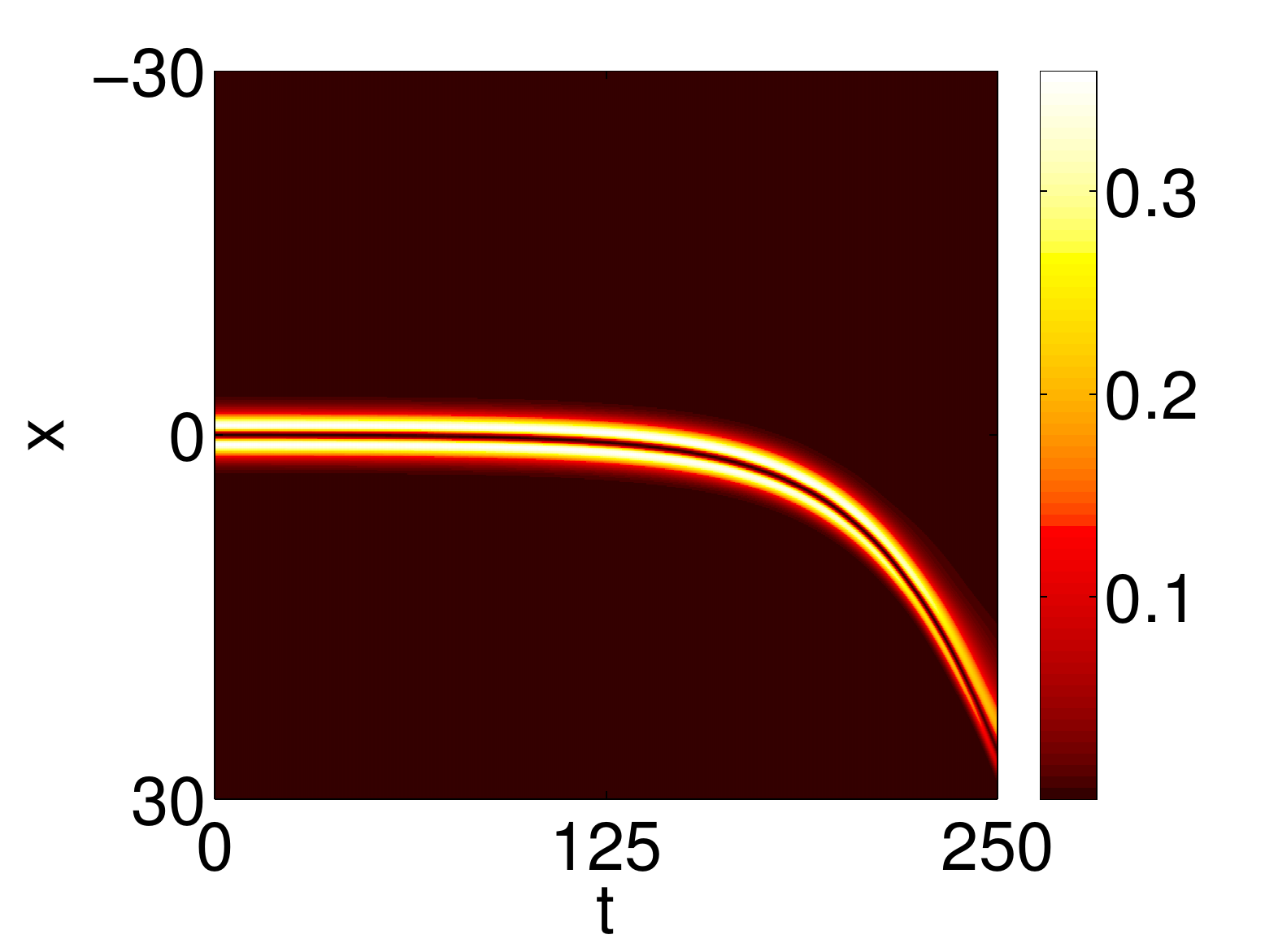}
\label{fig6e}
}
\subfigure[][]{\hspace{-0.3cm}
\includegraphics[height=.18\textheight, angle =0]{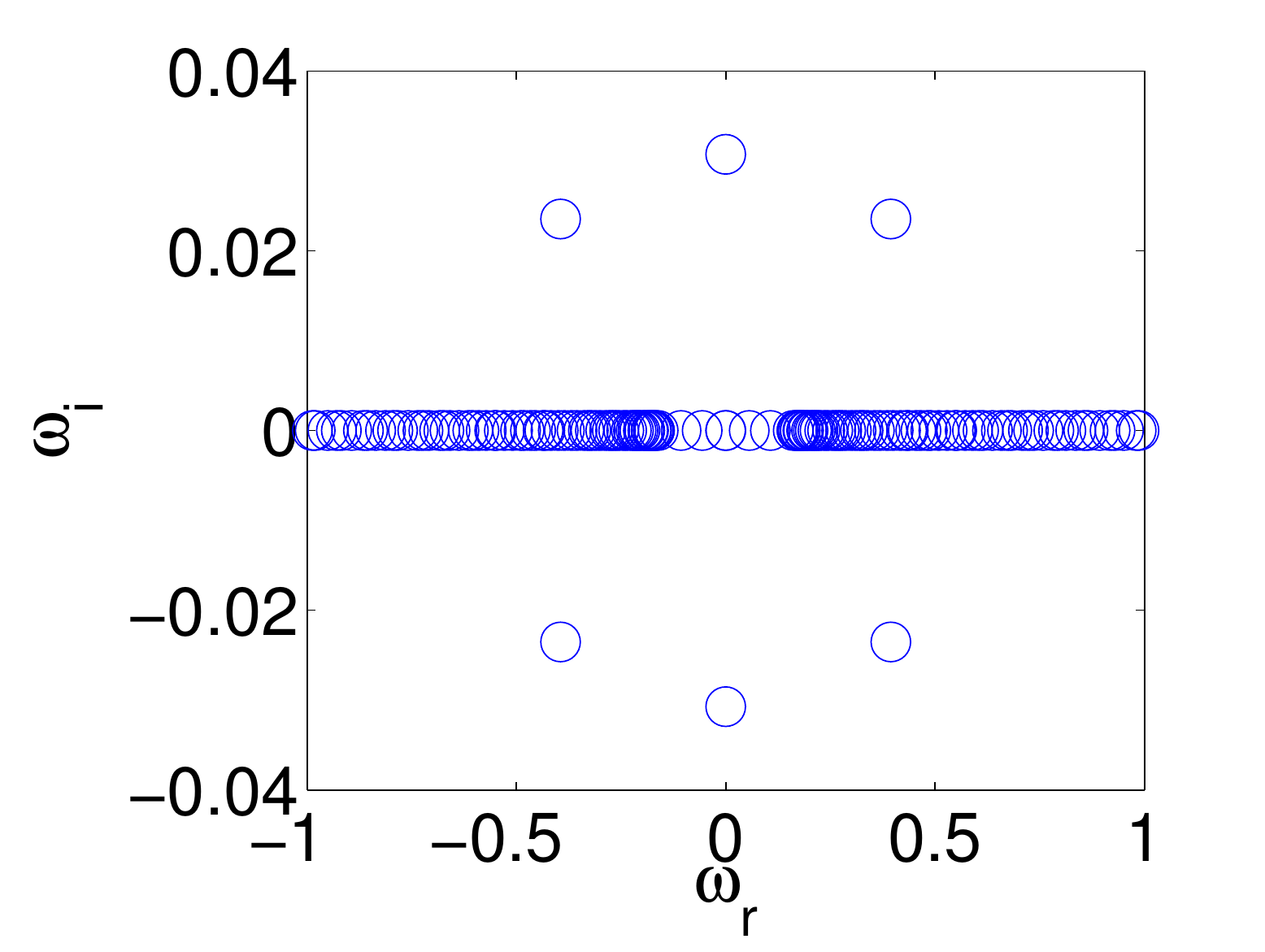}
\label{fig6f}
}
}
\mbox{\hspace{-0.1cm}
\subfigure[][]{\hspace{-0.3cm}
\includegraphics[height=.18\textheight, angle =0]{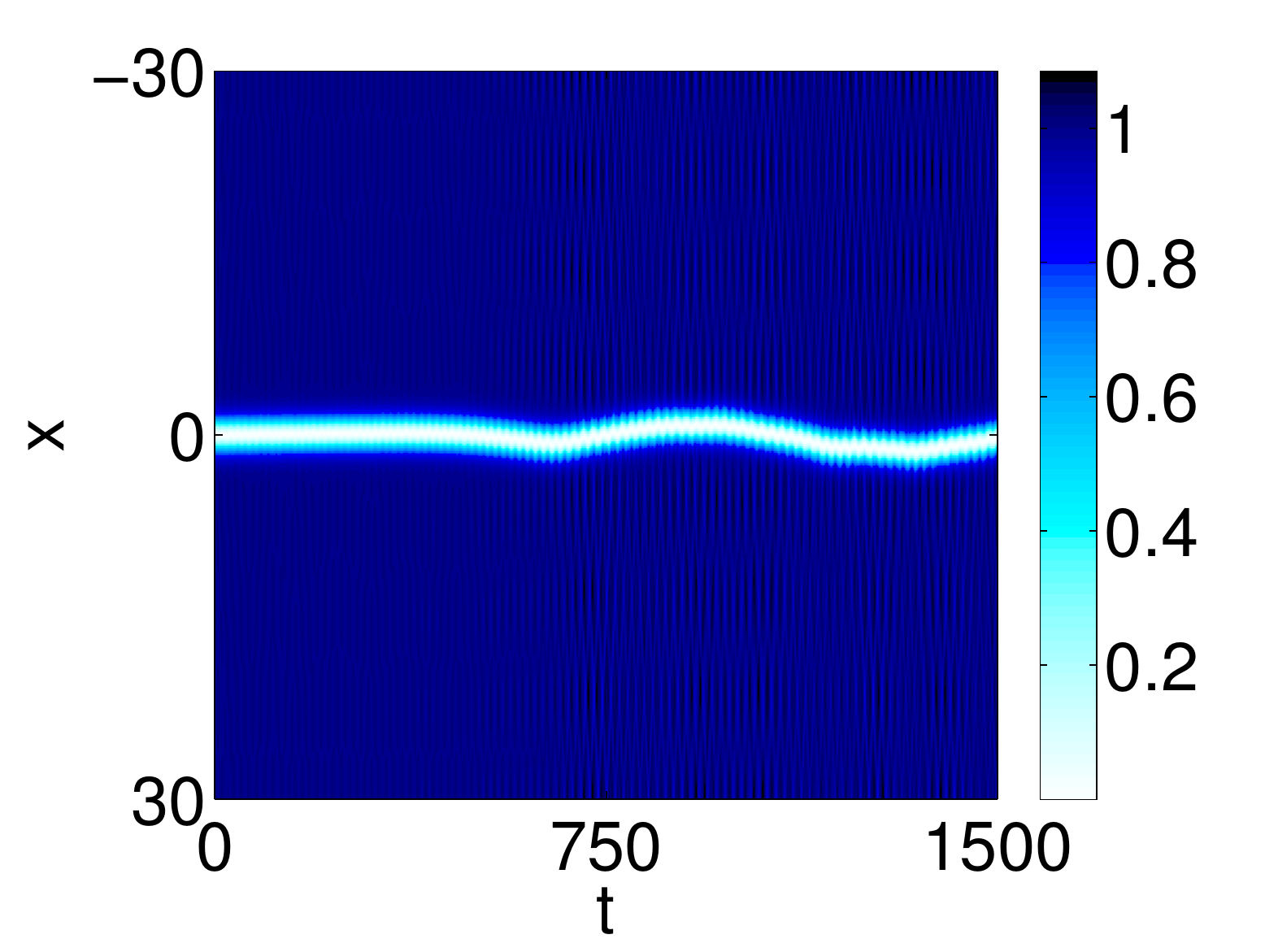}
\label{fig6g}
}
\subfigure[][]{\hspace{-0.3cm}
\includegraphics[height=.18\textheight, angle =0]{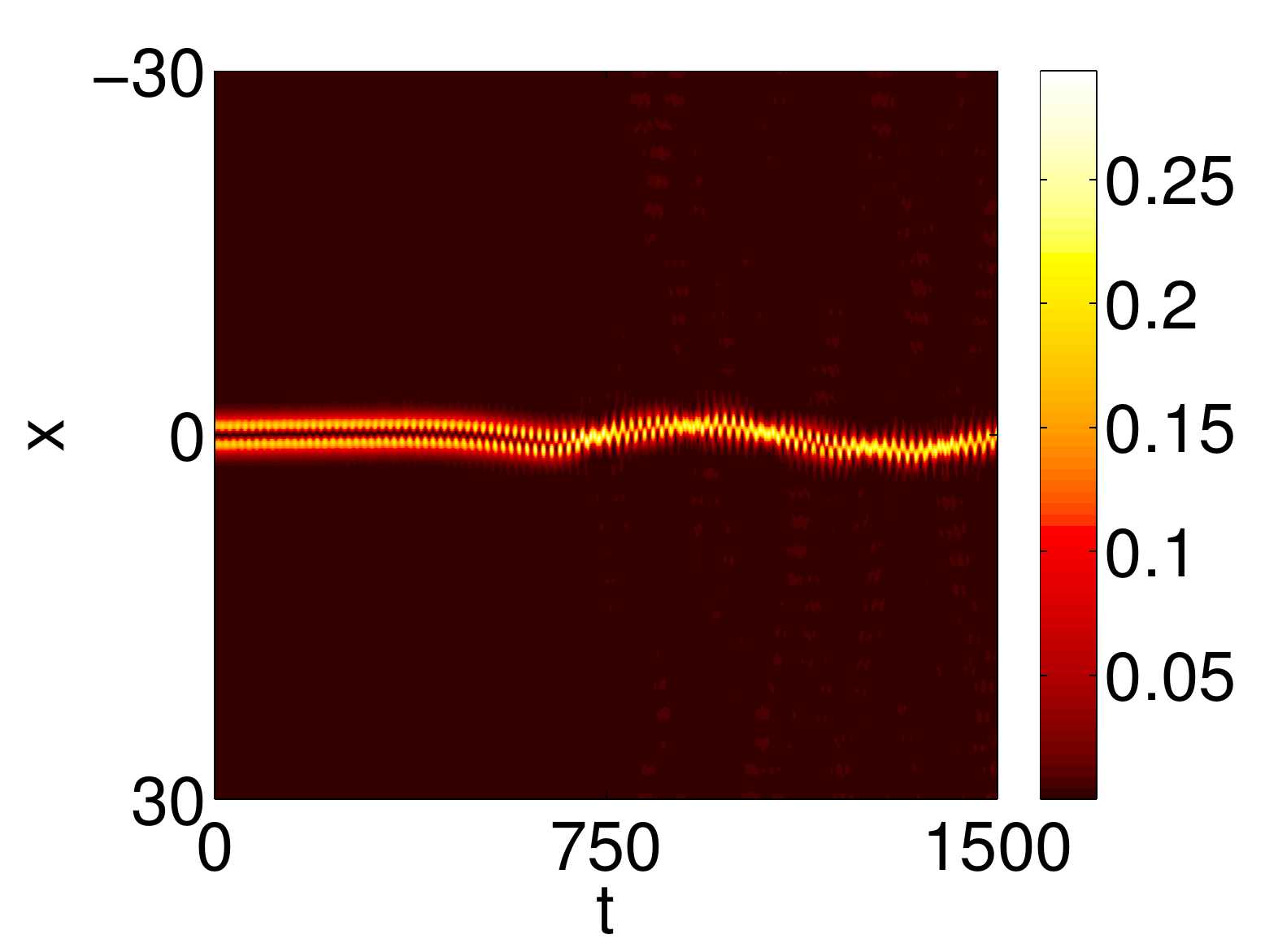}
\label{fig6h}
}
\subfigure[][]{\hspace{-0.3cm}
\includegraphics[height=.18\textheight, angle =0]{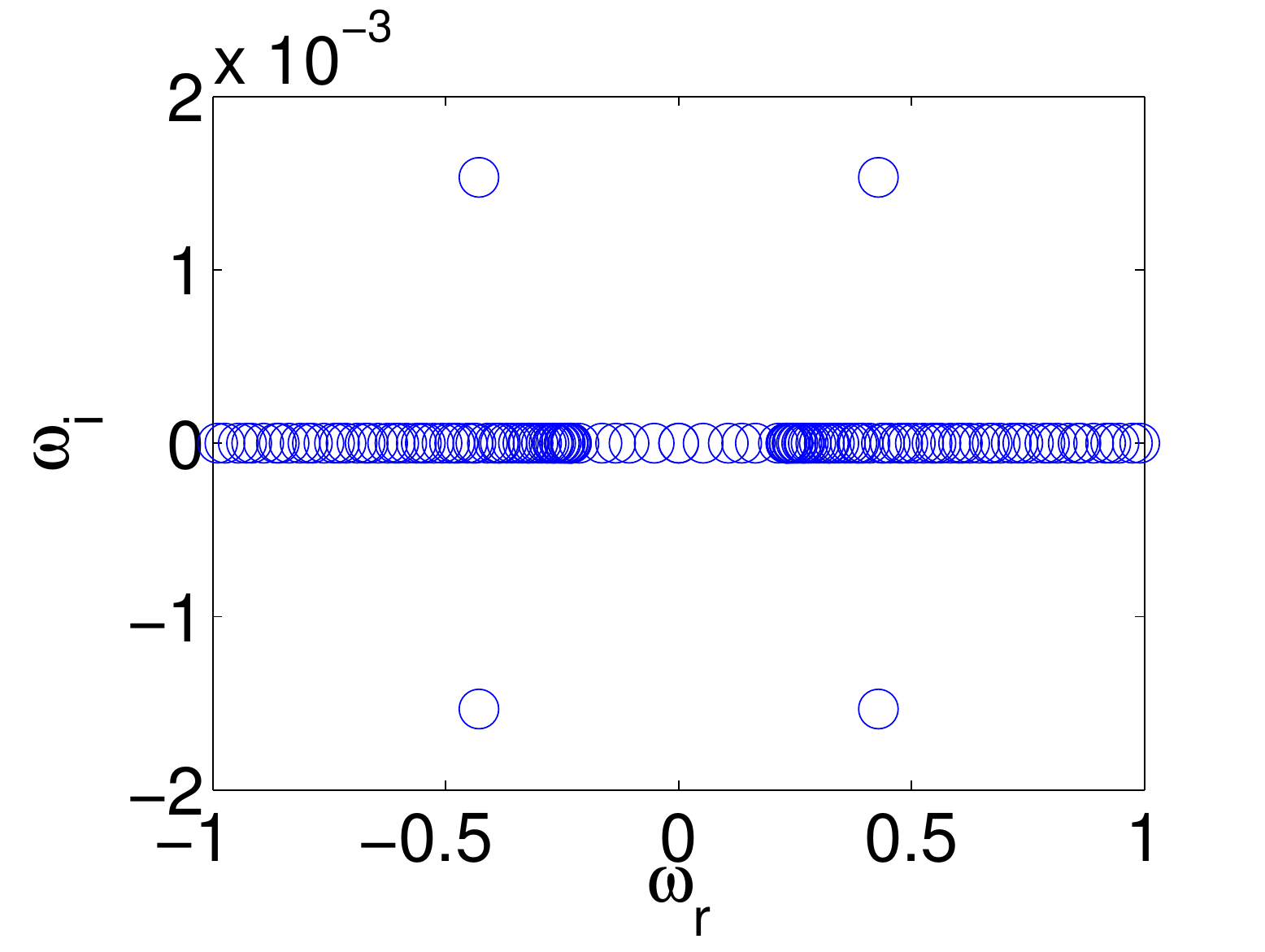}
\label{fig6i}
}
}
\end{center}
\par
\vspace{-0.7cm}
\caption{(Color online) Same as in Fig.~\protect\ref{fig5}, but for
soliton solutions of order $n=1$: spatiotemporal evolution of densities
$|\Phi _{-}(x,t)|^{2}$ (a), (d), and (g), and $|\Phi_{+}(x,t)|^{2}$ (b), 
(e), and (h). Panels (c), (f), and (i) display the eigenfrequency spectra
of the steady-state solutions for (a)-(c) $D=0.6$ and $\protect\mu _{+}=0.96$, 
(d)-(f) $D=0.2$ and $\protect\mu _{+}=0.83$, and (g)-(i) $D=0.2$ and $0.77$.}
\label{fig6}
\end{figure}

In a similar fashion, we also present results on steady-state solutions for
bound states of order $n=0$ in Figs.~\ref{fig5a_trap} and \ref{fig5b_trap} in
the presence of the trapping potential in Eqs.~(\ref{1}) and (\ref{2}) with 
$\Omega=0.1$. Specifically, Fig.~\ref{fig5a_trap} displays trapped DB soliton
solutions (dash-dotted black and blue lines, respectively) which, according 
to our stability analysis, are fairly stable in the absence of the trap
[cf. Fig.~\ref{fig1b}]. 
The naive intuition here would be that the trap would only contribute to the
stability of the configuration. Yet exactly the opposite is happening here. 
In particular, deeper intuition suggests that the trap contributes to the 
breaking of the translational invariance of the system, releasing a negative-energy
(negative-Krein-signature) mode along the imaginary axis of perturbation
eigenvalues (see also the discussion in Ref.~\cite{peter2}). Upon variation of
parameters, such as the chemical potential, this eigenvalue collides with other 
ones that correspond to positive energy, giving rise to instability quartets. 
Thus, while the $D=1$ case is, as is well known from previous studies of DB 
solitons in BECs (see, e.g., ~Refs.\cite{buschanglin,peter2}), generally stable, 
for other values of $D$, the setting with the trap is considerably less robust 
than in the homogeneous limit, where the translational invariance absorbs this 
potentially dangerous eigendirection.
This scenario is depicted in Fig.~\ref{fig5b_trap}, which complements the existence 
results presenting stability characteristics, namely, the dominant unstable
eigenfrequency. Note that in Fig.~\ref{fig5b_trap} we include the $D=0.2$ 
branch of Fig.~\ref{fig1b} denoted by a dash-dotted red line, for comparison.

\begin{figure}[th]
\begin{center}
\vspace{-0.1cm}
\mbox{\hspace{-0.1cm}
\subfigure[][]{\hspace{-0.3cm}
\includegraphics[height=.18\textheight, angle =0]{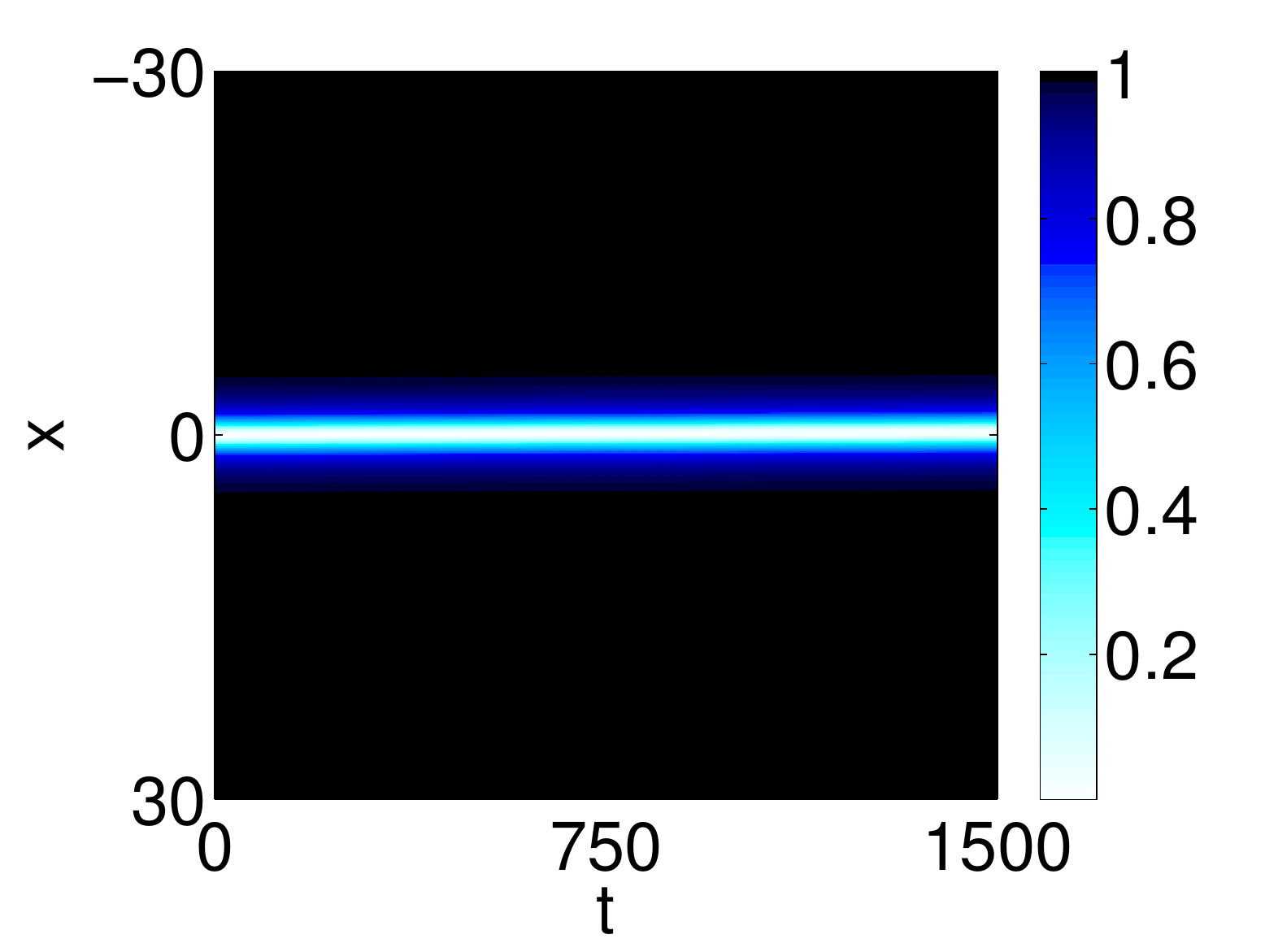}
\label{fig7a}
}
\subfigure[][]{\hspace{-0.3cm}
\includegraphics[height=.18\textheight, angle =0]{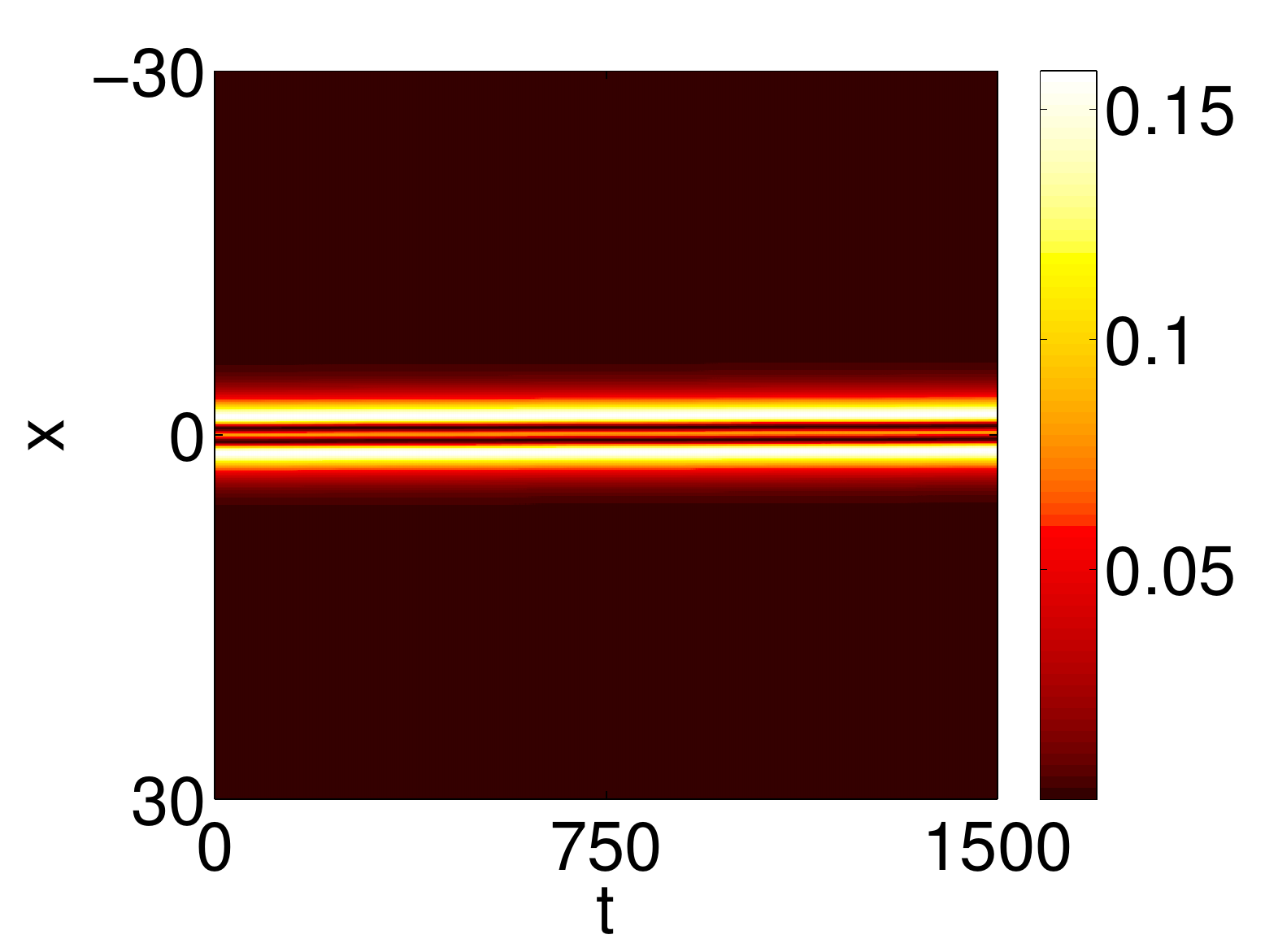}
\label{fig7b}
}
\subfigure[][]{\hspace{-0.3cm}
\includegraphics[height=.18\textheight, angle =0]{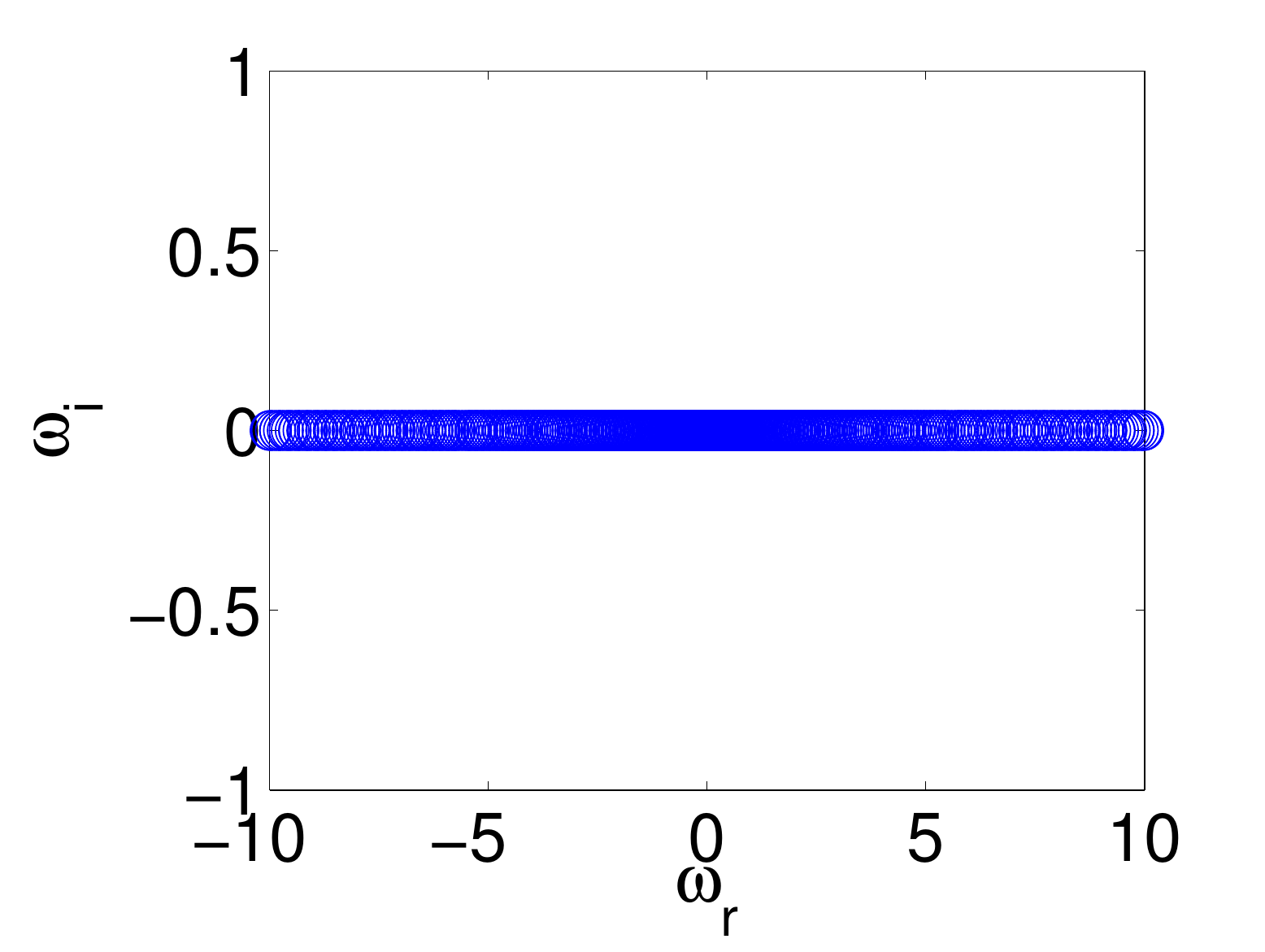}
\label{fig7c}
}
}
\mbox{\hspace{-0.1cm}
\subfigure[][]{\hspace{-0.3cm}
\includegraphics[height=.18\textheight, angle =0]{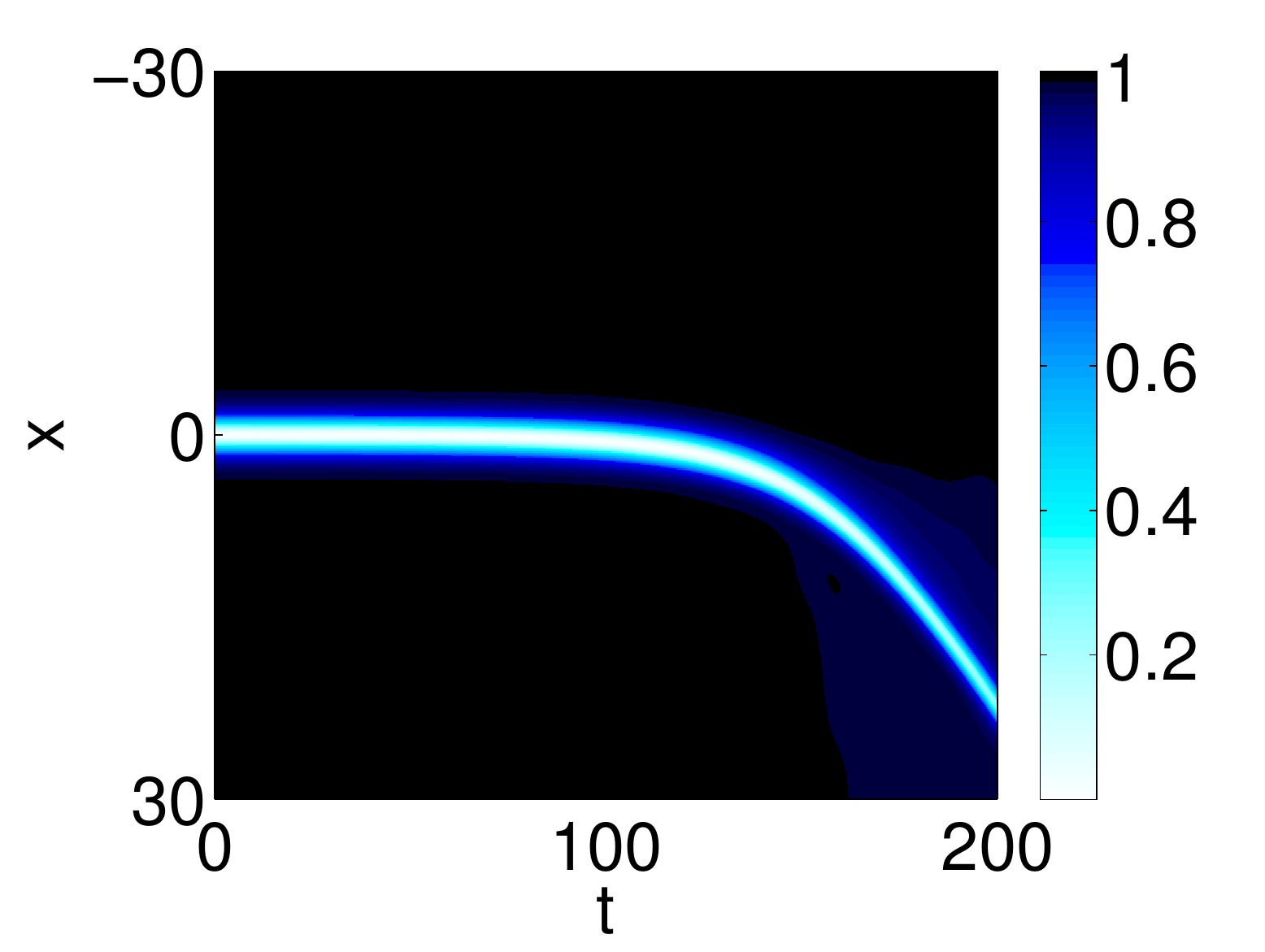}
\label{fig7d}
}
\subfigure[][]{\hspace{-0.3cm}
\includegraphics[height=.18\textheight, angle =0]{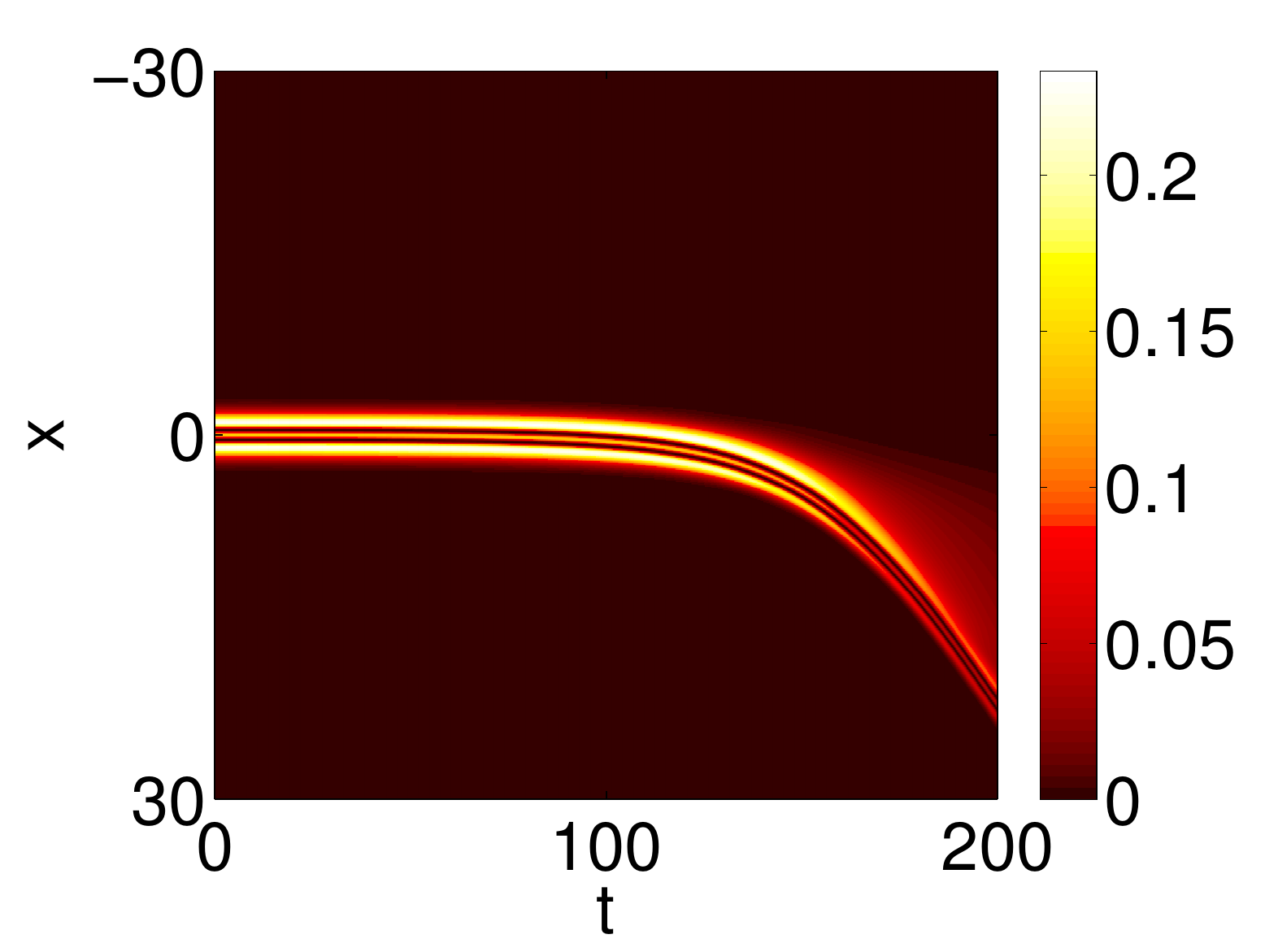}
\label{fig7e}
}
\subfigure[][]{\hspace{-0.3cm}
\includegraphics[height=.18\textheight, angle =0]{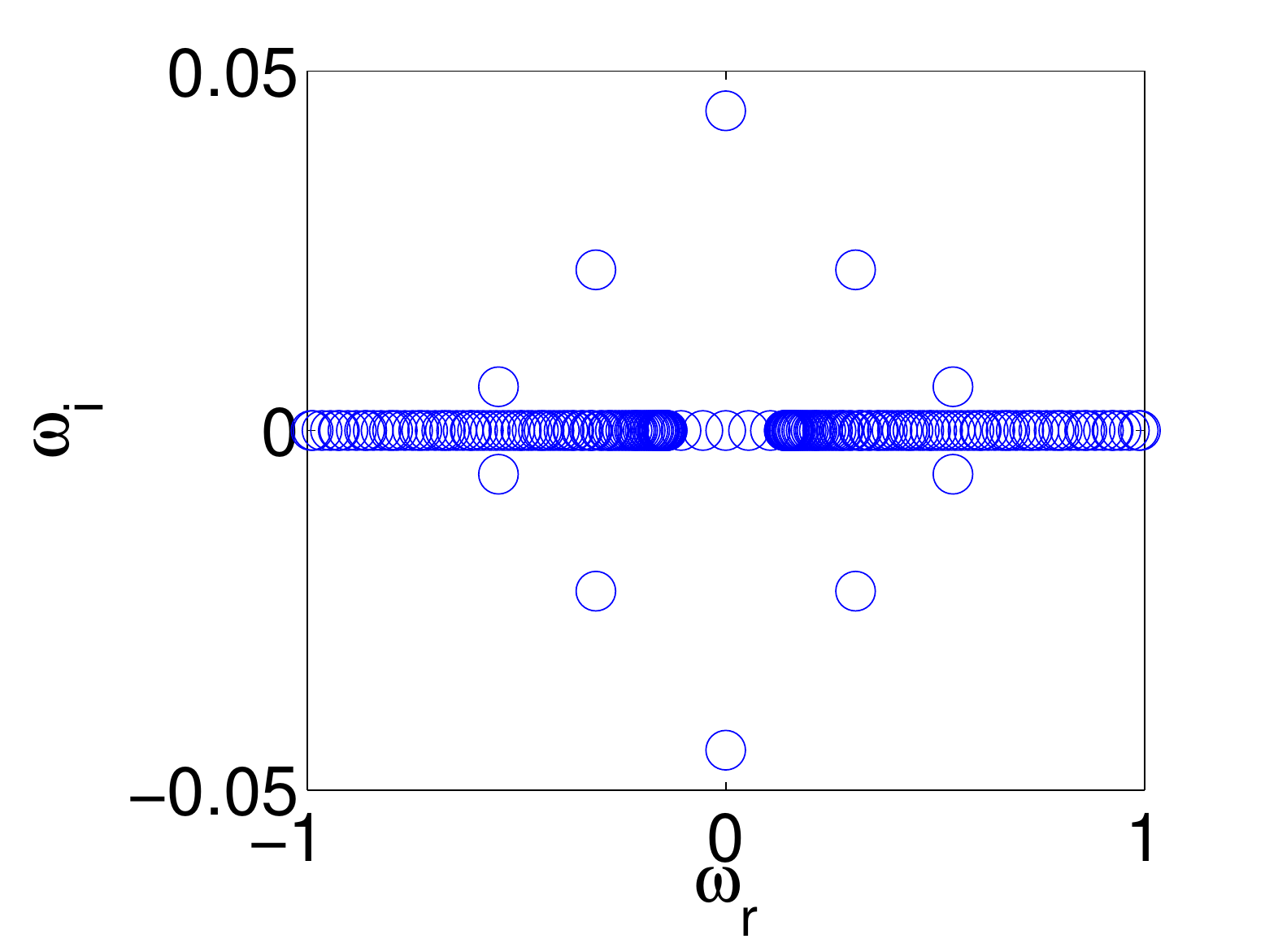}
\label{fig7f}
}
}
\mbox{\hspace{-0.1cm}
\subfigure[][]{\hspace{-0.3cm}
\includegraphics[height=.18\textheight, angle =0]{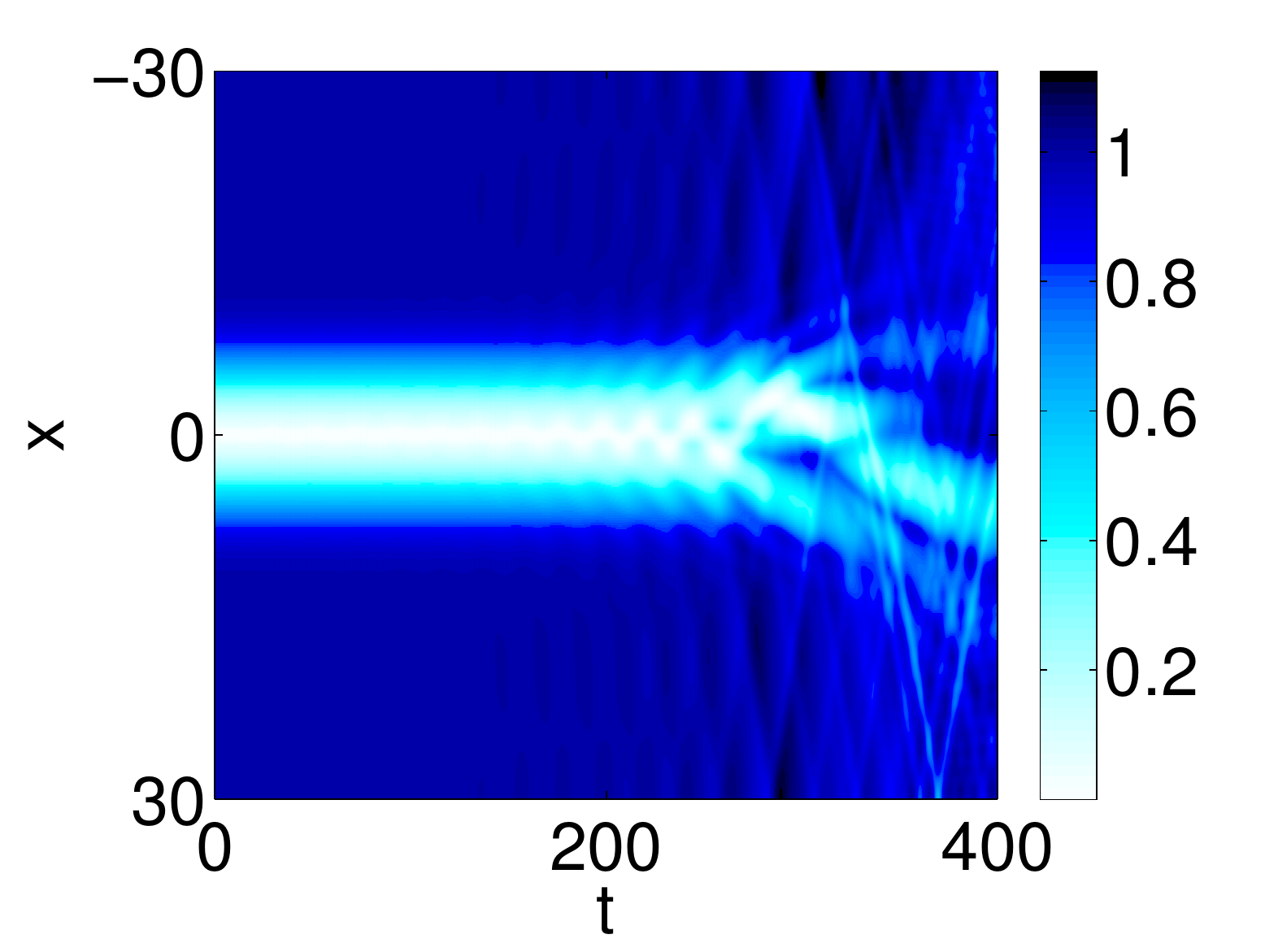}
\label{fig7g}
}
\subfigure[][]{\hspace{-0.3cm}
\includegraphics[height=.18\textheight, angle =0]{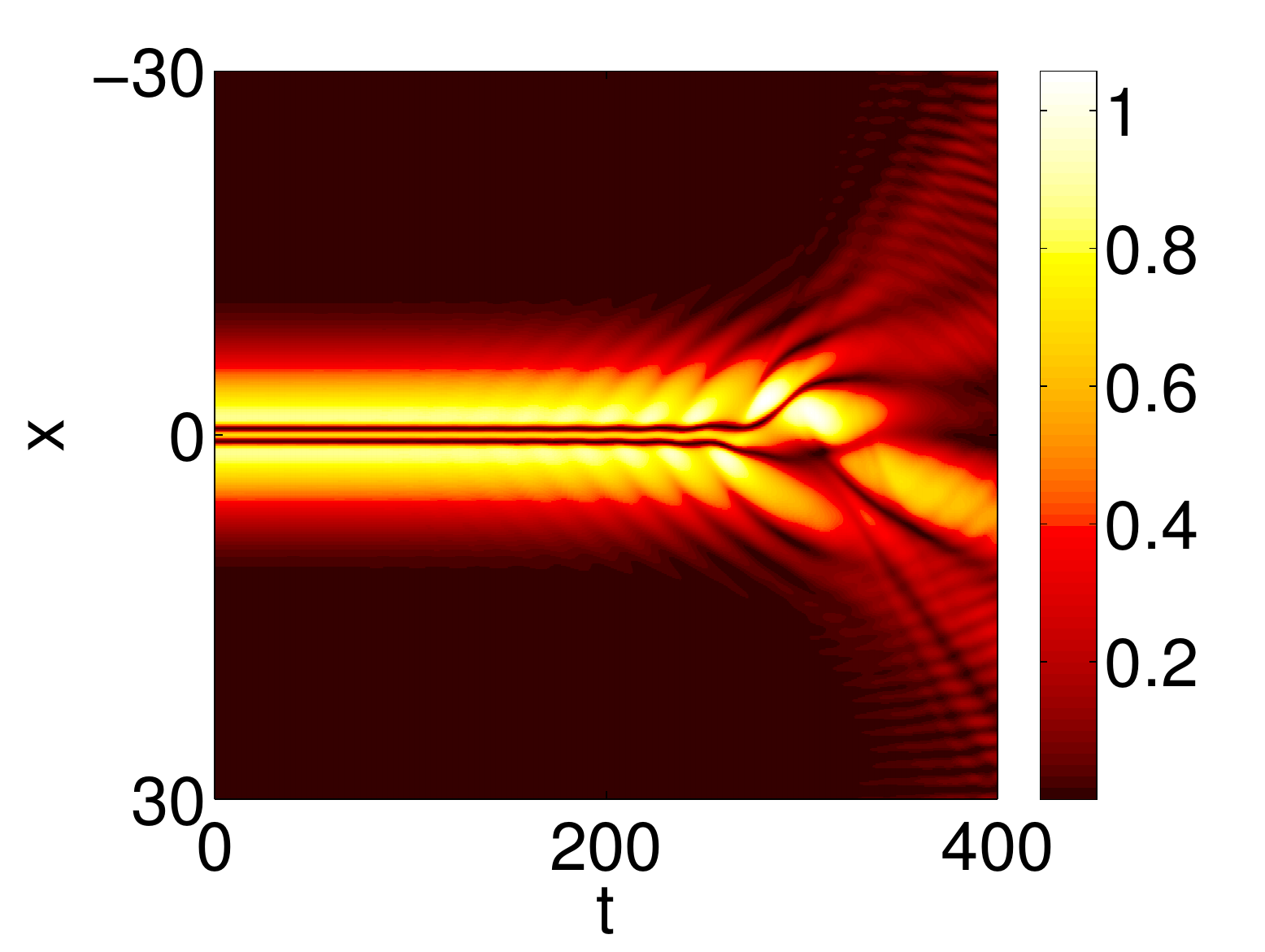}
\label{fig7h}
}
\subfigure[][]{\hspace{-0.3cm}
\includegraphics[height=.18\textheight, angle =0]{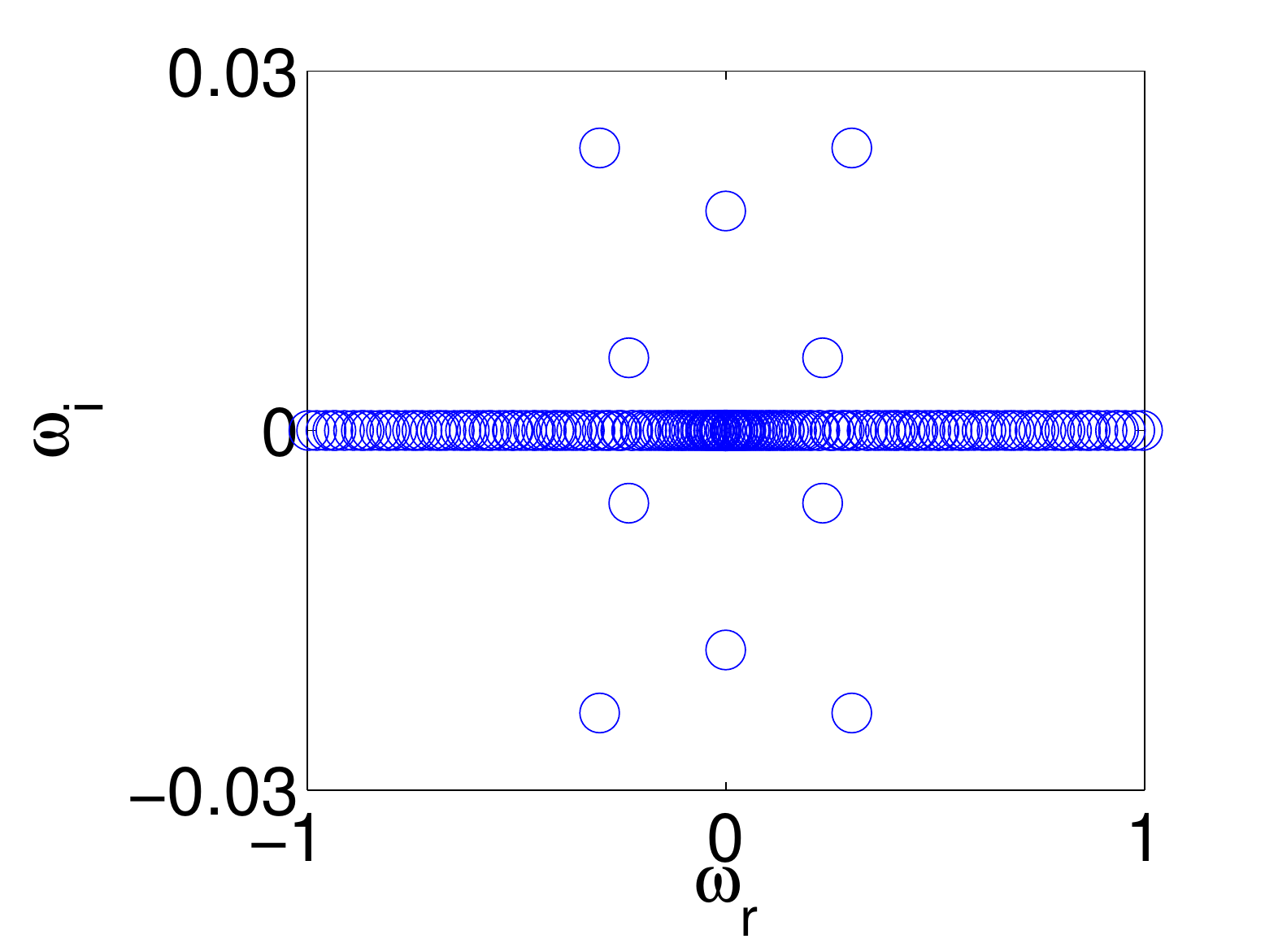}
\label{fig7i}
}
}
\mbox{\hspace{-0.1cm}
\subfigure[][]{\hspace{-0.3cm}
\includegraphics[height=.18\textheight, angle =0]{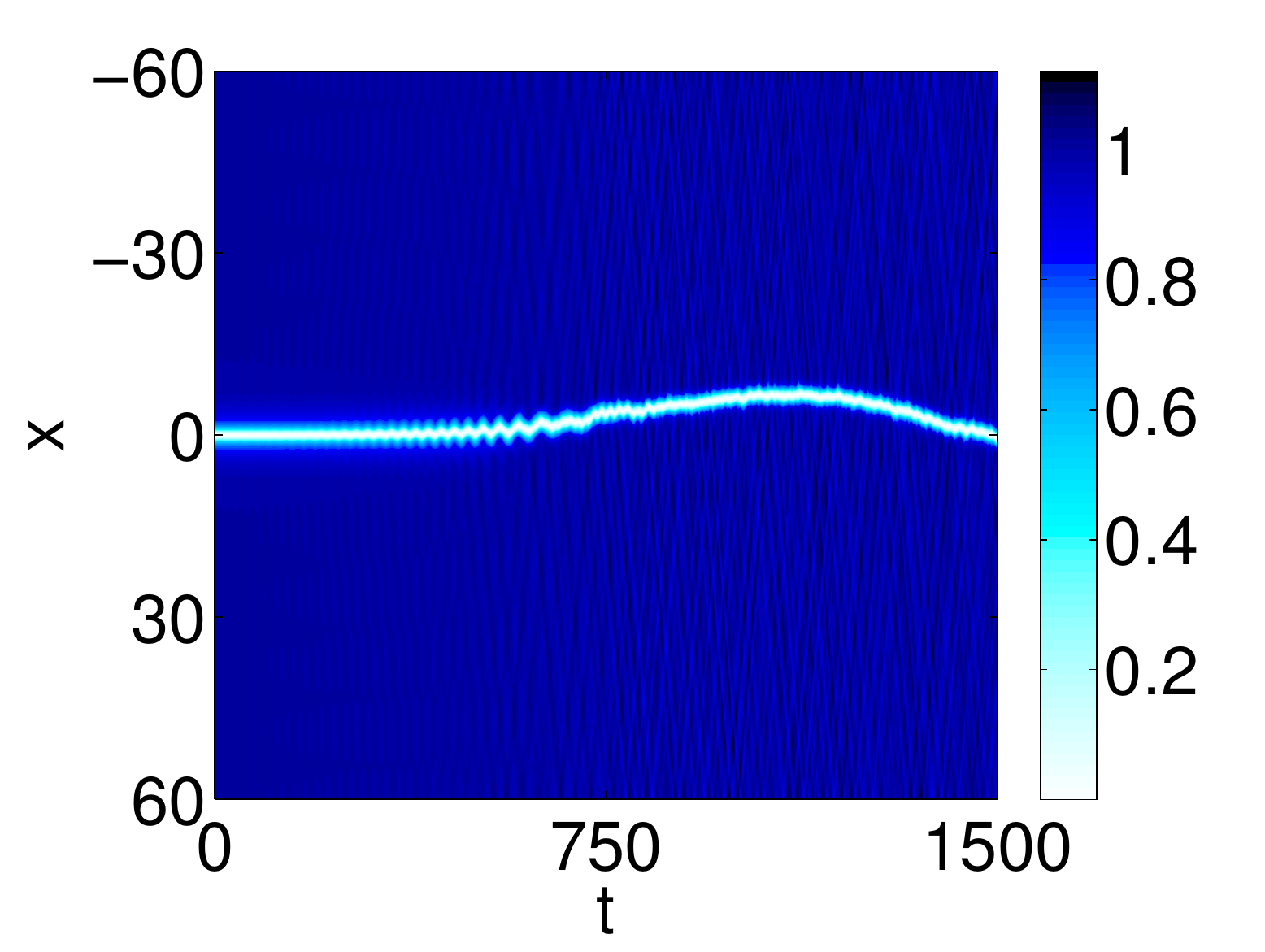}
\label{fig7j}
}
\subfigure[][]{\hspace{-0.3cm}
\includegraphics[height=.18\textheight, angle =0]{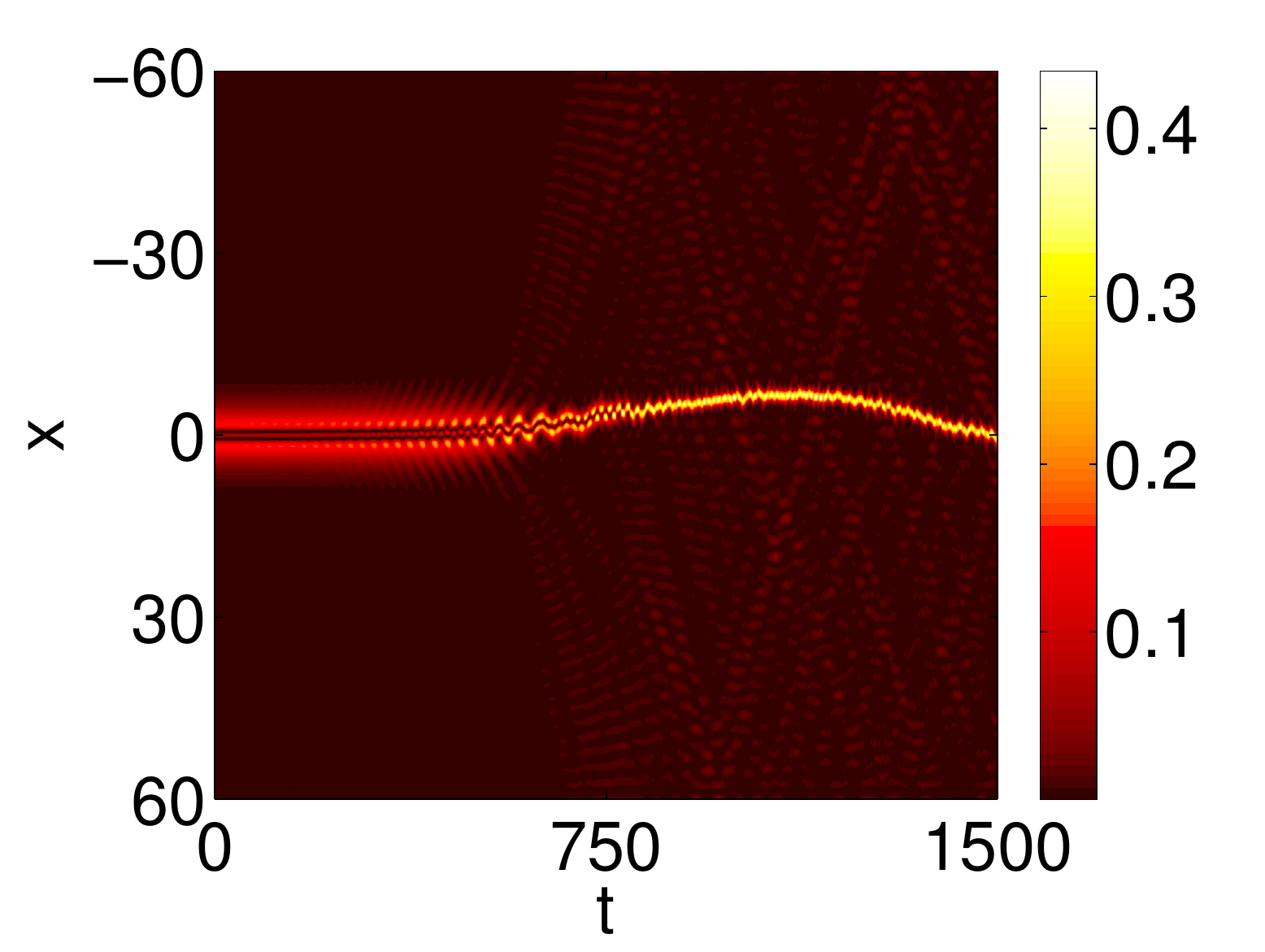}
\label{fig7k}
}
\subfigure[][]{\hspace{-0.3cm}
\includegraphics[height=.18\textheight, angle =0]{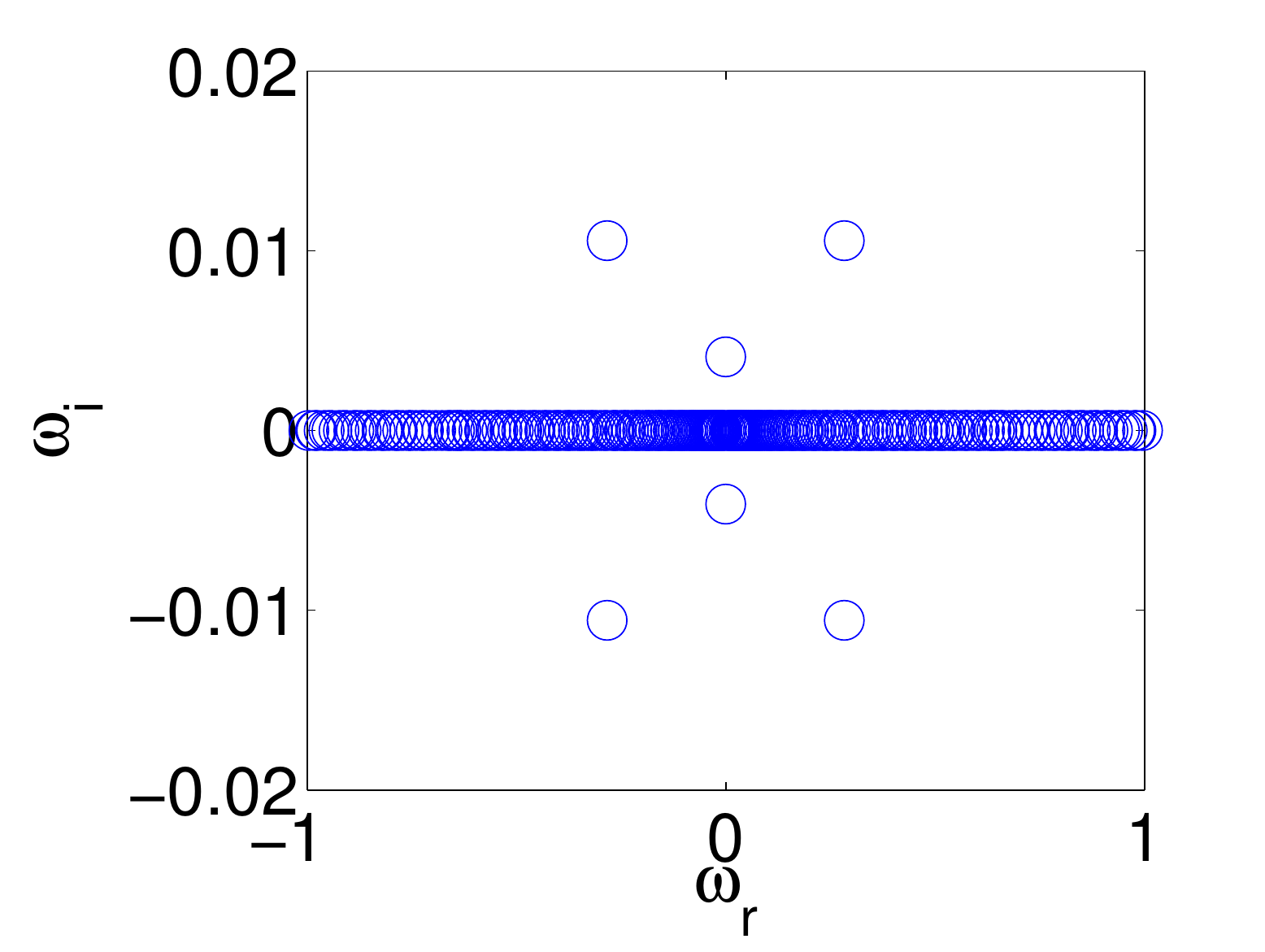}
\label{fig7l}
}
}
\end{center}
\par
\vspace{-0.7cm}
\caption{(Color online) Same as in Fig.~\protect\ref{fig5}, but for soliton 
solutions of order $n=2$: spatiotemporal evolution of densities $|\Phi _{-}(x,t)|^{2}$ 
(a), (d), (g), and (j), and $|\Phi_{+}(x,t)|^{2}$ (b), (e), (h), and (k). 
Panels (c), (f), (i), and (l) display the eigenfrequency spectra of the steady-state 
solutions for (a)-(c) $D=0.2$ and $\protect\mu _{+}=0.9655$, (d)-(f) $D=0.1$ and $\protect\mu _{+}=0.86$,
(g)-(i) $D=0.1$ and $\protect\mu _{+}=0.99$, and (j)-(l) $D=0.25$ and $\protect\mu _{+}=0.99$.}
\label{fig7}
\end{figure}

\begin{figure}[tph]
\begin{center}
\vspace{-0.7cm}
\mbox{\hspace{-0.1cm}
\subfigure[][]{\hspace{-0.3cm}
\includegraphics[height=.18\textheight, angle =0]{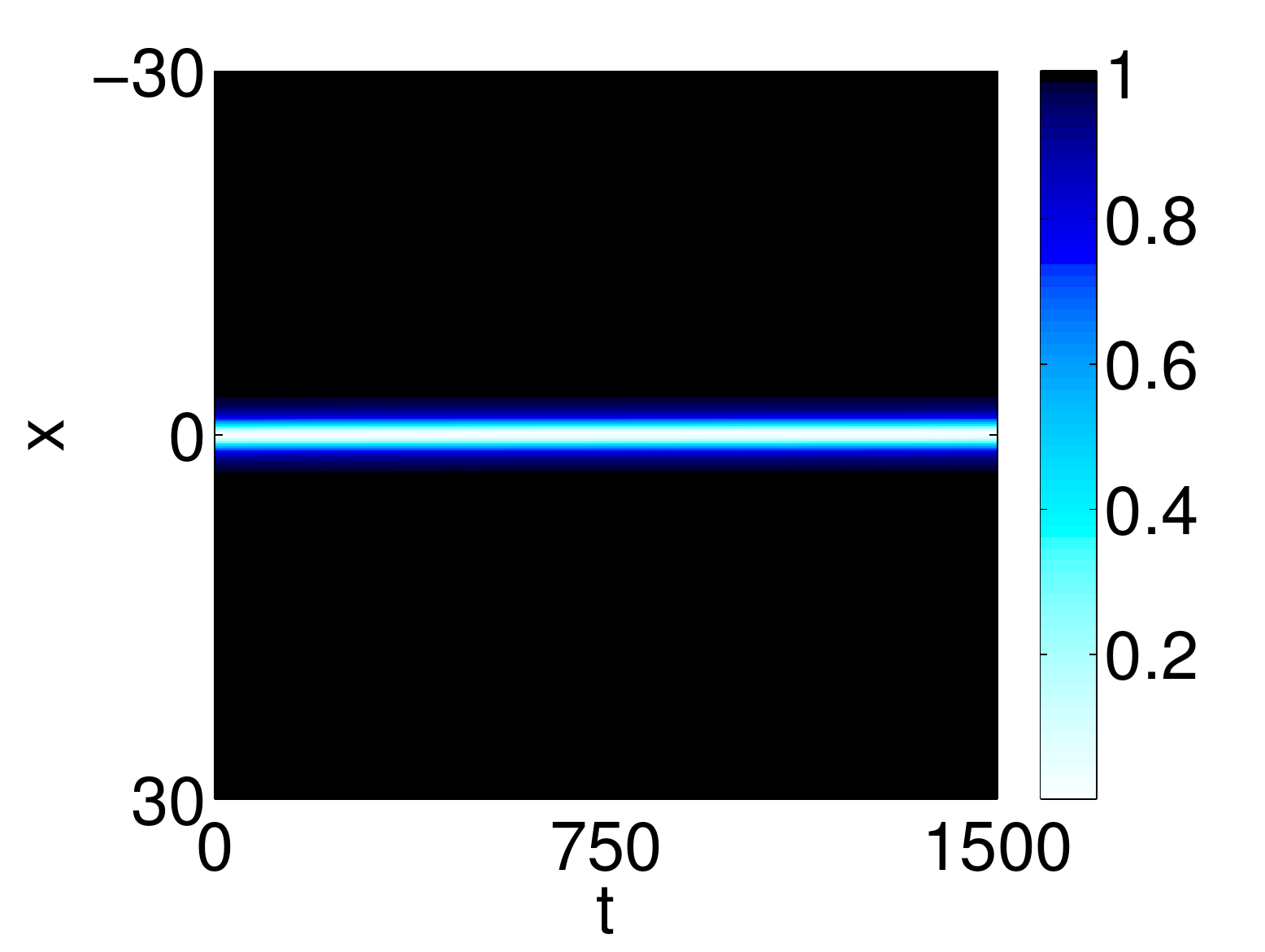}
\label{fig8a}
}
\subfigure[][]{\hspace{-0.3cm}
\includegraphics[height=.18\textheight, angle =0]{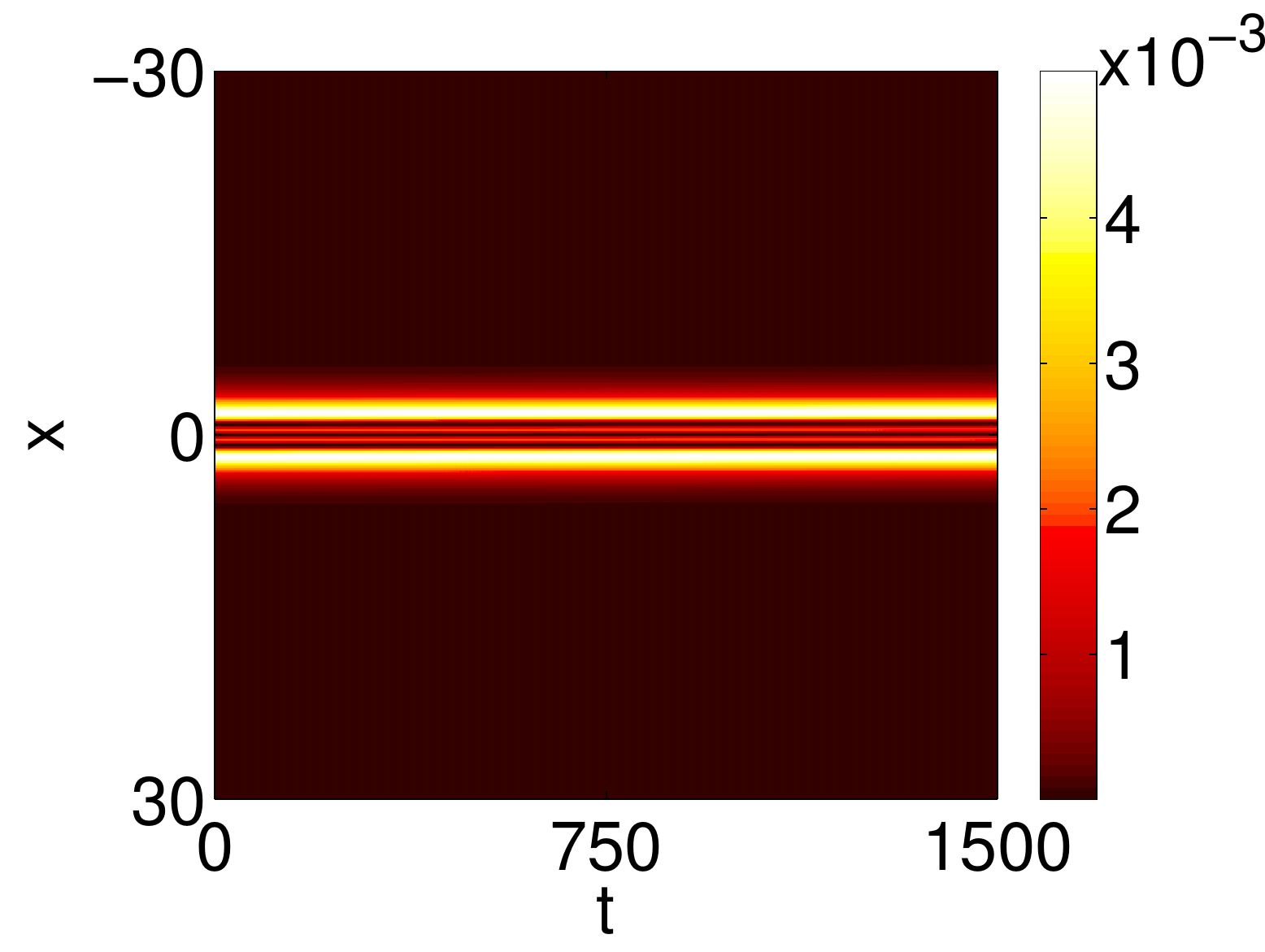}
\label{fig8b}
}
\subfigure[][]{\hspace{-0.3cm}
\includegraphics[height=.18\textheight, angle =0]{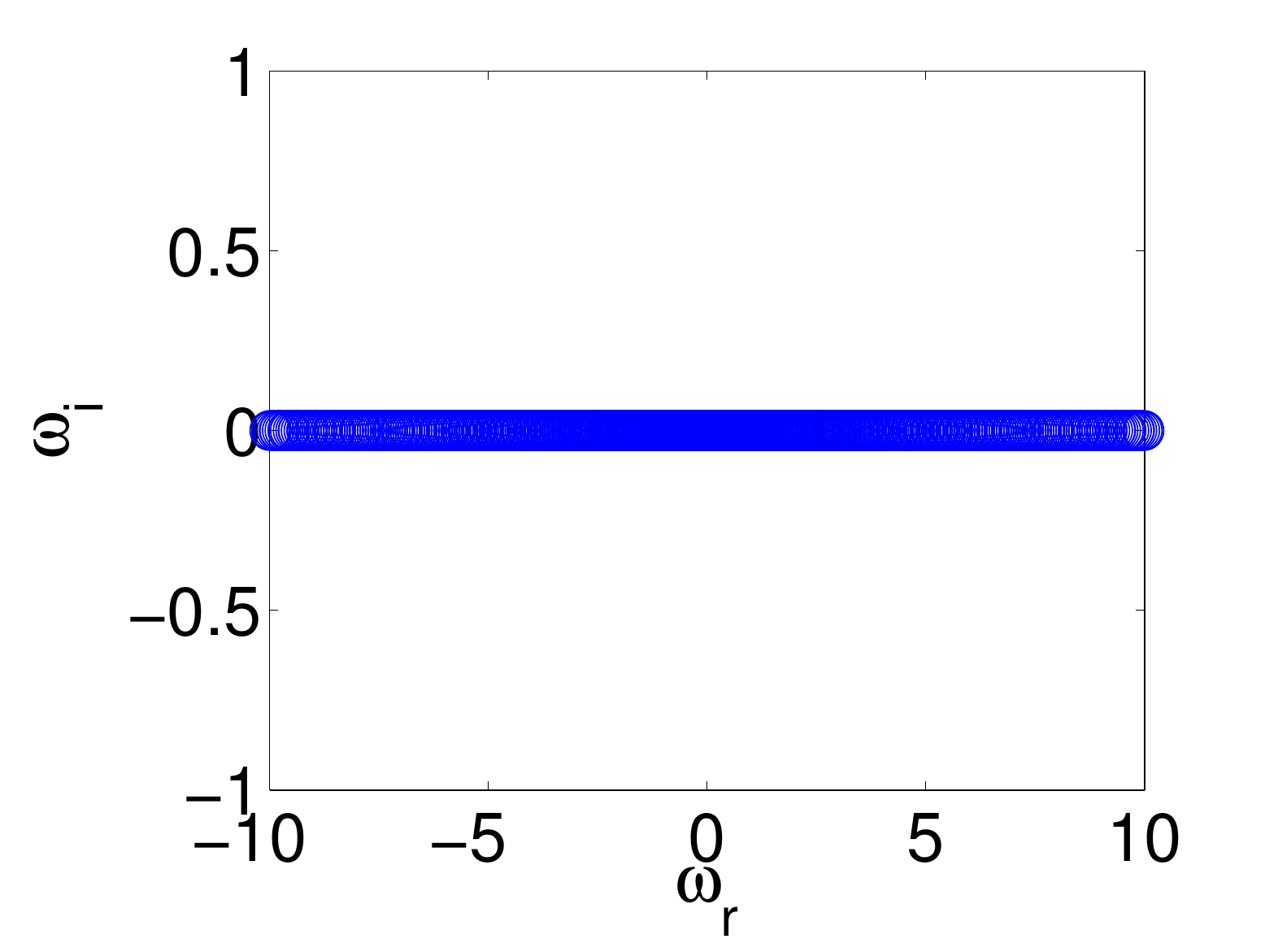}
\label{fig8c}
}
} \vspace{-0.20cm}
\mbox{\hspace{-0.1cm}
\subfigure[][]{\hspace{-0.3cm}
\includegraphics[height=.18\textheight, angle =0]{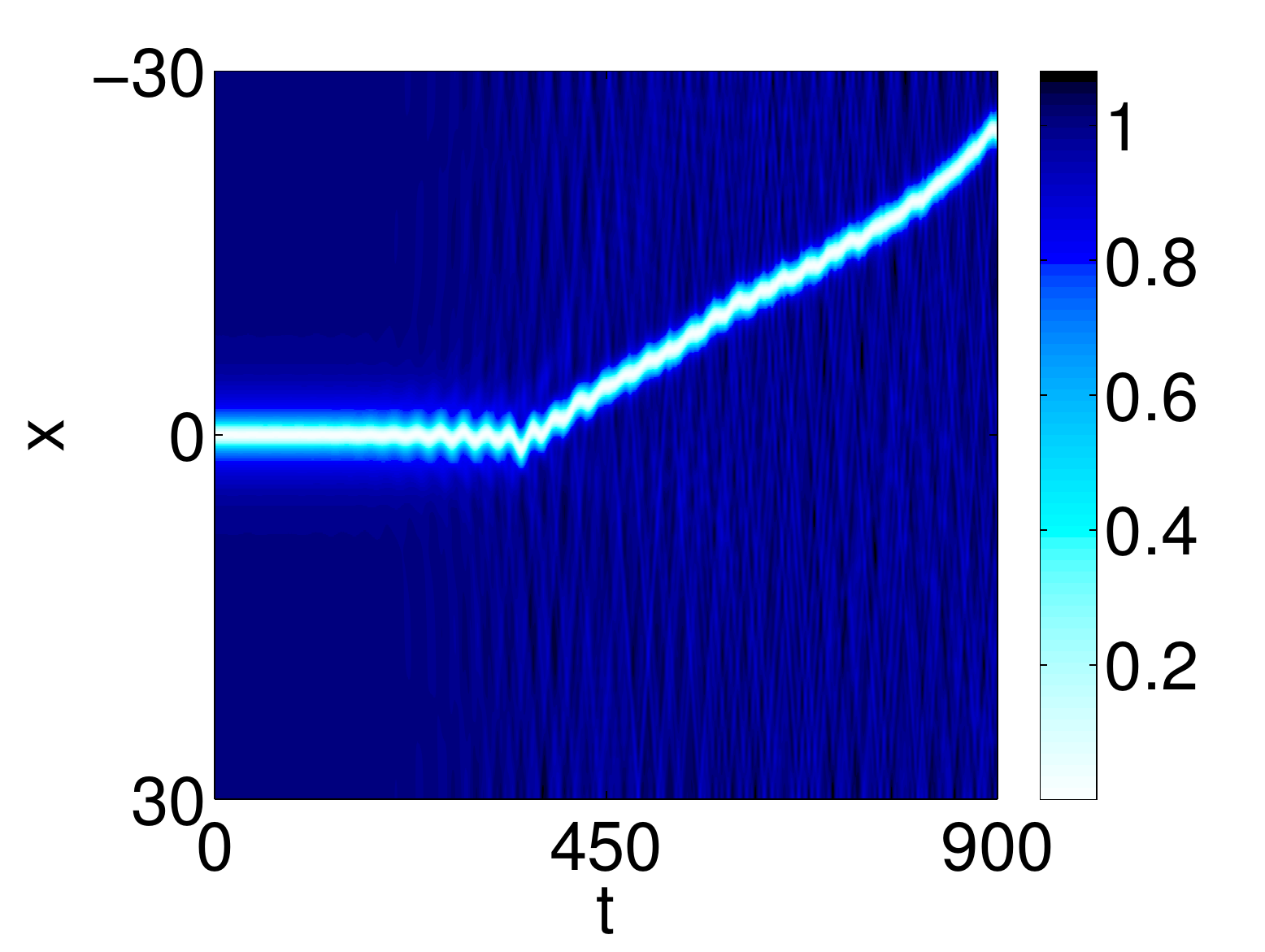}
\label{fig8d}
}
\subfigure[][]{\hspace{-0.3cm}
\includegraphics[height=.18\textheight, angle =0]{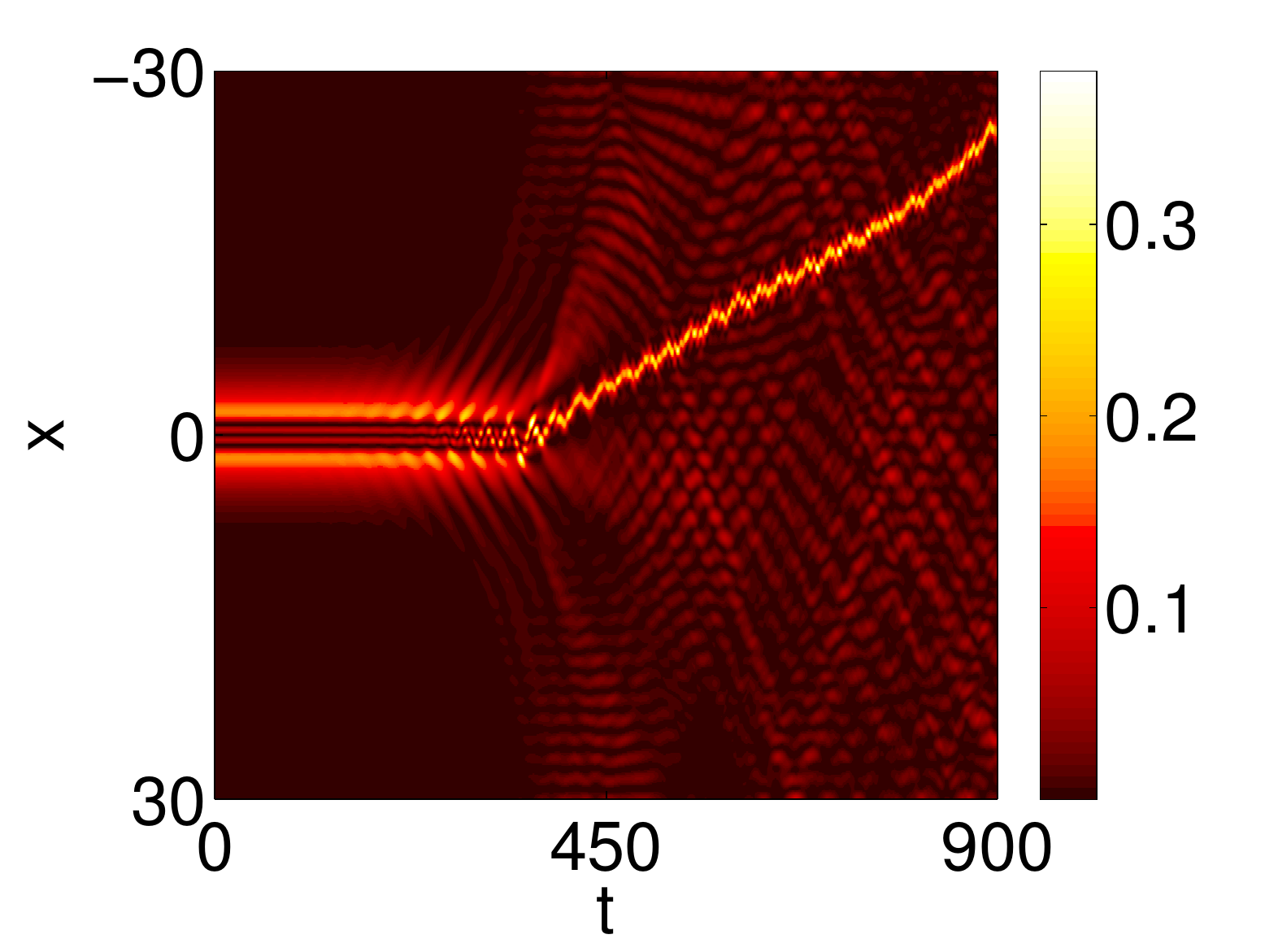}
\label{fig8e}
}
\subfigure[][]{\hspace{-0.3cm}
\includegraphics[height=.18\textheight, angle =0]{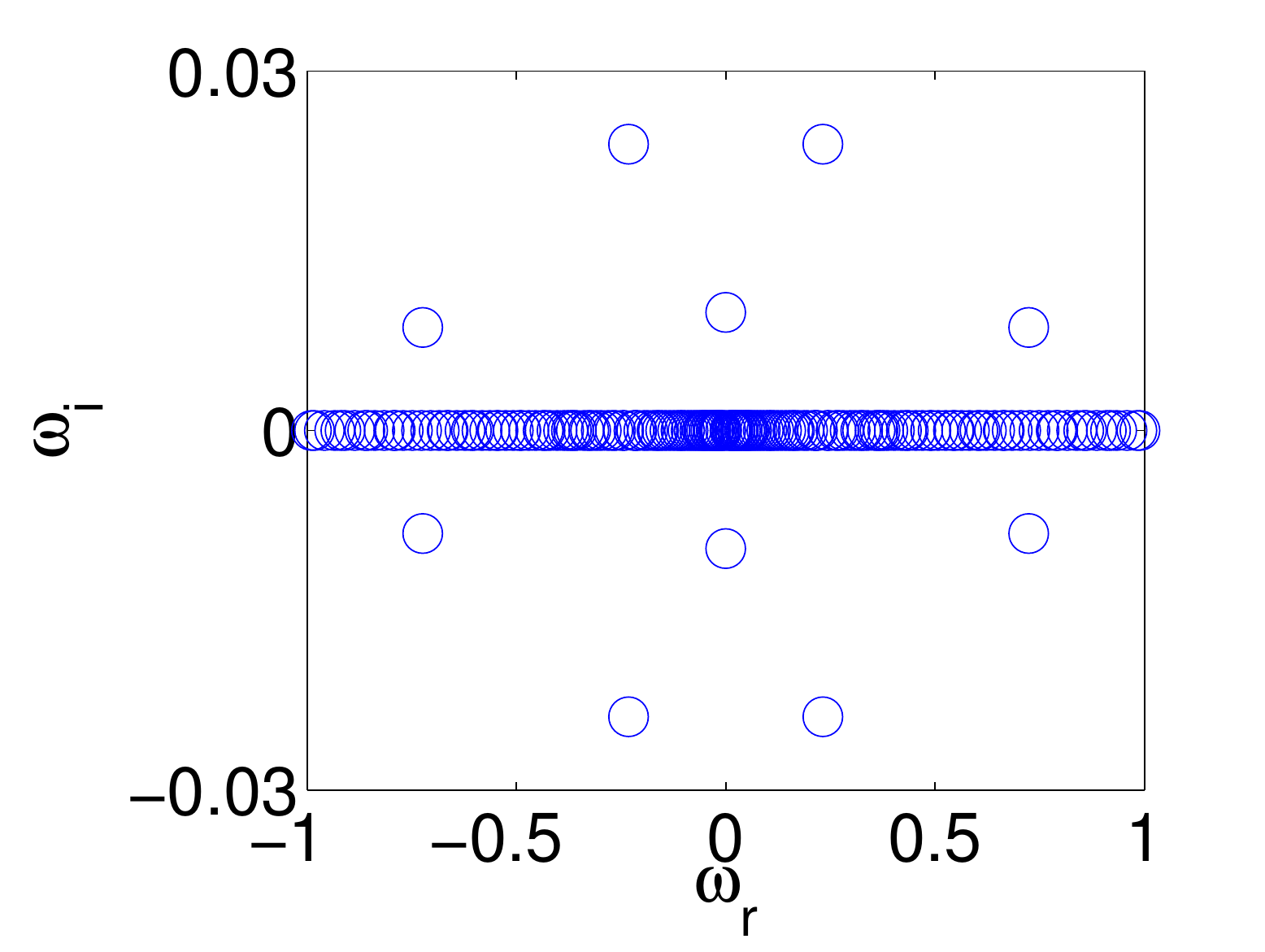}
\label{fig8f}
}
} \vspace{-0.20cm}
\mbox{\hspace{-0.1cm}
\subfigure[][]{\hspace{-0.3cm}
\includegraphics[height=.18\textheight, angle =0]{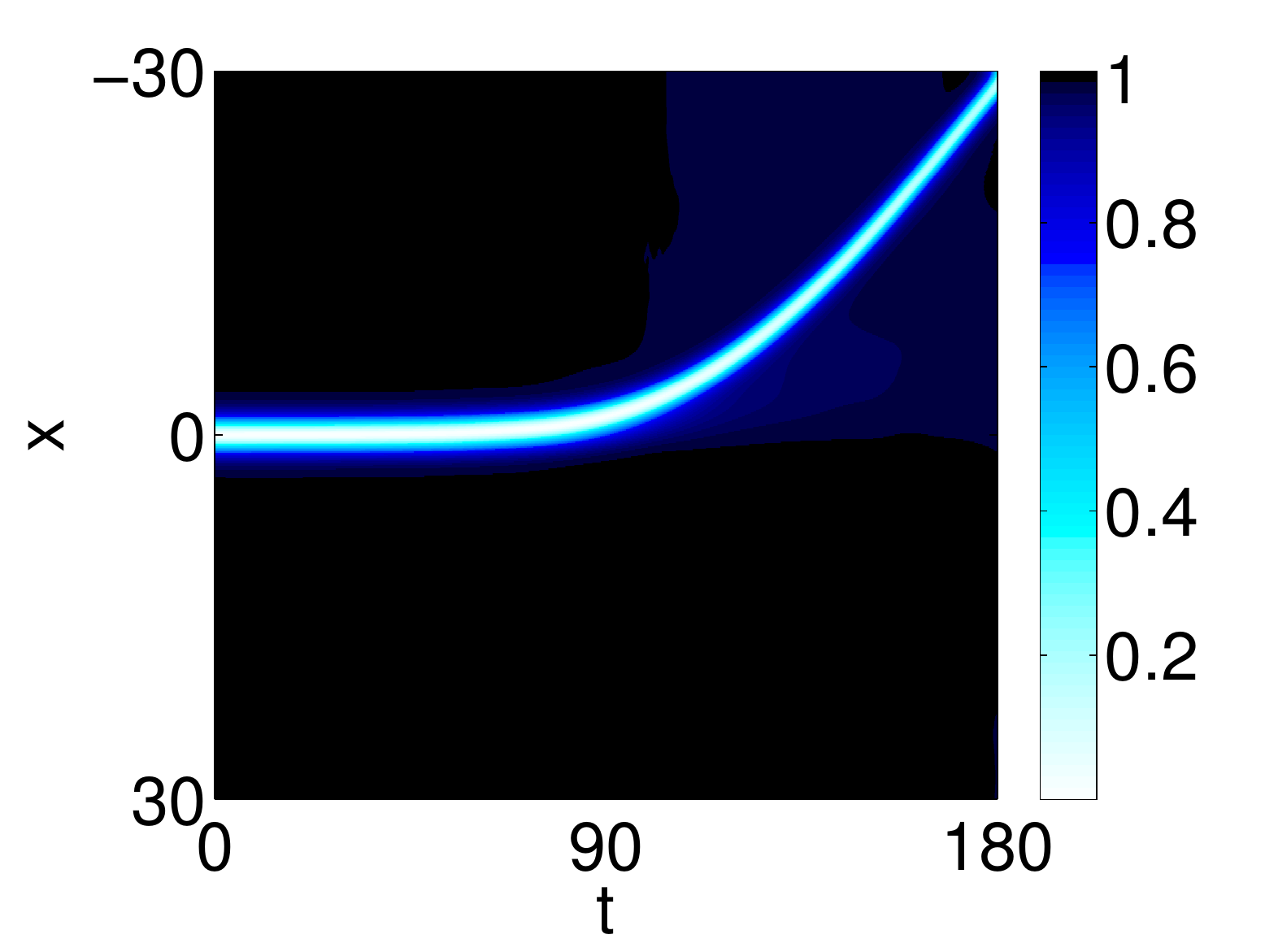}
\label{fig8g}
}
\subfigure[][]{\hspace{-0.3cm}
\includegraphics[height=.18\textheight, angle =0]{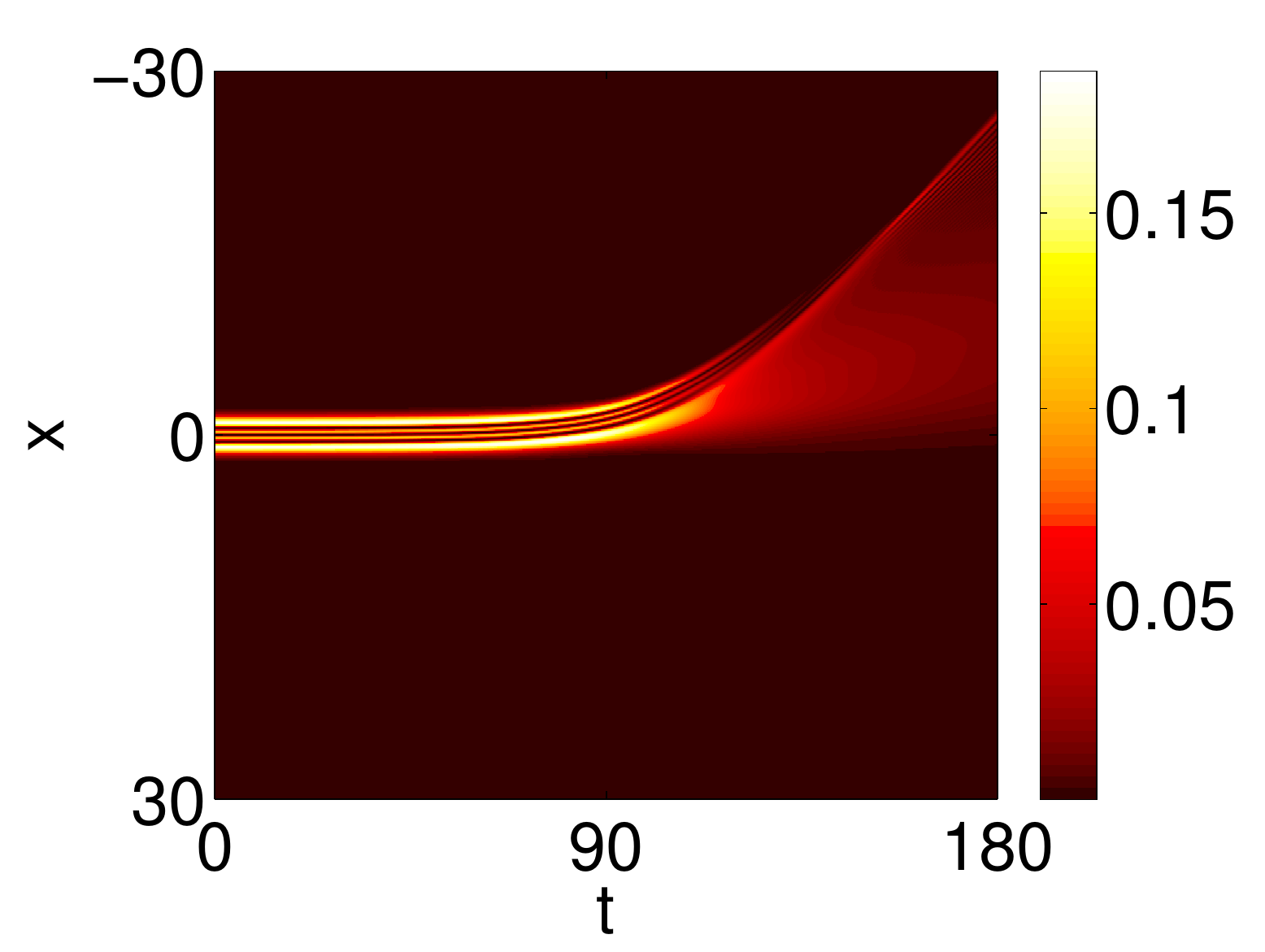}
\label{fig8h}
}
\subfigure[][]{\hspace{-0.3cm}
\includegraphics[height=.18\textheight, angle =0]{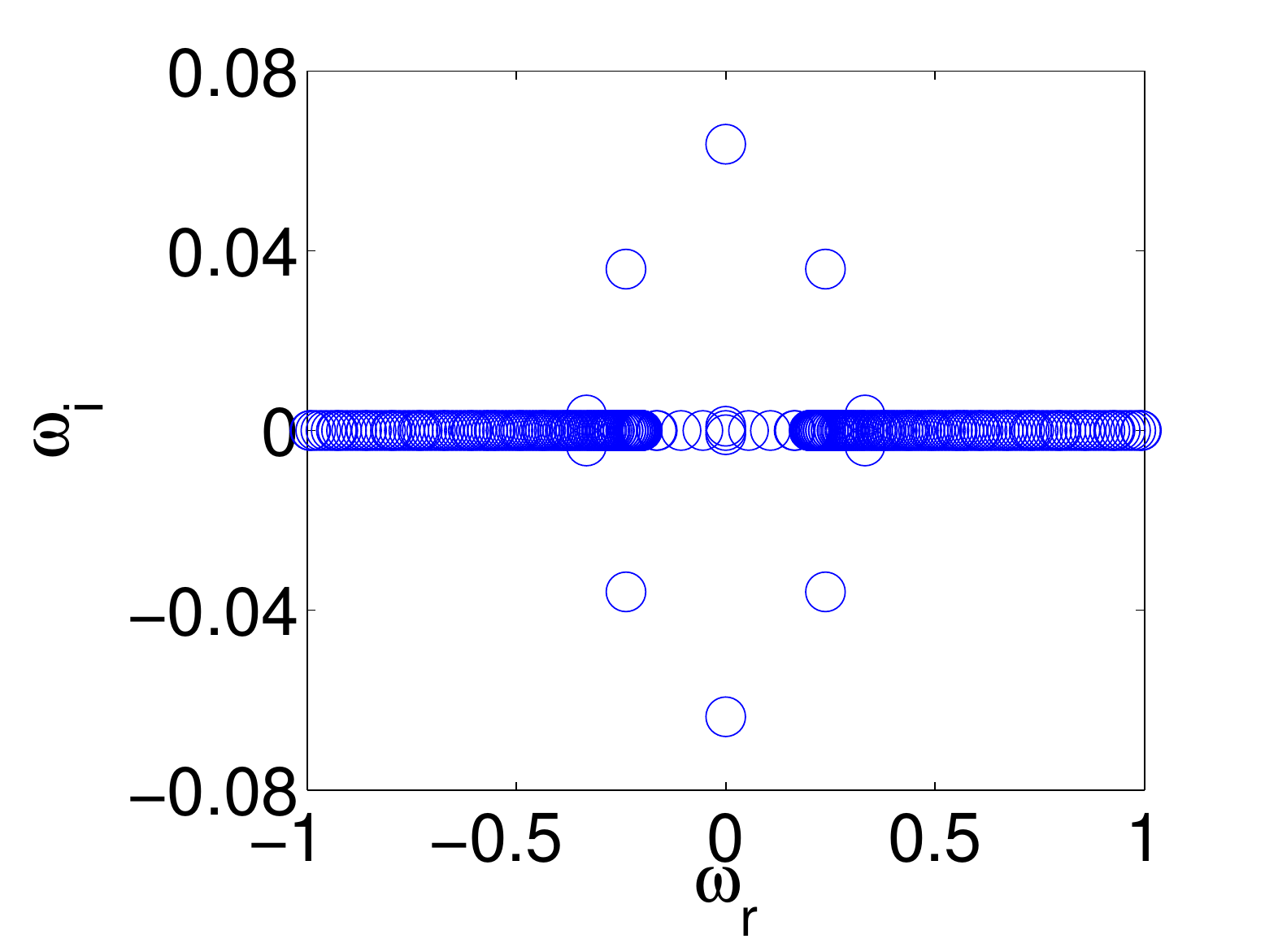}
\label{fig8i}
}
} \vspace{-0.20cm}
\mbox{\hspace{-0.1cm}
\subfigure[][]{\hspace{-0.3cm}
\includegraphics[height=.18\textheight, angle =0]{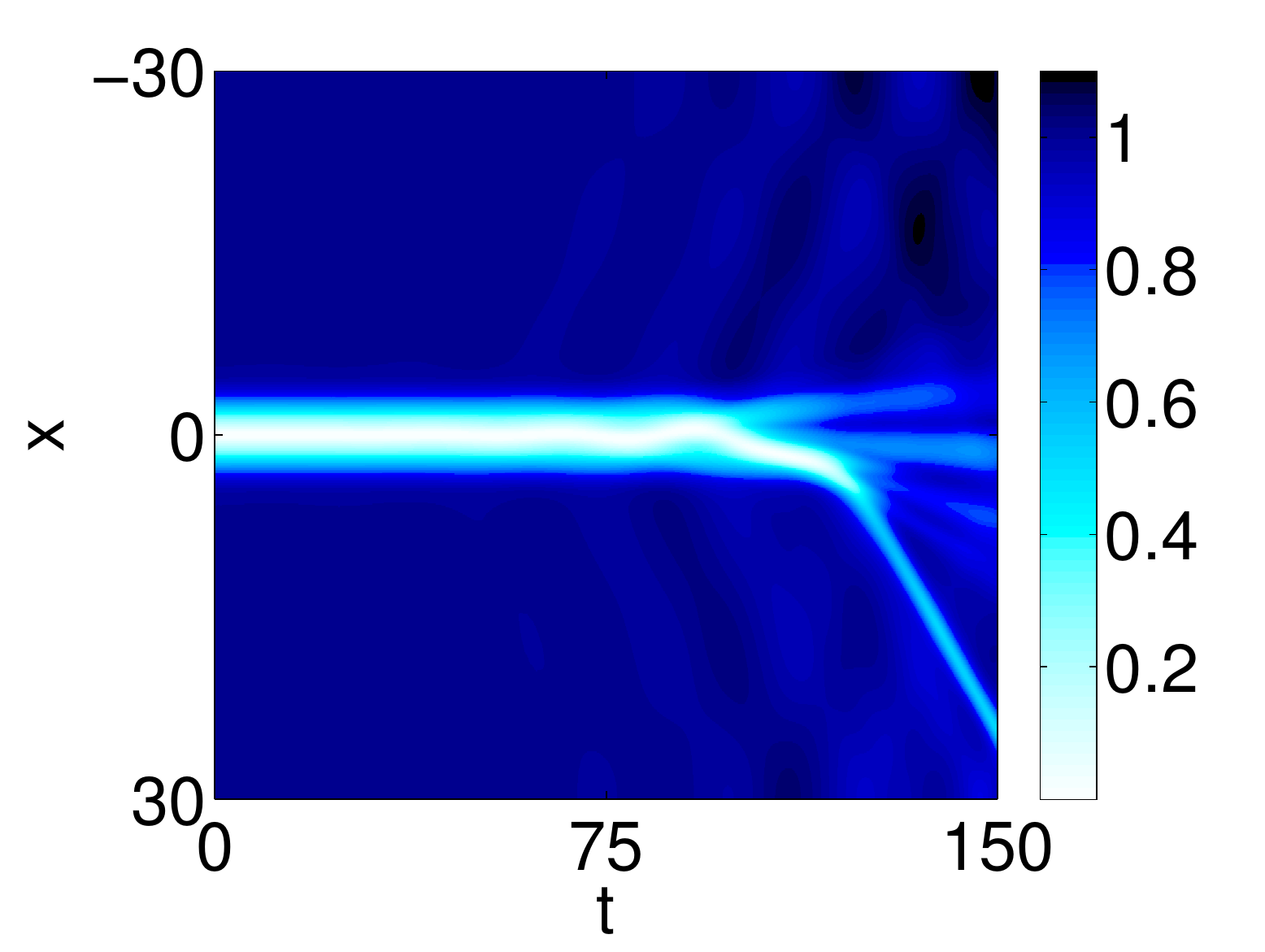}
\label{fig8j}
}
\subfigure[][]{\hspace{-0.3cm}
\includegraphics[height=.18\textheight, angle =0]{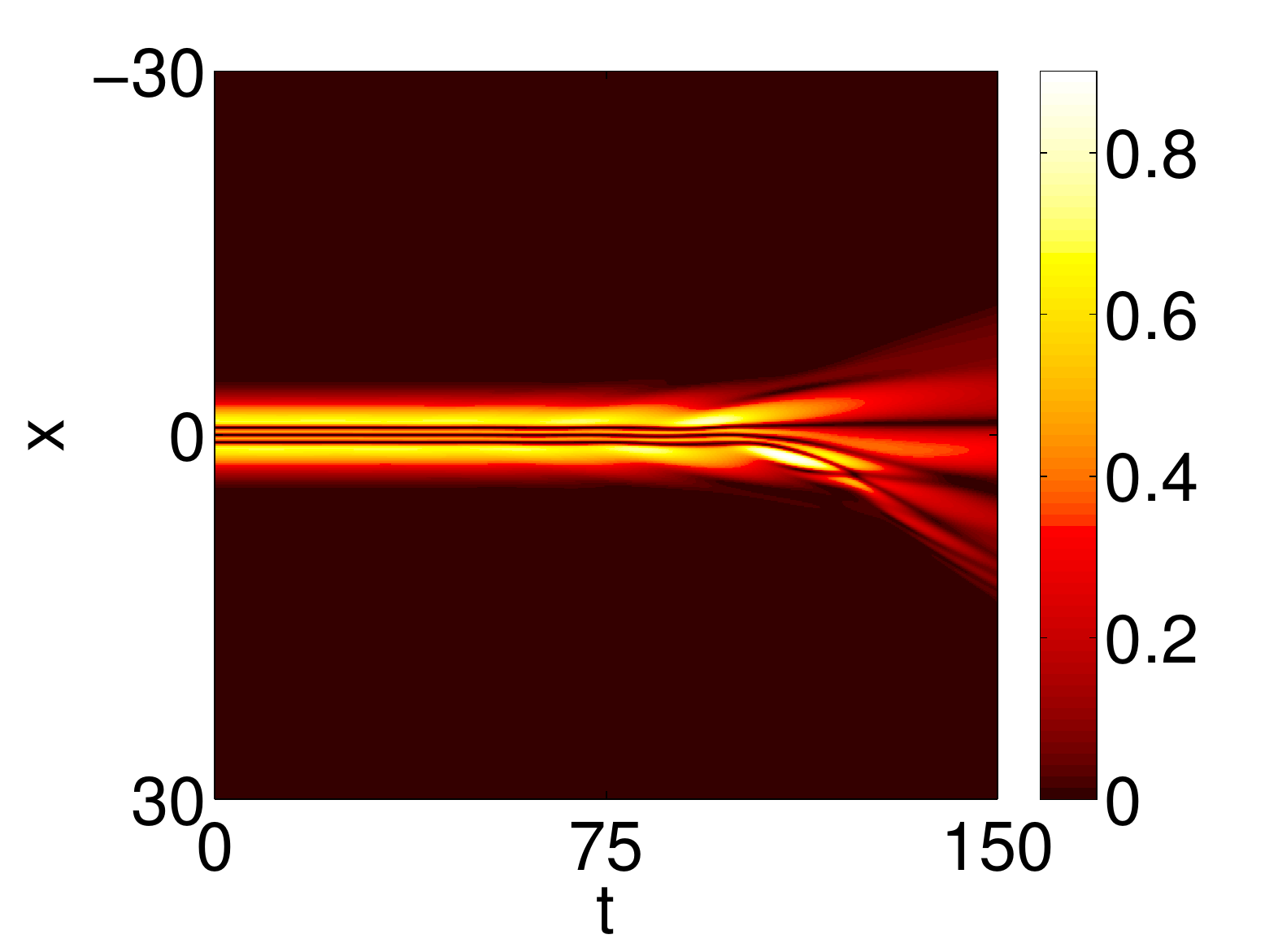}
\label{fig8k}
}
\subfigure[][]{\hspace{-0.3cm}
\includegraphics[height=.18\textheight, angle =0]{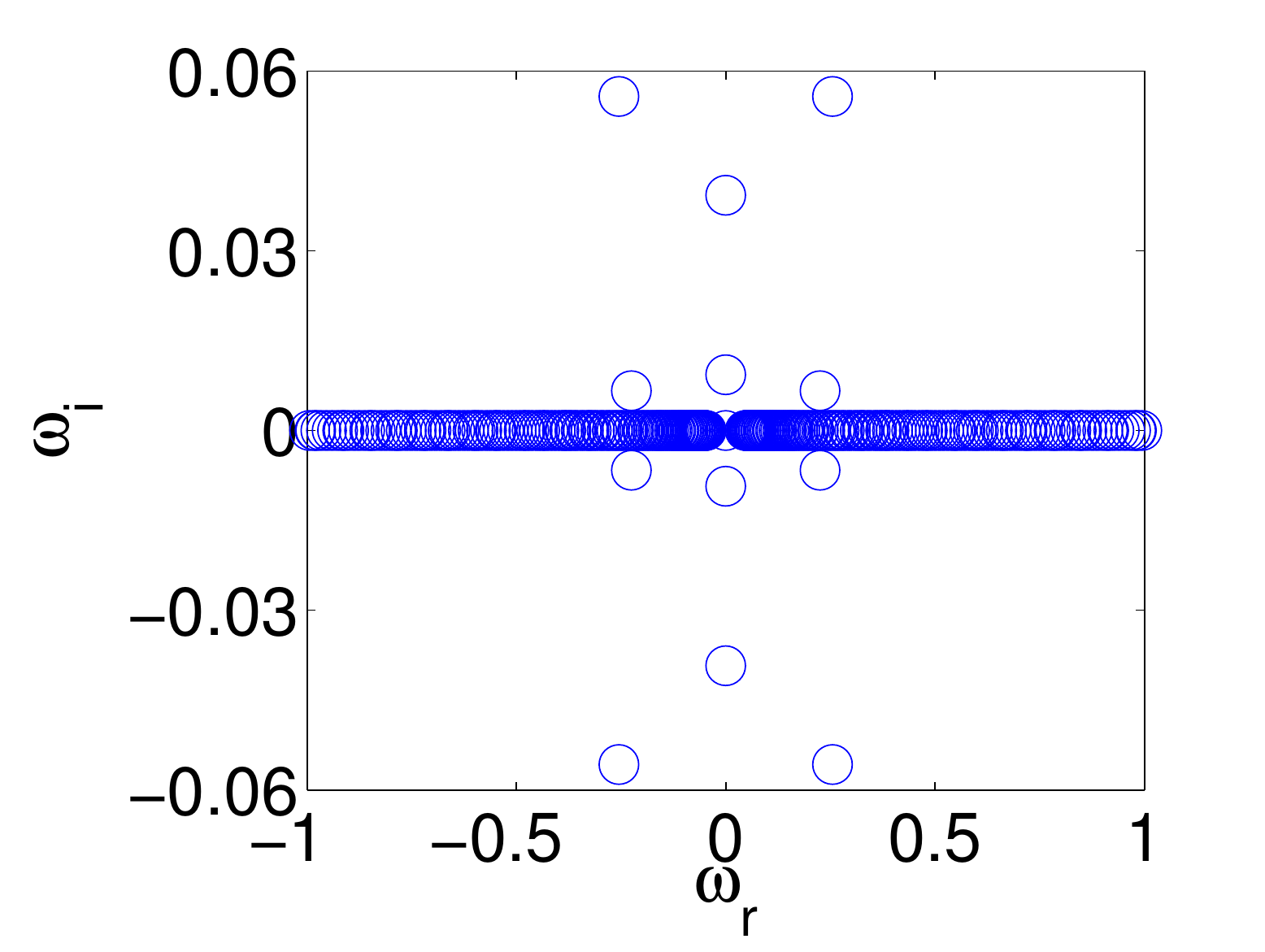}
\label{fig8l}
}
} \vspace{-0.20cm}
\mbox{\hspace{-0.1cm}
\subfigure[][]{\hspace{-0.3cm}
\includegraphics[height=.18\textheight, angle =0]{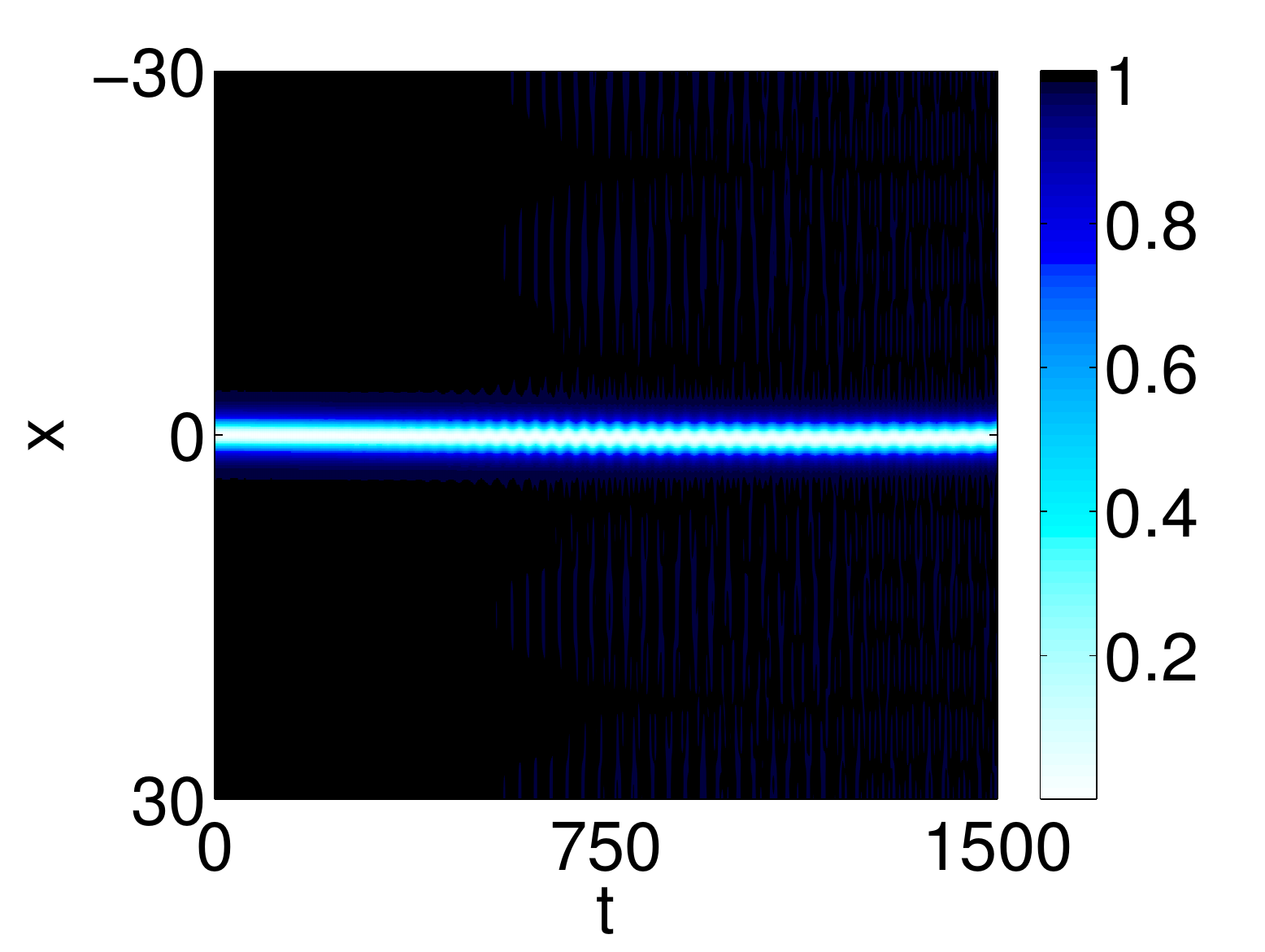}
\label{fig8m}
}
\subfigure[][]{\hspace{-0.3cm}
\includegraphics[height=.18\textheight, angle =0]{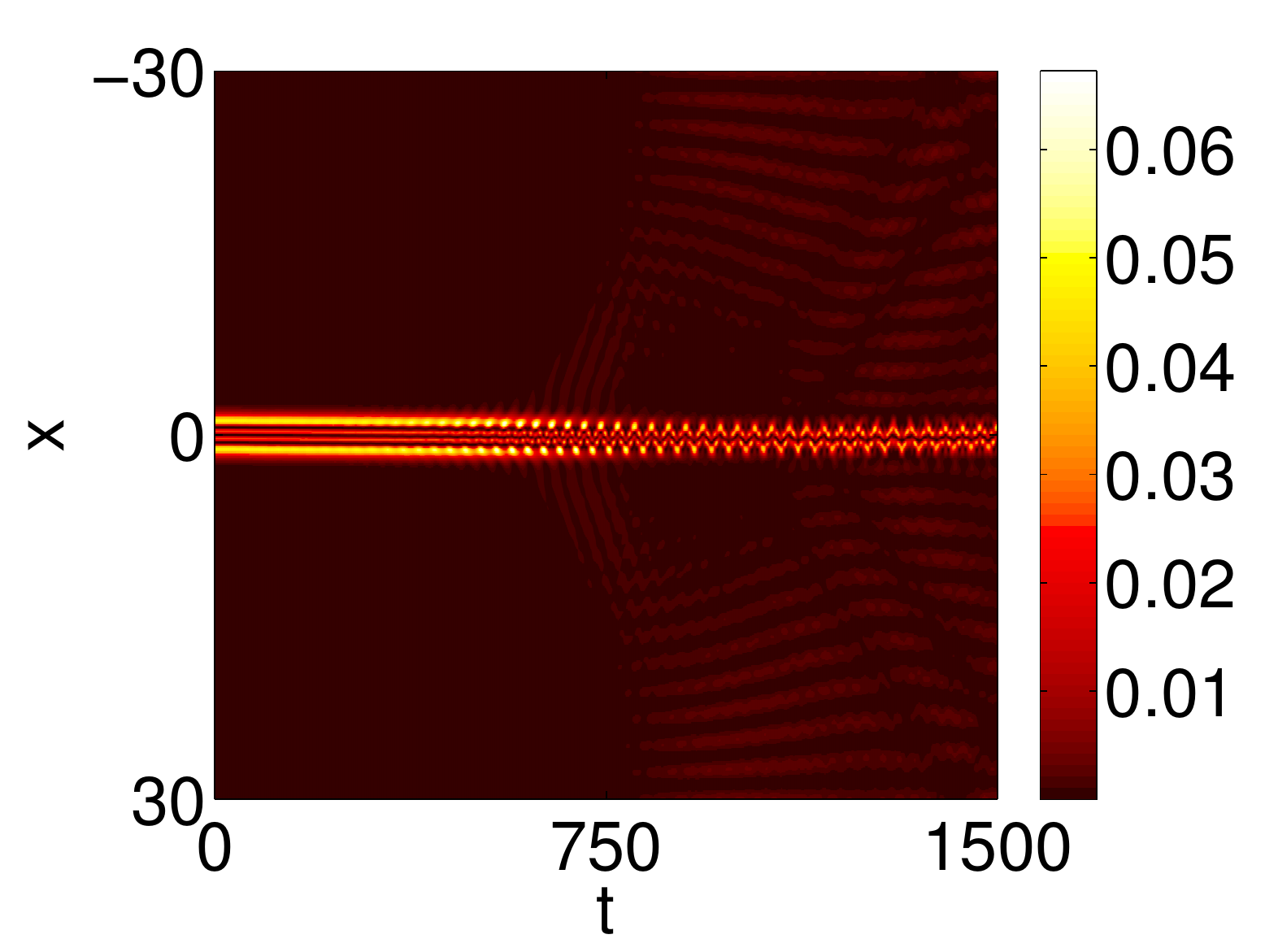}
\label{fig8n}
}
\subfigure[][]{\hspace{-0.3cm}
\includegraphics[height=.18\textheight, angle =0]{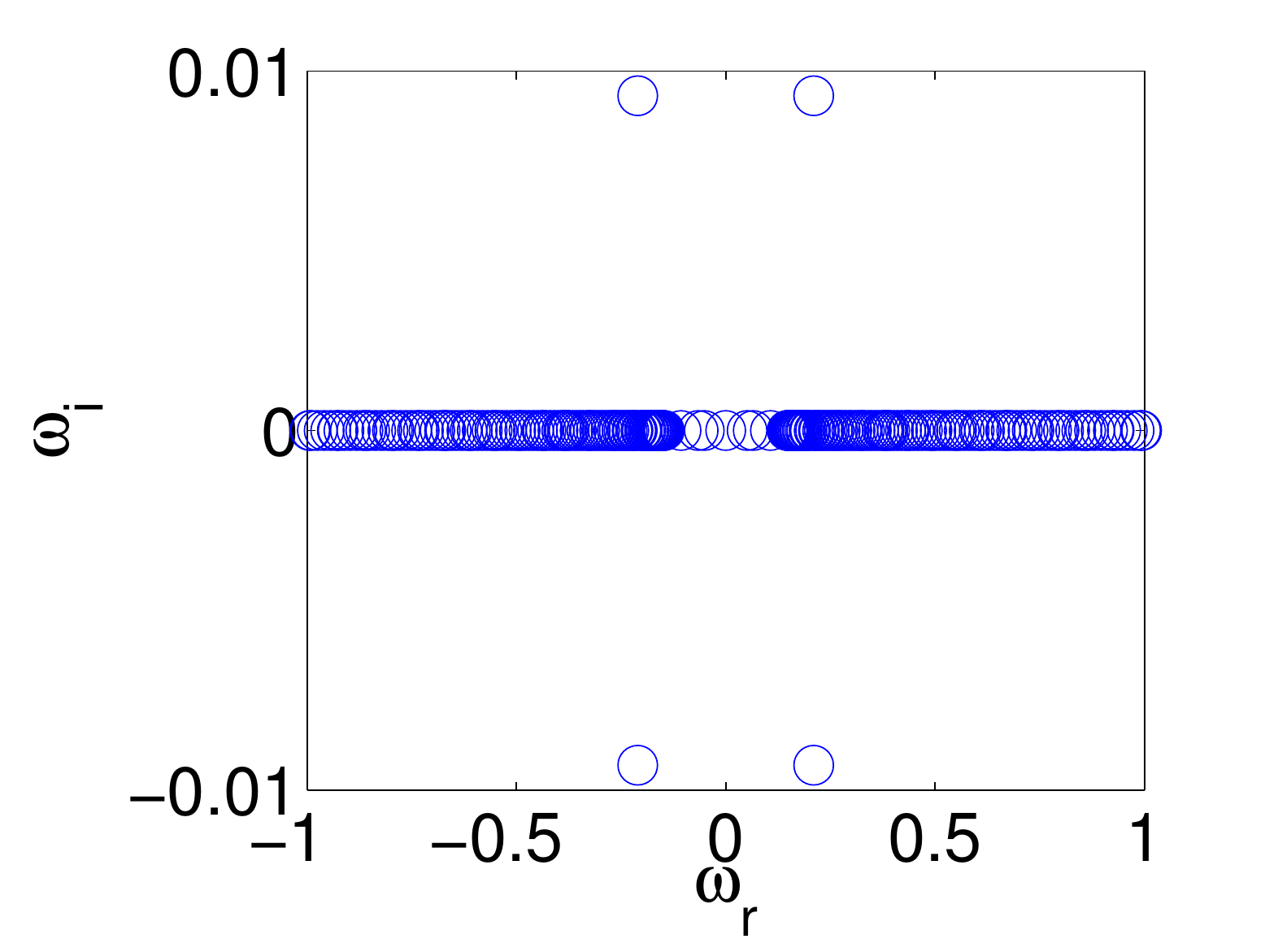}
\label{fig8o}
}
}
\end{center}
\vspace{-0.6cm}
\caption{(Color online) Same as in Fig.~\protect\ref{fig5}, but for soliton
solutions of order $n=3$ and (a)-(c) $D=0.12$ and $\mu_{+}=0.977$, (d)-(f)
$D=0.12$ and $\mu_{+}=0.99$, (g)-(i) $D=0.04$ and $\mu_{+}=0.8$, (j)-(l) $D=0.04$ 
and $\mu_{+}=0.95$, and (m)-(o) $D=0.06$ and $\mu_{+}=0.851$.}
\label{fig8}
\end{figure}

Finally, we present results on the evolution of perturbed steady-state
solutions of orders $n=0,1,2,3$ (for various values of $D$ and $\mu _{+}$) 
in Figs.~\ref{fig5}-\ref{fig8}, respectively, while results corresponding 
to the bound state of order $n=0$ (with $D=0.2$) in the presence of the 
trapping potential are presented in Figs.~\ref{fig5c_trap}-\ref{fig5e_trap} 
($\mu_{+}=0.36$) and \ref{fig5f_trap}-\ref{fig5h_trap} ($\mu _{+}=0.8$). 
In particular, Figs.~\ref{fig5c_trap} and \ref{fig5f_trap} and Figs.~\ref{fig5d_trap}
and \ref{fig5g_trap} correspond to the spatiotemporal evolution of the dark and 
bright components, respectively, while the corresponding eigenfrequency spectra
of perturbations around the steady states (for which the evolution is examined)
are presented in Figs.~\ref{fig5e_trap} and \ref{fig5h_trap}. For the stable 
steady states at hand [see Figs.~\ref{fig1a}-\ref{fig4a}], the corresponding 
dynamical evolution of the (a) dark and (b) bright components is depicted in 
the top rows of Figs.~\ref{fig5}-\ref{fig8}, respectively. In addition, the 
dynamical evolution of stable steady states in the presence of a harmonic trap
is presented in Figs.~\ref{fig5c_trap}-\ref{fig5e_trap} [see, in particular,  
Figs.~\ref{fig5c_trap} and \ref{fig5d_trap}]. 
Clearly, the stable solutions are indeed persistent, in the presence of small
random perturbations, within the time range of the simulations.

In contrast, a number of different scenarios are observed for unstable
solutions, depending upon the corresponding dominant unstable eigenmode, 
as predicted by computations of the eigenfrequencies. Specifically, 
perturbations applied along the unstable eigendirection corresponding to
an exponential eigenmode typically lead to solitons' mobility. This is the
case in Figs.~\ref{fig5d}, \ref{fig5e}, \ref{fig6d}, \ref{fig6e}, \ref{fig7d},
\ref{fig7e}, \ref{fig8g}, and \ref{fig8h} where motion of the solitons is 
observed. While for the fundamental branch this type of mobility may be
persistent, for the higher excited states the acceleration induced by the 
instability eventually leads to a breakdown of the solution (an apparent 
merging of the dark-in-bright solitons therein~\cite{njp_yan}), after a 
sufficiently long time has elapsed. In the presence of the trapping 
potential, in Fig.~\ref{fig5_trap}, the oscillatory instability displaces 
once again the solitary wave from its equilibrium position.
However, here the large difference of $D$ from $1$ does not allow the resulting
moving DB to oscillate in the trap (as it would at $D$ close to $1$~\cite%
{buschanglin,peter1,peter2}) but rather contributes to its rapid destruction
upon interaction with the background.

A number of additional possibilities emerge when unstable steady states are
perturbed by oscillatory eigenmodes. This leads to oscillatory growth
eventually translating into an apparent jerky motion of the corresponding
dark and bright components. This behavior is presented in Figs.~\ref{fig6g}
and \ref{fig6h} (for $n=1$) [here it is clear that the instability wipes out the
initial dark-in-bright solitary wave, transforming the bright component into
a fundamental single-peak mode that is breathing in time]. Note also Figs.~%
\ref{fig7g} and \ref{fig7h} (for $n=2$), where an explosion breaks apart the
entire solitary wave. In Figs. \ref{fig7j} and \ref{fig7k} (i.e., for $n=2$), we
observe a progression through the instability from a two-node solution in
the bright component to a single-node one, and, eventually, to a fundamental
state that is again breathing in time. In Figs.~\ref{fig8d} and \ref{fig8e}, 
a rapid destruction of the $n=3$ state occurs again, this time directly 
transforming it into a fundamental traveling-wave structure. In Figs. \ref{fig8j} 
and \ref{fig8k}, a more complex scenario arises, with the dark soliton splitting
off into apparently gray ones, the fastest of which is not accompanied by a
bright counterpart. As a result, the bright component disperses, maintaining, 
however, some of its nodal structure. Finally, in Figs. \ref{fig8m} and \ref{fig8n} 
(once again, for $n=3$), the third excited state (the second spatially odd one) 
transforms itself into the corresponding first excited state maintaining, again, 
breathing oscillations.

It is worth noting that, for the fundamental branch, there is at most a 
translational (imaginary) eigenfrequency responsible for the instability.
It is thus rather natural that its manifestation in direct simulations
involves mobility. However, as we progressively move to higher excited
states, the number of potentially unstable modes increases, creating an
oscillatory instability for $n=1$, two instabilities for $n=2$, and so on, in accord
with Figs.~\ref{fig6}, ~\ref{fig7}, etc. It is the intricate interplay of
these distinct dynamical instabilities (often with comparable growth rates)
that is responsible for the resulting complex dynamics.

\section{Conclusions}


In the present work we have revisited the model based on two-component
nonlinear Schr{\"{o}}dinger equations with defocusing nonlinearity, which
plays a fundamental role in nonlinear optics, as well as in the description
of repulsively interacting binary BEC mixtures. We have examined the 
fundamental dynamical features of dark-bright solitons, namely, the formation
of the effective potential well for the bright component induced by the dark 
one via the XPM interaction. Using the dispersion coefficient of the bright 
component as a control parameter, we modified the depth of the effective 
potential well, enabling the formation of higher-order excited bound states 
in this well, including those with $n=1,2$, and $3$ nodes. These may be considered 
as dark-in-bright solitary structures~\cite{njp_yan} or as excited states 
trapped in the potential well. We have shown that, while the ground single-peak
state is generally very robust (except for the case of large difference between
the dispersion coefficients of the two components), this is not true for the 
excited states, which are subject to progressively wider intervals of both
exponential and oscillatory instabilities. The instability of the
ground state leads to motion of the DB soliton, but does not destroy it. For
the excited states with progressively increasing $n$, the complexity of the
evolution scenarios also increases, resulting from the interplay of the
increasing number of instability modes. Exotic scenarios involve the fusion
of the dark-in-bright solitary waves, the explosion of the waveforms into
multiple splinters, and relaxation, either abruptly or gradually, into less
excited states, possibly accompanied by breathing.

The present analysis suggests a number of paths for future studies. One
possibility might be to expand on the exact solutions identified herein for $%
D=0$ beyond this special limit, using a perturbative expansion to explore
both their existence and, potentially, also their stability in the limit of
small $D$. Moreover, one can extend the present considerations to the quite
important (e.g., in BEC) and widely studied class of spinor systems,
involving more than two components~\cite{ueda_spin}. Following this
possibility, one can envision, in the spirit of Ref.~\cite{DDB}, one dark
component creating a potential for the other two bright components, which
could be found either in the same or, for suitable parametric regimes, 
possibly in different states of their respective potential wells.
This would be a particularly intriguing setup to explore, as concerns its
existence and stability properties. Furthermore, we note that the
possibility of one component forming a well for another one is independent
of the spatial dimension. For instance, in two dimensions the notion of
vortex-bright solitons~\cite{VB,pola} is a by-product of the same type of
potential approach (the topological charge of the vortices is not experienced by
the bright component when the interaction is incoherent, i.e., the
potential well is solely determined by the density distribution in the
vortex). Here it would be interesting to explore what type of excited
states could be created, such as a dark-in-bright ring and associated
multi ring states, \textit{inter alia}. 
\vspace{0.5cm}

\noindent\begin{center}\textbf{ACKNOWLEDGMENTS}\end{center}
E.G.C. gratefully acknowledges financial support from the FP7-People Grant No.
IRSES-605096 and thanks Hans Johnston (UMass Amherst) for providing computing 
facilities. The work of D.J.F. was partially supported by the Special Account 
for Research Grants of the University of Athens. P.G.K. acknowledges support 
from the National Science Foundation under Grants No. CMMI-1000337 and No. 
DMS-1312856, from FP7-People under Grant No. IRSES-605096, and from the U.S. AFOSR 
under Grant No. FA9550-12-10332. The work of P.G.K. and B.A.M. was supported 
in part by the U.S.-Israel Binational Science Foundation through Grant No. 2010239.

\appendix

\section{The linearization ansatz and stability matrix}

\label{app_lin} In this appendix we briefly discuss the linearization
ansatz employed for the investigation of the stability of the stationary
solutions, together with the formulation of the stability matrix. We start
with the perturbation ansatz around stationary solutions $\phi _{\pm
}^{0}(x) $ (which may be complex, in principle)
\begin{subequations}
\begin{eqnarray}
\widetilde{\Phi }_{-} &=&e^{-i\mu _{-}t}\Big\lbrace \phi _{-}^{0}+\varepsilon
\left[ a(x)e^{i\omega t}+b^{\ast }(x)e^{-i\omega ^{\ast }t}\right]\Big\rbrace ,
\\
\widetilde{\Phi }_{+} &=&e^{-i\mu _{+}t}\Big\lbrace \phi _{+}^{0}+\varepsilon
\left[ c(x)e^{i\omega t}+d^{\ast }(x)e^{-i\omega ^{\ast }t}\right]\Big\rbrace,
\end{eqnarray}
\label{lin_ansatz}
\end{subequations}
where $\omega $ is the (complex) eigenfrequency, $\varepsilon $ is a
small amplitude of the perturbation, and the asterisk stands for complex
conjugation. Then we insert Eqs.~(\ref{lin_ansatz}) into Eqs.~(\ref%
{start_manakov}) and thus obtain, at order $\varepsilon$, an eigenvalue 
problem in the following matrix form:

\begin{equation}
\rho
\begin{pmatrix}
a \\
b \\
c \\
d%
\end{pmatrix}%
=%
\begin{pmatrix}
A_{11} & A_{12} & A_{13} & A_{14} \\
-A_{12}^{\ast } & -A_{11} & -A_{14}^{\ast } & -A_{13}^{\ast } \\
A_{13}^{\ast } & A_{14} & A_{33} & A_{34} \\
-A_{14}^{\ast } & -A_{13} & -A_{34}^{\ast } & -A_{33}%
\end{pmatrix}%
\begin{pmatrix}
a \\
b \\
c \\
d%
\end{pmatrix}%
,  \label{eig_prob}
\end{equation}%
with eigenvalues $\rho =-\omega $, eigenvectors $\mathcal{V}=(a,b,c,d)^{T}$,
and matrix elements given by
\begin{subequations}
\begin{eqnarray}
A_{11} &=&-\frac{D_{-}}{2}\frac{\partial ^{2}}{\partial x^{2}}+\gamma \left(
2g_{11}|\phi _{-}^{0}|^{2}+g_{12}|\phi _{+}^{0}|^{2}\right) +V-\mu _{-},
\label{A11} \\
A_{12} &=&\gamma \,g_{11}\,\left( \phi _{-}^{0}\right) ^{2}, \\
A_{13} &=&\gamma \,g_{12}\,\phi _{-}^{0}\left( \phi _{+}^{0}\right) ^{\ast },
\\
A_{14} &=&\gamma \,g_{12}\,\phi _{-}^{0}\phi _{+}^{0}, \\
A_{33} &=&-\frac{D_{+}}{2}\frac{\partial ^{2}}{\partial x^{2}}+\gamma \left(
g_{12}|\phi _{-}^{0}|^{2}+2g_{22}|\phi _{+}^{0}|^{2}\right) +V-\mu _{+},
\label{A33} \\
A_{34} &=&\gamma \,g_{22}\,\left( \phi _{+}^{0}\right) ^{2}.
\end{eqnarray}


\end{subequations}

\end{document}